\documentclass[twocolumn]{aa}

\usepackage{graphicx}
\usepackage{txfonts}
\usepackage{aalongtable}
\usepackage{natbib}
\usepackage{supertabular}
\usepackage{psfig}
\usepackage{mathrsfs}
\usepackage{fontenc}

\begin{document}
\title{Differential rotation in rapidly rotating early-type stars. I. Motivations for 
combined spectroscopic and interferometric studies}

\author{J. Zorec \inst{1}        \and
Y. Fr\'emat \inst{2}             \and
A. Domiciano de Souza \inst{3}   \and
O. Delaa \inst{3}                \and
P. Stee \inst{3}                 \and
D. Mourard \inst{3}              \and
L. Cidale \inst{4,5}\fnmsep\thanks{Member of the Carrera del Investigador 
Cient\'{\i}fico, CONICET, Argentina}        \and
C. Martayan \inst{6}             \and
C. Georgy \inst{7}               \and
S. Ekstr\"om  \inst{7}            
}

\institute{Institut d'Astrophysique de Paris, UMR 7095 du CNRS, Universit\'e 
Pierre \& Marie Curie, 98bis bd. Arago, 75014 Paris, France                     \and
Royal Observatory of Belgium, 3 av. Circulaire, 1180 Brussels, Belgium          \and
Laboratoire Fizeau, UNS-OCA-CNRS UMR6203, Parc Valrose, 06108 Nice Cedex 02, France  \and
Facultad de Ciencias Astron\'omicas y Geof\'{i}sicas, Universidad Nacional de 
La Plata, Paseo del Bosque S/N, La Plata, Buenos Aires, Argentina               \and
Instituto de Astrof\'{i}sica de La Plata, (CCT La Plata - CONICET, UNLP), 
Paseo del Bosque S/N, La Plata, Buenos Aires, Argentina                         \and
European Organization for Astronomical Research in the Southern  Hemisphere, Alonso de Cordova 3107, Vitacura, Santiago de Chile, Chile \and
Observatoire de Gen\`eve, Universit\'e de Gen\`eve, 51 Chemin des Maillettes, CH-1290 Sauverny, Suisse
}    

\offprints{J. Zorec: \email{zorec@iap.fr}}
\date{Received ..., ; Accepted ...,}

\abstract
{Since the external regions of the envelopes of rapidly rotating early-type stars are unstable to convection, a coupling may exist between the convection and the internal rotation.} 
{We explore what can be learned from spectroscopic and interferometric observations about the properties of the rotation law in the external layers of these objects.}
{Using simple relations between the entropy and specific rotational quantities, some of which are found to be efficient at accounting for the solar differential rotation in the convective region, we derived analytical solutions that represent possible differential rotations in the envelope of early-type stars. A surface latitudinal differential rotation may not only be an external imprint of the inner rotation, but induces changes in the stellar geometry, the gravitational darkening, the aspect of spectral line profiles, and the emitted spectral energy distribution.}
{By studying the equation of the surface of stars with non-conservative rotation laws, we conclude that objects undergo geometrical deformations that are a function of the latitudinal differential rotation able to be scrutinized both spectroscopically and by interferometry. The combination of Fourier analysis of spectral lines with model atmospheres provides independent estimates of the surface latitudinal differential rotation and the inclination angle. Models of stars at different evolutionary stages rotating with internal conservative rotation laws were calculated to show that the Roche approximation can be safely used to account for the gravitational potential. The surface temperature gradient in rapid rotators induce an acceleration to the surface angular velocity. Although a non-zero differential rotation parameter may indicate that the rotation is neither rigid nor shellular underneath the stellar surface, still further information, perhaps non-radial pulsations, is needed to determine its characteristics as a function of depth.}
{} 
\keywords{Stars: early-type; Stars: rotation; Stars: spectroscopy; Stars: interferometry}
\titlerunning{Spectroscopy and interferometry of early-type differential rapid rotators} 
\authorrunning{Zorec et al.}

\maketitle

\section{Introduction}
\label{intr}
\subsection{Review of observational approaches}
\label{rooa}

 One of the most enduring unknowns in stellar physics has been the inner distribution of the angular momentum in a star. In the past few decades, significant progress has been made in describing theoretically the evolution of rotating stars. This has required an understanding of numerous hydrodynamic and magnetic instabilities triggered by the rotation, as well as the mixing processes of chemical elements unleashed by these instabilities \citep{tass78,tass00,zahn83,
zahn92,iau215,mae09}. However, apart from the Sun, reliable observational information about the internal rotation of stars remains scarce or non-existent.\par
 Nevertheless, many attempts have been made to obtain information on the internal rotation from detailed studies of: a) the position of stars in the HR diagram; b) the evolution of the $V\!\sin i$ parameter during the main sequence (MS) phase; c) the shape of absorption lines, whose characteristics can depend upon the rotational law in layers close to the stellar surface; d) the global stellar geometry described with interferometric data.\par  
 We briefly review these efforts:\par
\medskip
  {\bf a}) The most numerous efforts among those just mentioned are the statistical analysis of photometric data on the rotational spread of the MS, which can be described by \citep{rox66,mae68,coll77,coll91}

\begin{equation}
\label{mvrot}
\Delta M_{\rm V} = k(n)V^n,
\end{equation}

\noindent where $\Delta M_{\rm V}$ is the deviation in absolute magnitude from the zero-rotation MS, $V$ is the true equatorial rotational velocity, and $k(n)$ is a constant whose value depends on the power $n$. When $n\!=\!2$,  $k(n\!=\!2)$ is on the order of $k_o10^{-5}$ mag/(km~s$^{-1})^2$, so that for $k_o\!\lesssim\!1$ the deviations may indicate that the internal rotation is uniform, while for $k_o\!\gtrsim\!1$  the internal rotation can be differential \citep{cott83}. This type of analysis found that stars do not seem to rotate uniformly. However, owing to the measurement uncertainties and difficulties in defining the MS of zero rotation, the available data could not provide any firm evidence of a particular law of non-uniform rotation \citep{strsar66,golay68,mae68,maepe70,smi71,smiwo74,mosmi81}. Furthermore, using detailed model atmospheres for differentially rotating stars, \citet{coll85} concluded that photometry alone can place albeit rather weak constraints on the degree of differential rotation within the stars.\par
\medskip
 {\bf b}) Depending upon the internal angular momentum redistribution and evolutionary
rearrangements of the inertial momentum, the surface equatorial rotational velocity of stars changes accordingly. Thus, the study of the variation in the true rotational velocity, $V$, as a function of time was studied by several authors using the ratio 

\begin{equation}
\label{vsinlc}
R_{\rm LC} = \frac{\langle V\!\sin i\rangle_{\rm LC}}{\langle V\!\sin i\rangle_{\rm ZAMS}} = \frac{\langle V\rangle_{\rm LC}}{\langle V\rangle_{\rm ZAMS}},
\end{equation}

\noindent where $\langle V\!\sin i\rangle_{\rm LC}$ is the average of the $V\!\sin i$ parameters of stars with in principle the same mass and luminosity class (LC), $\langle V\!\sin i\rangle_{\rm ZAMS}$ is the average of $V\!\sin i$ for stars with the same mass, but  located near the zero-age-main sequence (ZAMS). These ratios were compared with similar ones predicted theoretically for stars evolving as rotators in two different and extreme ways. On the one hand, the stars were assumed to evolve all their way as uniform rotators, which implies that the angular momentum is entirely redistributed at each evolutionary step. On the other hand, it was assumed that each stellar layer conserved its initial specific angular momentum, i.e., the stars did not undergo any redistribution of its internal angular momentum. Since in many cases the observed ratios $R_{\rm LC}$ were found to be situated in-between the two extreme theoretical predictions, it was suggested that stars should be differential rotators. However, these studies could not provide any information about the characteristics of the internal rotational law \citep{san55,danfa72,zo87col92,zo04iau215}. Somewhat related to this category of inquiries is the study of the evolution of the total angular momentum of B and Be stars carried by \citet{zo90nato316}, who concluded that these objects should undergo some internal angular momentum redistribution to explain the observed evolution of the $V\!\sin i$ parameters.\par
\medskip
 {\bf c$_1$}) The study of the absorption line profiles of MS B-type stars found some evidence of possible surface differential rotation. The angular velocity in the surface of stars was assumed to depend on the colatitude angle, $\theta$, as

\begin{equation}
\label{omsto}
\Omega(\theta) = \Omega_o[1-S\times(1-R(\theta)\sin^2\theta)] \ ,
\end{equation}

\noindent where $\Omega_o$ is the equatorial angular velocity, $R(\theta)$ is the equation of the stellar surface, $S$ is the parameter that testifies to the differential rotation. Using stellar models that are more or less gravity darkened, \citet{sto68} and \citet{sto87} found that in most cases $S\!<\!0$, which suggested that the angular velocity tends to increase from the equator to the pole. Nevertheless, a dependence of the surface angular velocity on the latitude $\Omega(\theta)$ could be due either to an actual differential rotation present under the stellar surface, or simply to zonal atmospheric currents, which could appear in rapidly rotating early-type stars, as speculated by \citet{cra93}. In Sect.~\ref{egosetfr}, we recall that an acceleration of the angular velocity towards the equator, i.e. $S\!>\!0$, can be promoted by a temperature gradient induced by the gravity darkening effect. \par
\medskip
{\bf c$_2$}) The possibility of detecting surface differential rotation by means of the Fourier analysis of spectral line profiles was discussed by \citet{hua61}, \citet{gray77},
\citet{brun81}, \citet{gar82}, and \citet{rei02}. Evidence of surface differential rotation in late-type stars with $V\!\sin i\!<\!50$ km~s$^{-1}$ were given by \citet{rei03b}, \citet{rei04}, and \citet{reiro04}, but no differential rotation for late-type stars with $V\!\sin i\!>\!50$ km~s$^{-1}$ and A-type stars with $V\!\sin i\!>\!150$ km~s$^{-1}$ were reported by \citet{rei03a} and \citet{gray77}, respectively. It is possible that modest differential rotation is difficult to detect with the Fourier transform technique in slowly rotating A-type stars, because the rotational broadening is not large compared with the broadening caused by other mechanisms such as thermal turbulence and pressure effects \citep{gray77}. However, in those cases where there is some evidence of differential rotation, the parameter $S$ cannot be differentiated from the unknown inclination angle factor $\sin i$. Nevertheless, its sign indicates acceleration of the angular velocity towards the equator. To our knowledge, the Fourier technique for differential rotation has not yet been applied to early-type stars.\par
\medskip
{\bf c$_3$}) \citet{and80} proposed a method to probe the inner angular velocity of stars based on the use of the rotational splitting of non-radial oscillations. However, owing to uncertainties in the identification of pulsation modes and rough determinations of stellar fundamental parameters, particularly their evolutionary stage, this method has not yet been able to be applied with reliable success. Nevertheless, from the analysis of pulsation modes derived from photometric variations \citep{deup04} and COROT data \citep{degr09}, constraints on the internal rotation of $\beta$~Cep stars have been inferred.\par
\medskip
{\bf d}) In the past few years, interferometric methods have helped provide remarkable insights into not only the rotational distortion of stars \citep{arm03,vanbel04}, but also the induced gravity darkening effect by means of imaging techniques \citep{mcalis05,
auf06,vanb06,zhao09}. New instruments with higher spectral resolutions of up to the 10000 attained by VLTI/AMBER in the J and K bands and an angular resolution of about 1 mas in the K band \citep{petrov07}, or spectral resolution reaching 30000 and angular resolutions 
as high as 0.3 mas in the visible using the VEGA/CHARA interferometric array \citep{mourard09}, will not only probably enable us to determine with greater detail than in previous studies the global geometry of stars deformed and gravity darkened by the rotation, but also carry out differential interferometry.\par
 A method based on differential interferometry that requires high spectral and spatial resolution was presented by \citet{arm04a,arm04b,arm04c} to distinguish observationally the parameter controlled by the degree of the surface differential rotation from the inclination angle factor $\sin i$.\par 

\subsection{Aims of the present attempt}
\label{aotpa}

  Most theoretical predictions about the evolution of rotating early-type stars and the mixing of chemical species triggered by the instabilities set up by the rotation, come from calculations performed in the framework of two significant assumptions: a) the global rotational energy stored by the stars in the ZAMS is lower than the limit allowed by the rigid rotation in the critical regime; b) the internal angular velocity undergoes an instantaneous ``shellular" redistribution at each evolutionary step. However, \citet{clem79} using a cylindrical (conservative) distribution of the angular velocity, and \citet{mae08} basing their calculation on a ``shellular" distribution, showed in a more detailed way that in rapidly rotating early-type stars the envelope layers beneath the surface, may have wider convective zones in radius than in non-rotating stars. In the Sun, only the layers unstable to convection rotate differentially with a non-shellular pattern. This motivates the inquiry of whether in massive and intermediate mass stars some coupling may also exist between convection and rotation beneath their surface. In this case, the characteristics of the rotational law in the external stellar layers should differ from those currently assumed in the above evoked stellar models.\par
  As demonstrated by many authors, the global geometry of a star depends not only on the total amount of angular momentum stored by the star, but also on its internal distribution \citep{boden71,Zo86,smi92,uryu94,uryu95}. This geometry mostly relies on the stellar surface rotation, which acts as an imprint of its properties in the layers beneath the surface. In this case, we should not exclude the resulting mixing of chemical elements in the stellar atmosphere being more or less dependent on the characteristics of the external rotational law, upon which the description of the stellar structure, based on the abundance determination of chemical elements, should also rely. Therefore, to provide new information and/or constraints to test the global assumptions currently made to calculate models of stellar structure with rotation and thus help deepen our understanding of the properties of early-type fast rotators, we might ask: 1) what can be deduced, using first principles, about the properties the rotation laws can have beneath the surface as a consequence of the coupling between rotation and convection; 2) what are the parameters needed to characterize these stars that may be accessible to observations; 3) whether the combined interpretations of spectroscopic and interferometric data of rapidly rotating early-type stars enable us to determine these parameters. \par
  In this attempt, the most interesting information might be the indication of some differential rotation in the stellar surfaces and the sign of its latitudinal gradient. Both pieces of data can be obtained, as much as possible, in a consistent way by taking into account the stellar geometrical deformation produced by this rotation and the concomitant gravitational darkening effect that responds to possible non-conservative rotation laws.\par 
\medskip
 The present paper is organized as follows. In Sect.~\ref{rlie}, we use first principles to infer possible rotation laws in the convective layers beneath the stellar surface of early-type rapid rotators. Sect.~\ref{eotss} presents the equation of the stellar surface of stars with non-conservative rotation laws. A discussion of the gravity darkening effect for 
non-conservative rotation laws is presented in Sect.~\ref{tdss}. The discussion about the validity of the Roche approximation in representing the gravitational potential is presented in Sect.~\ref{mrots}. This discussion is based on 2D models of rotating stars where the evolutionary stages are taken into account in a simplified way. We briefly comment on the determination of the rotational profile in the stellar envelope in Sect.~\ref{trpe}. In Sect.~\ref{attain}, we summarize the attainable information on rapidly rotating early-type stars with external differential rotation from the combined analysis of spectroscopic and interferometric data. Our conclusions are presented in Sect.~\ref{conclus}.\par

\section{Rotational law in the stellar envelope}
\label{rlie}

\subsection{The angular velocity distribution beneath the stellar surface}
\label{tavdbtss}

 The effects of rotation are generally introduced in the structure equations of rotating stars by replacing the spherical stratification of non-rotating star-models by a rotationally distorted stratification, which keeps the whole calculation problem to one dimension \citep{kiptho70,endsof76,pinson90,flilan95,chab95,mey97}. This procedure is justified if the internal differential rotation has a shellular distribution law because it comes from theoretical inferences made by \citet{zahn83,zahn92}, which rely on the assumption that the horizontal turbulence is much stronger than the vertical one. The limiting case of a shellular rotational profile is rigid rotation. From the calculations by \citet{mae05}, it follows that magnetic fields created by the Pitts \& Tayler instability \citep{pity85,spru99,spru02} can lock the stellar layers to each other and force the star to evolve as a rigid rotator. However, \citet{zahn07} concluded that the dynamo action can be less efficient as previously expected \citep{spru99,spru02} and that the magnetic fields created contribute little or less to the transport of the angular momentum. In addition, if the magnetic field in early-type rapid rotators is created, it could perhaps has some effect on their rotational profile in the convective parts of the envelope as in the Sun \citep{bal09a,bal09b}.\par 
  The differential rotation in the surface of the Sun is a direct consequence of the differential rotation in the convective layers beneath the surface layers. Therefore,
it is important to stress that this differential rotation in depth is not shellular, in spite of the strong turbulent viscosity that according to the above-mentioned theoretical assumptions, might otherwise cause shellular-like rotation.\par   
  \citet{mae08} demonstrated that rotation does not inhibit convection as could be thought from Solberg-H\o iland's stability criterion, but it changes the thermal gradient so as
to enhance convection. Hence, in rapidly rotating massive stars the two external convection zones, associated with increased opacities due to He- and Fe-ionization, respectively, are both considerably enlarged in depth so that the entire convective zone covers a non-negligible external region, which ranges from some 1/8 of the stellar radius in the pole to nearly 1/4 in the equator.\par
  However, the angular momentum distribution in the convective regions remains a puzzling question. To account for it in the stellar model calculations, two extreme approximations have been used: a) rigid rotation, angular velocity $\Omega\!=$ constant, which is supposedly promoted by the turbulent viscosity \citep{mae08}; b) constant specific angular momentum $j$ ($j\!=\!\Omega\varpi^2$; $\varpi$ is the distance to the axis of rotation), possibly due to the redistribution produced by the convective plumes \citep{tay73}. Nevertheless, the solar convective regions, characterized by significant turbulence, are the only ones with significant differential rotation \citep{sch98}, even though the solar rotational profile in the convective region, revealed by the helioseismological data, does not fall between these two extreme possibilities \citep{deup01}. Global insights into the rotational law in the convective regions could be obtained by exploring the solutions to the thermal wind balance relation under imposed conditions between the entropy and rotation. This procedure successfully explains the Solar rotation law in the convective regions \citep{bal09a,bal09b}. However, we pay attention here to our dealing with rapidly rotating stars. Detailed physical justifications of the assumptions made in the present paper are postponed to future contributions.\par

\subsection{Rotation inferred from the baroclinic balance relation}

\label{tbbr}
\begin{figure}[t!]
\centerline{\psfig{file=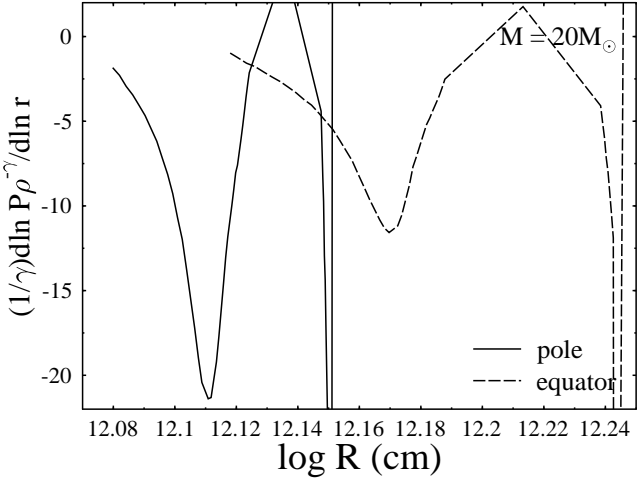}}
\caption{\label{adgrad} Adiabatic gradient $d\ln P\rho^{-\gamma}/d\ln r$ in the two rotationally enlarged convective regions in the envelope of a $20M_{\odot}$ star. The gradients are shown as a function of the logarithm of the stellar radius in the polar and equatorial directions.}
\end{figure}

 The curl of the time-independent momentum equation of an inviscid, axisymmetric rotating star with negligible magnetic fields, yields the baroclinic balance relation. Using the cylindrical coordinates ($\varpi,\phi,z$), this balance condition reads \citep{tass78}

\begin{equation}
\label{baroc1}
\frac{1}{\varpi^3}\frac{\partial j^2}{\partial z} = \frac{1}{\rho^2}(\nabla P\times\nabla\rho).\hat{{\rm e}}_{\phi}\ ,
\end{equation}

\noindent where $j$ is the specific angular momentum ($j=\Omega\varpi^2$), $P$ and $\rho$ are pressure and density, respectively, and $\hat{{\rm e}}_{\phi}$ is the azimuthal unit vector. Taking into account the equation of hydrostatic equilibrium

\begin{equation}
\label{moteq}
\frac{1}{\rho}\nabla P = {\rm\bf g_{\rm eff}},
\end{equation}

\noindent where ${\bf g_{\rm eff}}$ is the effective gravity, and using for the specific entropy $S$ the expression

\begin{equation}
\label{entr}
S = \frac{k}{\gamma-1}\ln P\rho^{-\gamma} + constant,
\end{equation}

\noindent where $k$ is the Boltzmann constant and  $\gamma$ is the ratio of specific heats at constant pressure and constant volume per unit mass, the wind equation in Eq.~(\ref{baroc1}) can be rewritten as

\begin{equation}
\label{baroc2}
\frac{1}{\varpi^3}\frac{\partial j^2}{\partial z}\!=\!\frac{1}{C_{\rm P}}(\nabla S\times{\rm\bf g_{\rm eff}}).\hat{{\rm e}}_{\phi},
\end{equation}

\noindent where $C_{\rm P}$ is the constant pressure specific heat.\par 
   We can attempt a discussion of Eq.~(\ref{baroc1}) by seeking solutions for the stellar internal rotation under at least three different conditions where for the moment the effects carried by the meridional circulation are neglected: $a$) marginal stability imposed by the Solberg-H\o iland criterion; $b$) state enforced by parallel surfaces of specific entropy and specific angular momentum; $c$) frame where the specific entropy parallels the local specific kinetic rotational energy.\par
  $a$) The Solberg-H\o iland stability criterion states that ``a baroclinic star in permanent rotation is dynamically stable against axisymmetric perturbations if two conditions are satisfied: (i) the specific entropy $S$ increases outwards, and (ii) on each surface S = $constant$, the specific angular momentum increases from the pole to the equator". The second condition is written mathematically as \citep{tass78}

\begin{equation}
\label{marg0}
-g_z[\nabla j^2\times\nabla S] \geq 0,
\end{equation}

\noindent where $g_z$ is the $z-$component of the effective gravity. Making the ansatz for strict marginal equilibrium, we have

\begin{equation}
\label{marg1}
-g_z[\nabla j^2\times\nabla S] = 0.
\end{equation}
  
\noindent Equation (\ref{marg1}) suggests then that surfaces $j^2\!=\!constant$ and $S\!=\!constant$ should be parallel, $S\!=\!S(j^2)$, i.e. a displaced fluid element in baroclinic turbulence retains both entropy and angular momentum. We do not consider here the conditions that might render possible the balance implied by Eq.~(\ref{marg1}).\par
  $b$) The assumption that the surfaces of constant specific entropy and constant angular velocity coincide, i.e. $S\!=\!S(\Omega^2)$, brings another alternative solution to the baroclinic equilibrium equation (\ref{baroc1}). This coupling can be enforced by magnetic fields \citep{bal09a}, although hydrodynamic constraints in the Sun can justify it entirely \citep{bal09b}.\par
  $c$) Only as an extrapolation to the conditions $S\!=\!S(j^2)$ and $S\!=\!S(\Omega^2)$ may we also consider $S\!=\!S(\varpi^2\Omega^2)$, since two energy-related quantities are parallel: specific entropy $S$ and specific rotational kinetic energy $\epsilon_{\Omega}\!=\!\varpi^2\Omega^2$.\par
  We now use $\mathscr{H}$ to represent in turn $j^2$, $\Omega^2$, and $\epsilon_{\Omega}$. We have then $S=S(\mathscr{H})$ and $\nabla S\!=\!(dS/d\mathscr{H})\nabla\mathscr{H}$. Rewriting Eq.~(\ref{baroc2}) in spherical coordinates $(r,\theta,\phi)$ and knowing that in these coordinates the $r-$ and $\theta-$components of the effective gravity ~${\bf g_{\rm eff}}$ are

\begin{eqnarray}
\label{grav}
g_r       & = & -\frac{\partial\Phi_{\rm G}}{\partial r}+\Omega^2(r,\theta)r\sin^2\theta\ , 
\nonumber \\
g_{\theta} & = & -\frac{1}{r}\frac{\partial\Phi_{\rm G}}{\partial\theta}+\Omega^2(r,\theta)r\sin\theta\cos\theta\ ,
\end{eqnarray}

\noindent we readily obtain

\begin{equation}
\label{baroc4}
\frac{\partial\mathscr{H}}{\partial r}(1\!-\!\alpha_{\mathscr{H}}\mathscr{H})\!-\!\frac{\tan\theta}{r}\frac{\partial\mathscr{H}}{\partial\theta}\left[1\!-\!\alpha_{\mathscr{H}}\mathscr{H}\!+\!\alpha_{\mathscr{H}}GM\mathscr{F}
\right] = 0\ ,
\end{equation}

\noindent where we have used the notations

\begin{equation}
\label{alfaj} 
\alpha_{\mathscr{H}} = \frac{1}{C_{\rm P}}\frac{dS}{d\mathscr{H}}
\end{equation}

\noindent and

\begin{equation}
\mathscr{F} =
\left\{
\begin{array}{rl}
r\sin^2\theta & \mbox{  if  } \mathscr{H} = j^2 \\
\frac{1}{r^3\sin^2\theta} & \mbox{  if  } \mathscr{H} =  \Omega^2 \\
\frac{1}{r} & \mbox{  if  } \mathscr{H} = \epsilon_{\Omega} 
\end{array}
\right.
\label{hh}
\end{equation}

\noindent The characteristic equation of Eq.~(\ref{baroc4}) is

\begin{equation}
\label{carj}
\frac{dr}{1-\alpha_{\mathscr{H}}\mathscr{H}} = -\frac{d\theta}{\frac{\tan\theta}{r}\left[(1-\alpha_{\mathscr{H}}\mathscr{H})+\alpha_{\mathscr{H}}GM\mathscr{F}
\right]}\ .
\end{equation}

\noindent As for the assumption $S\!=\!S(\mathscr{H})$ made here, $\mathscr{H}$ and $S$ are constant along the characteristic curves of Eq.~\ref{baroc4}, and the differential equation Eq.~(\ref{carj}) integrates immediately to give

\begin{eqnarray}
\frac{1}{r^2\sin^2\theta} & = & \frac{1}{r}\Upsilon_{\rm J}+C_{\rm J}  
\ \ \ \ \ \ \ \ \ \ \ \ \mbox{  if  } \mathscr{H} = j^2 \nonumber \\
r\sin^2\theta & = & -\frac{1}{r}\Upsilon_{\Omega}+C_{\Omega}  
\ \ \ \ \ \ \ \mbox{  if  } \mathscr{H} = \Omega^2 \nonumber \\
\frac{1}{r^2\sin^2\theta} & = & C_{\epsilon_{\Omega}}\exp(\frac{\Upsilon_{\epsilon_{\Omega}}}{r})\ \ \ \ \   \mbox{  if  } \mathscr{H} = \epsilon_{\Omega} 
\label{soluhh}
\end{eqnarray}

\noindent with the $C_{\mathscr{H}}$ as the integration constants (where $\mathscr{H}$ has different meanings). Once the function giving the angular velocity $\Omega_{\rm s}(\theta)$ in the stellar surface is specified, the iso-rotation curves can be obtained easily everywhere inside the star using the relations in Eq.~(\ref{soluhh}). Each iso-rotation curve depends on an integration constant $C_{\mathscr{H}}$ obtained as $C_{\mathscr{H}}\!=\!C_{\mathscr{H}}(\Upsilon_{\mathscr{H}},\theta_s)$, where $\theta_s$ is the colatitude angle at which a given iso-rotation contour intersects the stellar surface. The constants $\Upsilon_{\mathscr{H}}$ are defined as

\begin{equation}
\label{short}
\Upsilon_{\mathscr{H}} = -\left(\frac{2\alpha_{\mathscr{H}}GM}{1-\alpha_{\mathscr{H}}}\right)\ . \\
\end{equation}

\noindent Since we consider convective regions where $\nabla S\leq0$, we find that $\Upsilon_{\mathscr{H}}\geq0$. We note that depending on the meaning of $\mathscr{H}$, the expressions $\Upsilon_{\mathscr{H}}=$ $\Upsilon_{\rm J}R_{\rm e}$, $\Upsilon_{\Omega}/R^2_{\rm e}$ and $\Upsilon_{\epsilon_{\Omega}}$ are dimensionless constants of the order of unity ($R_{\rm e}$ is the stellar radius at the equator). Defining

\begin{equation}
A_{\mathscr{H}} = -\frac{1}{\gamma}\left(\frac{d\ln P\rho^{-\gamma}}{d\ln r}\right)\left(\frac{d\ln r}{d\ln\mathscr{H}}\right), 
\label{defa}
\end{equation}

\noindent which in convection regions is $A_{\mathscr{H}}\geq0$, we have

\begin{equation}
\Upsilon_{\mathscr{H}} = 2\left(\frac{GM}{R_{\rm e}^3\Omega^2}\right)\left(\frac{A_{\mathscr{H}}}{1-A_{\mathscr{H}}}\right)\ . 
\label{defups}
\end{equation}

\noindent From the model of a rapidly rotating star with $M=20M_{\odot}$ given in \citet{mae08}, we derived the adiabatic gradient $\nabla = d\ln P\rho^{-\gamma}/d\ln r$ shown in Fig.~\ref{adgrad}. This quantity is shown as a function of stellar radius in the polar and equatorial directions. The value of $\Upsilon_{\mathscr{H}}$ depends also on the adopted value for the gradient $d\ln r/d\ln\Omega^2$ that represents the differential rotation. For the Sun, the term $d\ln r/d\ln\Omega^2$ comes from seismology data and its value ranges from 1 to 10. However, until now in early-type stars, the factor $d\ln r/d\ln\Omega^2$ has been a perfectly unknown quantity. For a rough estimate, we may assume that it can be at least of the same order of magnitude as in the Sun. Similar values are also obtained in the models calculated by \citet{meynet2000}. Using the values of the gradient $\nabla$ shown in Fig.~\ref{adgrad}, it can be shown that the ratio $A_{\mathscr{H}}/(1-A_{\mathscr{H}})$ is nearly constant over the entire convective region and that depending on the assumed value of $d\ln r/d\ln\Omega^2$, it changes from 0 to 1. Hence, we have that 

\begin{equation}
 0\lesssim\Upsilon_{\mathscr{H}}\lesssim2(GM/R_{\rm e}^3\Omega^2),
\label{estups}
\end{equation}

\noindent where $0.5\!\lesssim\!(R_{\rm e}^3\Omega^2/GM)\!\lesssim\!1$ for rapid rotators. For the sake of comparison, we note that in the convective region of the Sun, it is $|\nabla|\simeq10^{-6}$ and $(GM/R_{\rm e}^3\Omega^2)\simeq10^5$ \citep{bal09a}.\par
\begin{figure*}
\centerline{\psfig{file=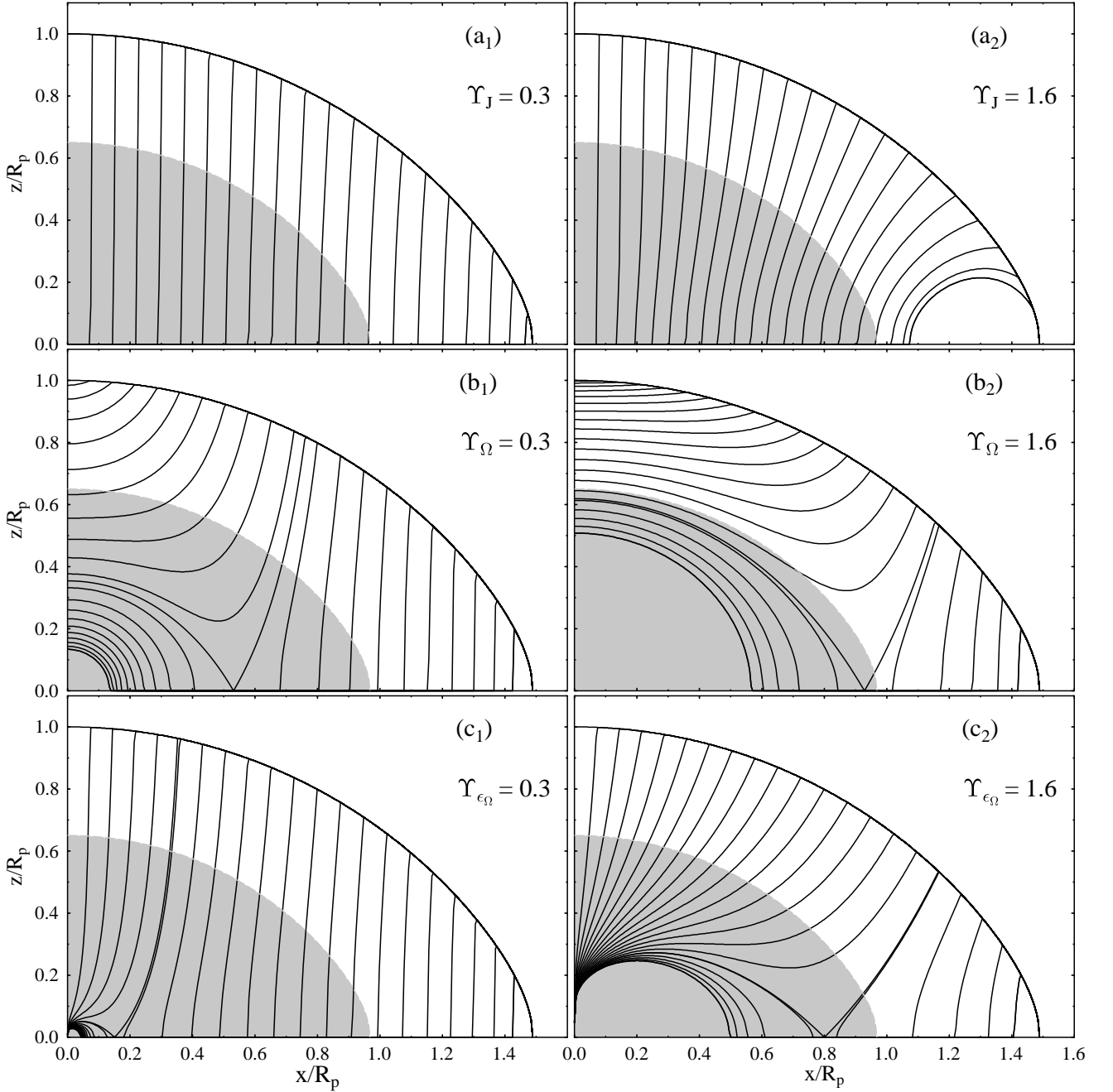}}
\caption{\label{constg} $\Omega(r,\theta)/\Omega_{\rm e}\!=$ constant curves inside the stellar envelope of a model star with a global shape generated with an equatorial rotational parameter $\eta_o\!=\!0.8$ and a surface differential rotation represented by a Maunder relation with parameter $\alpha=+0.3$. All curves were generated with parameters $\Upsilon_{\rm J,\Omega,\epsilon_{\Omega}}=0.3$ and 1.6. The iso-rotation curves in ({\bf a$_1$}) and ({\bf a$_2$}) correspond to $S=S(j^2)$; those in ({\bf b$_1$}) and ({\bf b$_2$}) are for $S=S(\Omega^2)$; the curves in ({\bf c$_1$}) and ({\bf c$_2$}) obey the condition $S=S(\varpi^2\Omega^2)$. The solutions are supposed to be valid in the convective zones above the shaded domain.}
\end{figure*}

 Examples of isorotation curves obtained from Eq.~(\ref{soluhh}) are shown in Fig.~\ref{constg}, where we used stellar external contours whose rotational deformation is characterized by the ratio of centrifugal to gravitational acceleration at the equator $\eta_o=\Omega^2_{\rm e}R^3_{\rm e}/GM=0.8$ and a Maunder surface differential rotation law with parameter $\alpha=0.3$. The equation of the stellar surface with latitudinal differential rotation is discussed in some detail in Sect.~\ref{eotss}. Solutions similar to those shown in Figs.~\ref{constg}a and \ref{constg}b were obtained by \citet{bal09a} for the Sun. This author assumed that $\partial P/\partial r\gg\partial S/\partial\theta$ and $\partial S/\partial\theta\gg\partial P/\partial r$, which according to our expressions are for the limits $\alpha_{\rm J}j^2\to0$ and $\alpha_{\Omega}\Omega^2\to0$ that suit slowly rotating objects like the Sun.\par

\subsection{Comments on the rotational profiles obtained}
\label{crpo}

 The rotational profiles $\Omega(r,\theta)$ in the convective layers beneath the surface of early-type rapidly rotating stars were inferred here using the marginal condition of the Solberg-H\o iland stability criterion against axisymmetric perturbations. The derived solutions are the consequence of a dominant thermal wind balance where the entropy is: a) a function of the specific angular momentum, $S\!=\!S(j^2)$; b) a function of the angular velocity, $S\!=\!S(\Omega^2)$, and c) a function of the specific rotational kinetic energy, $S\!=\!S(\epsilon_{\Omega})$. The first case was widely studied in the literature \citep[][ references therein]{tass78}, and a number of independent numerical attempts to obtain baroclinic stars inexorably ended up with results close to the category shown in Figs.~\ref{constg}a1, i.e. conservative-like \citep{uryu94,uryu95}. Two-dimensional, implicit hydrodynamic numerical simulations by \citet{deup98} and \citet{deup01} showed that in convective regions $\Omega$ tends to adopt a cylindrical-like profile rather than a shellular or $\Omega\!=\!$ constant form. Intermediate solutions to the classes shown in Figs.~\ref{constg}b$_1$ and b$_2$ are compatible with the seismological data of the Sun relative to its convective regions. The rotational profile obtained by \citet{esp07} for fully radiative stars curiously enters into the category of solutions shown in \ref{constg}c$_2$. We note that for all cases shown in Fig.~\ref{constg} we have used constant values of $\Upsilon_{\mathscr{H}}$ all over the star. However, they are valid only for the convective regions, which lie roughly above the shaded sector and where, as already noted, $\Upsilon_{\mathscr{H}}$ is fairly uniform.\par
 According to a strong claim that shear instabilities widely prevail over all other mechanisms that can act to redistribute the angular velocity, most if not all models of stellar evolution of early-type stars with rotation were carried out by assuming that $\Omega$ has a shellular nature. This strongly simplifies the numerical aspect of the study, but does not necessarily preclude that other distributions of $\Omega$ might coexist with the shellular one, or even dominate in wide domains of the stellar interior. The $S\!=\!S(\Omega^2)-$case with $\Upsilon_{\Omega}>1.6$ clearly depicts shellular-like rotation over a wide interior stellar region.\par
 Although the present discussion should not be considered an argument in favor of non-conservative-like profiles of the angular velocity in the envelopes of early-type rapid rotators, the success that the displayed arguments do have in explaining the solar rotational profile in the convective layers, strongly motivate us to address the study of the surface differential rotation in rapidly rotating early-type stars. Whatever the observational or theoretical indication that $\partial\Omega/\partial\theta\neq0$ can exist in the surface of rapidly rotating early-type stars, they would immediately imply that profiles of non-shellular rotation exist, perhaps of some type similar to that in Fig.~\ref{constg}. They would also immediately enable us to achieve clearer understandings of the transport of angular momentum in the stars and the related mixing processes of chemical elements in the stellar surface layers.\par 

\subsection{Induced gradient in $\Omega(\theta)$ in the surface of early-type rapid rotators}
\label{egosetfr}

 In rapidly rotating stars, there is a strong latitudinal temperature gradient, which can induce a latitudinal gradient on the surface angular velocity $\Omega_{\rm s}$. This effect can be estimated from Eq.~(\ref{baroc1}) by  calculating the latitudinal variation in the temperature over an isobar. Giving to the surface  temperature a functional form $T\!=\!T(P,\theta)$, where $P$ is the pressure and $\theta$ the colatitude, we obtain

\begin{equation}
\label{tpt}
\nabla T(P,\theta) = \frac{\partial T}{\partial P}\nabla P+\frac{\partial T}{\partial\theta}\frac{\hat{\rm e}_{\theta}}{r},
\end{equation}

\noindent where $\hat{\rm e}_{\theta}$ is the colatitude unit vector and $r$ is the radial spherical coordinate. Making use of the equation of state $P\propto\rho T$, Eq.~(\ref{tpt}) and the $r-$component of Eq.~(\ref{moteq}), Eq.~(\ref{baroc1}) transforms into \citep{esp07}

\begin{equation}
\label{tgradt}
\varpi\frac{\partial\Omega^2}{\partial z} = -\frac{g_r}{r}\left(\frac{\partial\ln T}{\partial\theta}\right)_{\rm P},
\end{equation}

\noindent where the notation $(...)_{\rm P}$ indicates that the partial $\theta-$derivative is along a barotropic surface. As in rapidly rotating stars, it is $\partial\ln T/\partial\theta\!<\!0$ and $g_r\!<\!0$, from Eq.~(\ref{tgradt}) it appears that $\partial\Omega/\partial z\!<0$, which means that $\Omega$ should increase from the pole to the equator. Although in rapid rotators there are convective zones in the envelope, the dominant radiation flux can still be estimated using the diffusion approximation $F_{\rm rad}\!=-\chi\nabla T$, where $\chi$ is the coefficient of radiative conductivity. For simplicity, we
adopt von Zeipel's approximation \citep{vzei24} to represent the stellar surface temperature distribution as a function of the colatitude

\begin{equation}
\label{vzeip}
T(\theta) = T_p\left(\frac{g}{g_p}\right)^{1/4},
\end{equation}

\noindent where $T_p$ is the polar temperature of the star, $g\!=\!\left[(g_r)^2\!+\!(g_{\theta})^2\right]^{1/2}$ with $g_r$ and $g_{\theta}$ given by Eq.~(\ref{grav}), and $g_p$ is the value of the gravity in the pole. Using the Roche approximation for the gravitational potential, Eq.~(\ref{tgradt}) written in spherical coordinates becomes

\begin{eqnarray}
\label{tgradts}
\lefteqn{\frac{\partial\ln\Omega^2}{\partial r}-\frac{\tan\theta}{r}[1\!-\!(1\!-\!\eta_r)f_1(\theta,\eta_r)]\frac{\partial\ln\Omega^2}{\partial\theta} =}
\nonumber \\ 
&& -\frac{1}{r}f_1(\theta,\eta_r)f_2(\theta,\eta_r)
\end{eqnarray}

\noindent with the following definitions

\begin{eqnarray}
\label{def1e2}
f_1(\theta,\eta_r) & = & \frac{1}{4}\left[\frac{1-\eta_r\sin^2\theta}{1-\eta_r(2-\eta_r)\sin^2\theta}\right] 
\nonumber \\
f_2(\theta,\eta_r) & = & \left[\frac{2-\eta_r(1+\eta_r)\sin^2\theta}{\sin\theta\cos\theta}
\right]\frac{\partial\ln r}{\partial\theta}+2-\eta_r \nonumber \\
\eta_r & = & \frac{\Omega^2r^3}{GM}.
\end{eqnarray}

\begin{figure}[t!]
\centerline{\psfig{file=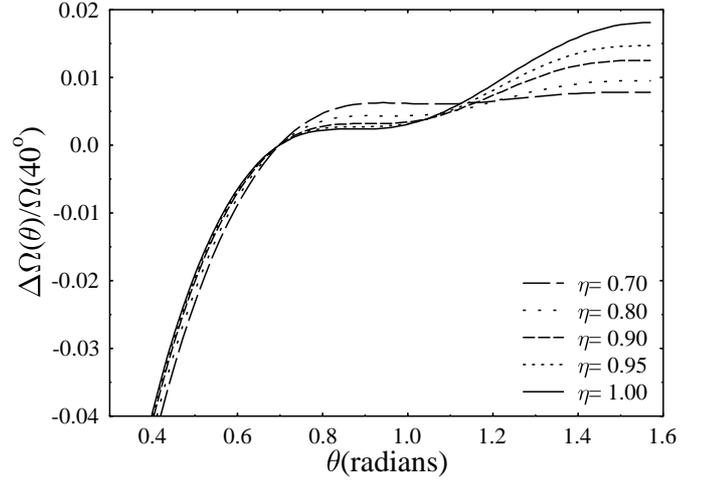}}
\caption{\label{oms}Normalized surface angular velocity $\Omega_s$ as a function of the colatitude $\theta$ and the rotational parameter $\eta$.}
\end{figure}

  The characteristic equation describing the dependence of the surface angular velocity with $\theta$ is then

\begin{equation}
\label{carot}
-\frac{d\theta}{\frac{\tan\theta}{r}[1\!-\!(1\!-\!\eta_r)f_1(\theta,\eta_r)]} = -\frac{d\ln\Omega^2}{\frac{1}{r}f_1(\theta,\eta_r)f_2(\theta,\eta_r)}\ ,
\end{equation}

\noindent which is solved by numerical integration. Figure~\ref{oms} shows the $\theta-$dependence of the angular velocity given as $\Delta\Omega(\theta)=[\Omega(\theta,\eta_r)-\Omega(40^o,\eta_r)]/\Omega(40^o,\eta_r)$, which was obtained from Eq.~(\ref{carot}) for different values of $\eta$. In this calculation, the gradient $\partial\ln r/\partial\theta$ in the stellar surface was estimated using the Roche potential of a rigid rotator characterized by average rotational parameters $\eta$. In Fig.~\ref{oms}, we see that in general it is $\partial\Omega/\partial\theta\!>\!0$, in particular for $\eta_o>0.7$ near the equator. This acceleration seems to be similar in character to the one previously calculated by \citet{esp07}. As $\eta\to0$, the curves $\Delta\Omega(\theta)\to0$.\par

\section{Equation of the stellar surface}
\label{eotss}
  
\subsection{The angular velocity distribution in the stellar surface}
\label{avdss}

 Since the shape of a rapidly rotating star depends explicitly on its surface angular velocity distribution, we adopt a law to study the possible effects it can produce. In Sect.~\ref{egosetfr}, we have seen that the temperature gradient induced by the gravity darkening effect introduces a small acceleration of the angular velocity from the pole towards the equator, which is not simple to represent analytically. The physical properties of the layers beneath the surface could perhaps enforce a stronger surface angular velocity gradient than inferred in Sect.~\ref{egosetfr}. If in rapidly rotating massive stars rotational profiles of the class obtained in Sect.~\ref{tbbr} actually existed, a first approach to describing their surface rotation could rely on the use of a solar-like surface angular velocity. The solar surface differential rotation depends on the colatitude $\theta$ as \citep{sno84}

\begin{equation}
\label{sol1}
\frac{\Omega_{\odot}(\theta)}{2\pi} = 451.5-65.3\cos^2\theta-66.7\cos^4\theta \ \ {\rm nHz}.
\end{equation}

\noindent This expression carries three coefficients that need to be determined empirically. At the moment, it seems unrealistic to use a similar expression for other stars since their surface cannot be resolved. However, within the errors smaller than 1.6\%, Eq.~(\ref{sol1}) can be reduced to the simplest one

\begin{equation}
\label{sol2}
\frac{\Omega_{\odot}(\theta)}{2\pi} = 459.3(1-0.29\cos^2\theta) \ \ {\rm nHz},
\end{equation}

\noindent where there are only two quantities to determine from observations. A first inquiry about the differential rotation on the surface of massive and intermediate mass stars can then be justified by using the simplified expression (Maunder formula)

\begin{equation}
\label{sol3}
\Omega(\theta) = \Omega_o(1+\alpha\cos^2\theta)\ .
\end{equation}

\noindent In Eq.~(\ref{sol3}), $\alpha<0$ indicates that the pole rotates slower
than the equator and vice-versa if $\alpha>0$. If massive and intermediate mass stars had surface rotations similar to the Sun, their differential rotation parameter would be expected to be at least of the order of $\alpha\sim-0.3$.\par
 In general, to study stars with unresolved surfaces we should probably use expressions of the type

\begin{equation}
\label{sol4}
\Omega(\theta) = \Omega_o[1+\alpha f(\theta)],
\end{equation}
 
\noindent where the function $f(\theta)$ needs to be somehow specified in advance.\par

\subsection{Formulation of the equation of the stellar surface}
\label{form}

\begin{figure}[t!]
\centerline{\psfig{file=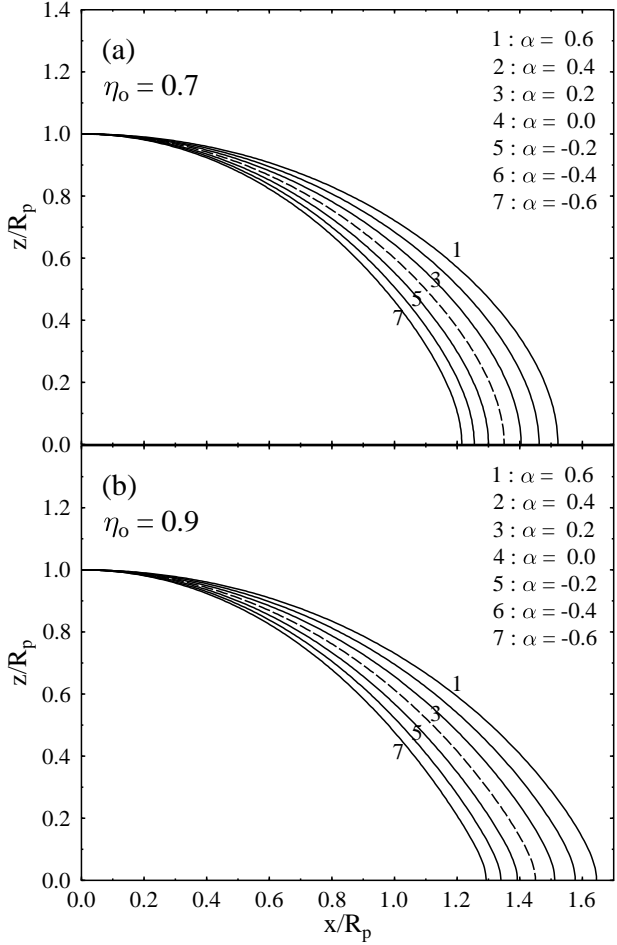}}
\caption{\label{forma} Shape of stars having average surface rotational parameters $\eta_o=0.7$ (a) and 0.9 (b), and latitudinal differential rotations given by Eq.~(\ref{sol3}) for several values of the parameter $\alpha$. The stellar shape for $\alpha=0$ is indicated by a dashed line to more clearly show the effect of $\alpha$ in other cases.}
\end{figure}

  The shape of a rapidly rotating star is generally described using the total potential of a self-gravitating system with a conservative rotation law, i.e., an internal cylindrical angular velocity, $\Omega\!=\!\Omega(\varpi)$, of which the rigid rotation is a particular case. When a star has a non-conservative internal rotational law $\Omega\!=\!\Omega(\varpi,z)$, as used in models of stellar evolution and/or suggested by the developments given in Sect.~\ref{tbbr}, it is no longer possible to define a rotational potential. However, according to \citet{mae09} the surface of a star with a non-conservative rotation law can be identified as the region where an arbitrary displacement $d${\bf s} does not imply any work done by the effective gravity {\bf g}$_{\rm eff}$

\begin{equation}
\label{work}
{\rm\bf g}_{\rm eff}.d{\rm\bf s} = 0\ .
\end{equation}

\noindent The effective gravity {\bf g}$_{\rm eff}$ written in cylindrical coordinates has the form

\begin{equation}
\label{work1}
{\rm\bf g}_{\rm eff} = -\nabla\Phi_{\rm G}+\Omega^2(\varpi,z)\varpi.\hat{\rm e}_{\varpi}\ ,
\end{equation}

\noindent where $\Phi_{\rm G}$ is the gravitational potential. Using the function $\Psi(\varpi,z)$ defined as

\begin{equation}
\label{work2}
\Psi(\varpi,z) = \Phi_{\rm G}-\frac{1}{2}\Omega^2(\varpi,z)\varpi^2\ ,
\end{equation}

\noindent the effective gravity can be expressed in the form

\begin{equation} 
\label{work3}
{\rm\bf g}_{\rm eff} = -[\nabla\Psi(\varpi,z)+\frac{1}{2}\varpi^2\nabla\Omega^2]\ ,
\end{equation}

\noindent so that the condition given in Eq.~(\ref{work}) can be rewritten as

\begin{equation}
\label{work4}
d\Psi(\varpi,z)+\frac{1}{2}\varpi^2(\nabla\Omega^2.d{\rm\bf s}) = 0 .
\end{equation}

  The function $\Psi(\varpi,z)$ can be a potential if and only if it is $\partial\Omega/\partial z=0$. It has been shown by \citet{mey97} that in stars with shellular internal rotational profiles, the surfaces $\Psi(\varpi,z)\!=$ constant are parallel to isobar surfaces. However, since $d\Psi(\varpi,z)$ is not an exact differential, the integration of Eq.~(\ref{work4}) depends on the chosen path. To define the shape of a star, it seems natural to integrate Eq.~(\ref{work4}) over a meridian curve, from the pole (spherical coordinate $\theta\!=\!0$) towards an arbitrary point $R_{\rm s}(\theta)$, where $R_{\rm s}(\theta)$ represents the function describing the stellar surface. Thus, by virtue of Eq.~(\ref{work2}) we obtain

\begin{equation}
\label{work5}
\Phi_{\rm G}(\theta)-\frac{1}{2}\Omega^2(\varpi,z)\varpi^2+\frac{1}{2}\!\int_0^{\theta}\!\!\varpi^2
(\nabla\Omega^2.d{\rm\bf s}) = \Phi_{\rm G}(0) \ ,
\end{equation}

\noindent which is in principle the sought equation to calculate the shape of a star having non-conservative rotational laws. Nevertheless, since the integration indicated in Eq.~(\ref{work5}) depends on the angular velocity distribution defined only over the stellar
surface, the following change in the coordinates describing the stellar surface

\begin{equation}
\label{nonc1}
\varpi_{\rm s}(\theta) = R(\theta)_{\rm s}\sin\theta\ ; \ \ \ z_{\rm s}(\theta) = R_{\rm s}(\theta)\cos\theta \ ,
\end{equation}

\noindent readily transforms $\nabla\Omega^2(\varpi,z).d{\rm\bf s}$ into a total differential that enables us to integrate Eq.~(\ref{work5}) by parts and derive the sought relation

\begin{equation}
\label{nonc3}
\Phi_{\rm G}(\theta)-\frac{1}{2}\int_0^{\theta}\!\!\Omega^2_{\rm s}(\theta)\left(\frac{d\varpi^2}{d\theta}\right)d\theta = \Phi_{\rm G}(0)\ .
\end{equation}

\noindent In Eq.~(\ref{nonc3}), we wrote $\Omega^2_{\rm s}$ to indicate explicitly that the angular velocity concerns only the stellar surface.  From Eq.~(\ref{nonc3}), we can see that the stellar shape depends on both the rotational profile in the surface $\Omega_{\rm s}(\theta)$ and the gravitational potential $\Phi_{\rm G}$, which in turn depends on the internal rotational law by means of the centrifugally distorted density distribution in the stellar interior. However, this last dependence is of second order or negligible, as will be shown in Sect.~\ref{totau}. \par

\begin{table}[]
\centering
\caption[]{\label{rerpdt}Radii ratios $R_{\rm e}/R_{\rm p}$ as a function of parameters
$\eta_o$ and $\alpha$}
\begin{tabular}{c|ccccccc}
\hline
\noalign{\smallskip}
$\eta_o$ & $\alpha\!=\!0.6$ & 0.4 & 0.2 & 0.0 & -0.2 & -0.4 & -0.6 \\
\noalign{\smallskip}
\hline
\noalign{\smallskip}
0.4 & 1.317  & 1.275  & 1.236  & 1.200  & 1.167  & 1.138  & 1.114 \\
0.5 & 1.388  & 1.339  & 1.293  & 1.250  & 1.211  & 1.176  & 1.146 \\ 
0.6 & 1.456  & 1.401  & 1.349  & 1.300  & 1.255  & 1.215  & 1.179 \\ 
0.7 & 1.522  & 1.462  & 1.404  & 1.350  & 1.300  & 1.255  & 1.215 \\
0.8 & 1.585  & 1.520  & 1.458  & 1.400  & 1.346  & 1.296  & 1.253 \\ 
0.9 & 1.645  & 1.577  & 1.512  & 1.450  & 1.392  & 1.340  & 1.293 \\
\noalign{\smallskip}
\hline
\noalign{\smallskip}
\end{tabular}
\end{table}

\subsubsection{Shape of stars with a Maunder rotation law in the surface}
\label{mrl}

According to the discussion in Sect.~\ref{egosetfr} and the simplified solar surface rotational law (\ref{sol3}), the differential rotation parameter in fast rotating early-type stars is expected to be $\alpha < 0$. However, because of the still poorly known physical characteristics of the convective external envelope layers of these objects, and possible effects linked to magnetic fields, we cannot exclude that $\alpha > 0$ in some objects. In this respect, we note that hydrodynamic calculations in stars with $8.75M_{\odot}$ carried out by \citet{deup98} show that the rotation law in the convective core becomes of conservative type $\Omega\sim\varpi^{-0.7}$ ($\varpi$ being the distance from the rotation axis). If convection in the envelope forced it to rotate in a similar way, we should expect an external imprint of this law given by an $\alpha > 0$. Many calculations of the stellar structure with fast differential rotation have been done in the past assuming that H\o ilnad's criterion for dynamical stability is satisfied and that radiative viscosity has significant effects on the surface layers. The resulting rotation law implies that the specific angular momentum increases with the mass contained in cylinders with radius $\varpi$ \citep{tass78}, which also leads to rotation laws that imply $\alpha>0$. However, the final shape of the rotation law in the envelope must certainly be the consequence of the interplay among many hydrodynamic and magneto-hydrodynamic instabilities, whose very final combined effect is still highly unknown. We adopt then the surface rotational law with the functional form given by Eq.~(\ref{sol2}), and assume that the parameter $\alpha$ can imply either equator-ward or polar-ward accelerated angular velocities.\par
 In a first step, we separate the effects produced on the external stellar shape by the surface differential rotation from those induced by the internal rotation on the mass distribution. This can simply be done, as we justify in Sect.~\ref{rocheapp}, by using a Roche approximation for the gravitational potential

\begin{equation}
\label{rgp}
\Phi_{\rm G}(\theta) = -GM/R_{\rm s}(\theta) \ .
\end{equation}

\noindent Equation~(\ref{nonc3}) can then be given the form

\begin{equation}
\label{iter1}
\frac{R_{\rm s}(\theta)}{R_{\rm e}} = \frac{1}{1+\eta_o[I(\pi/2)-I(\theta)]} \ ,
\end{equation}

\noindent where we have written

\begin{equation}
\label{iter1exp}
I(\theta) = \frac{1}{2}\int_0^{\theta}\!\!\left[\frac{\Omega_{\rm s}(\theta)}{\Omega_o}\right]^2\left(\frac{d\varpi^2}{d\theta}\right)d\theta\ .
\end{equation}

\noindent Since in relation (\ref{iter1}) the function of the stellar surface $R_{\rm s}(\theta)$ also appears in the integrand in terms of $\varpi$ by (\ref{nonc1}), $R_{\rm s}(\theta)$ is obtained by iteration. We start the iteration with a first estimate of $I(\pi/2)$ based on the shape of a star with a rigid rotation characterized by the rotational parameter~$\eta_o$

\begin{equation}
\label{etao}
\eta_o = \frac{\Omega^2_oR^3_{\rm e}}{GM}\ ,
\end{equation}

\noindent so that 

\begin{equation}
\label{rig}
\frac{R^{(o)}_{\rm s}(\theta)}{R_{\rm e}} = \frac{1}{1+\frac{1}{2}\eta_o\left\{1-\left[\frac{R^{(o)}_{\rm s}(\theta)}{R_{\rm e}}\right]^2\sin^2\theta\right\}} \ . 
\end{equation}

\noindent It is also obvious that in relation (\ref{iter1}) the equatorial-to-polar radii ratio $R_{\rm e}/R_{\rm p}$ is the solution of the iteration, which finally gives 

\begin{equation}
\label{rerpd}
\frac{R_{\rm e}}{R_{\rm p}} = 1+\eta_oI\left(\frac{\pi}{2},\frac{R_{\rm e}}{R_{\rm p}},\alpha\right) \ .
\end{equation}
 
\noindent We recall that for rigid rotation in the Roche approximation we have \par

\begin{equation}
\label{rerpr}
\left(\frac{R_{\rm e}}{R_{\rm p}}\right)_{\rm rigid} = 1+\frac{1}{2}\eta_o\ .
\end{equation}
 
    In Table~\ref{rerpdt}, we present the ratios of equator-to-polar radii $R_{\rm e}/R_{\rm p}$ as a function of $\eta_o$ and $\alpha$ used to obtain the stellar shapes shown in Fig.~\ref{forma}, which shows the shapes of stars computed with equatorial acceleration ratios $\eta_o=0.7$ and 0.9, whose surface angular velocity is given by Eq.~(\ref{sol3}), where the parameter $\alpha$ has several positive and negative values. In this figure, we can see that the stellar shape is a function of $\alpha$ that sensitively differs from that of homologous rigid rotators with the same parameter $\eta_o$. We note that for $\Omega(\theta)$ with $\alpha\!>\!0$ (acceleration from the equator towards the pole), we always find that

\begin{equation}
\label{rerpalf}
 \frac{R_{\rm e}(\eta_o,\alpha)}{R_{\rm p}(\eta_o,\alpha)} > \left[\frac{R_{\rm e}(\eta_o)}{R_{\rm p}(\eta_o)}\right]_{\rm rigid}\ .
\end{equation}

 {\it For modest $\alpha\!>\!0$ (recalling that in the Sun, it is $\alpha\!\simeq\!-0.3$), we can have $R_{\rm e}/R_{\rm p}\!>\!1.5$ even for $\eta_o<1$, although the stellar surface has no polar dimples as in models with high rotational energy content} (See Sect.~\ref{swpd}). This is due both to a rotational stretching of $R_{\rm e}$ and to the concomitant shrinkage of the polar radius $R_{\rm p}$, where $\Omega_{\rm pole}\!>\!\Omega_{\rm equator}$. When $\Omega_{\rm s}(\theta)$ is accelerated from the pole towards the equator ($\alpha\!<\!0$), we have $R_{\rm e}(\eta_o,-|\alpha|)/R_{\rm p}(\eta_o,-|\alpha|)\!<\!(R_{\rm e}(\eta_o)/R_{\rm p}(\eta_o))_{\rm rigid}$. Here, the equator/pole radii ratios respond mainly to the rotational stretching of $R_{\rm e}$.\par
 We note that in stars where $\alpha\!\neq\!0$, the $V\!\sin i$ parameter does not only depend on the rotation in the equator. A straightforward interpretation of the quantity $V\!\sin i$ can produce an incorrect value of the acceleration ratio $\eta_o$. Since we are dealing here with fast rotators ($\eta_o\gtrsim0.5$), the formation of spectral lines used to determine the $V\!\sin i$ needs to be treated properly by taking into account the surface velocity fields, the stellar deformation, and the gravity darkening effect, all dependent on $\eta_o$ and $\alpha$. We also note that if the interferometric measurements produce radii ratios $R_{\rm e}/R_{\rm p}\!>\!1.5$, it does not necessarily mean that the star is in a state of ``supercritical" rotation. From these comments and the results shown in Table~\ref{rerpdt}, we clearly see that the study of the surface differential rotation prefigures the need for two sources of information: a) spectroscopy, which can help us to estimate $\alpha$ and the inclination angle factor $\sin i$; b) interferometry, which provides information related to the stellar geometrical deformation and incidentally with $\alpha$. We briefly discuss these issues in Sects.~\ref{ifsl} and \ref{ifi}.\par

\subsubsection{Stars with polar dimples}
\label{swpd}

\begin{figure}
\centerline{\psfig{file=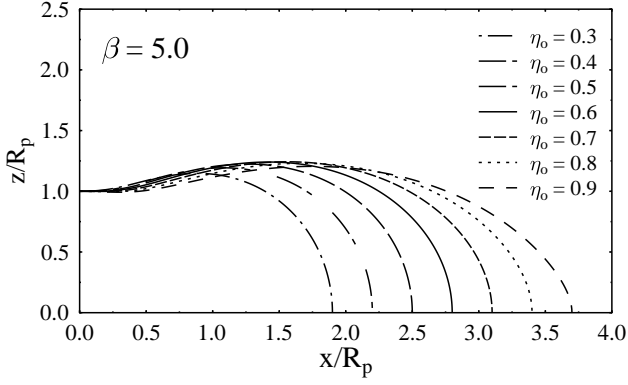}}
\caption{\label{fbetlaw} Shape of model stars having the surface angular velocity profile given by relation (\ref{betlaw}) for $\beta=5.0$ and several values of $\eta_{\rm e}$.}
\end{figure}

 Owing to the centripetal acceleration, the fraction of the gravitational force that must be supported by the pressure gradient increases towards the center of the stars. This means that the pressure gradient is more centrally concentrated in a rotating star than a non-rotating one. The rotation tends then to decrease the moment of inertia, which is accompanied by an increase in the density gradient towards the center. The gravitational attraction towards the pole is thus stronger, which produces a flattening of the pole or a dimple. In models of stars with high differential rotations in depth that enables the star to have a significant content of rotational energy, i.e. when the energy fraction $\tau\!=K/|W|$ ($K=$ rotational kinetic energy; $W=$ gravitational potential energy) becomes $\tau\gtrsim0.03$, the polar region is either strongly flattened or hollower \citep{boden71,clem74,clem79,Zo86,jack05}. Similar dimples can also be produced by a surface angular velocity accelerated enough towards the polar latitudes. To illustrate this effect, we adopt the following angular velocity profile, which is sometimes used in the literature to model internal conservative (cylindrical) laws \citep{jack05} for stars with high rotational energies ($0.02\lesssim\tau\lesssim0.22$)

\begin{equation}
\label{betlaw}
\Omega_{\rm s}(\theta) = \frac{\Omega_o}{1+\beta[R_{\rm s}(\theta)/R_{\rm e}]^2\sin^2\theta}\ ,
\end{equation}

\noindent where $R_{\rm s}(\theta)$ is the equation of the stellar surface and $\beta$ is a free parameter. A relation similar to Eq.~(\ref{betlaw}) may suit the rotational profiles shown in Fig.~\ref{constg}a if the angular velocity is accelerated from the equator to the pole. Figure~\ref{fbetlaw} shows the shape of model stars with a surface angular velocity profile given by expression (\ref{betlaw}) for $\beta\!=\!5.0$ and several values of ratio $\eta_o$ in the equator calculated with Eq.~(\ref{iter1}). It can be shown that using the Roche approximation and the rotational law in Eq.~(\ref{betlaw}), the ratio of equatorial to polar radii is then given by

\begin{equation}
\label{rerpbetlaw}
\frac{R_{\rm e}}{R_{\rm p}} = 1+\frac{\eta_o}{2}(1+\beta) \ .
\end{equation}

\noindent Writing $R^2_{\rm s}(\theta)=x^2+z^2$, the condition for a flattened/hollowed polar region to appear is $dz/dx\!\geq\!0$ at $x\!=\!0$. The condition to obtain a hollowed polar region is then given by

\begin{equation}
\label{minbet}
\eta_o(1+\beta)^2 \geq \left[1+\eta_o\left(\frac{1+\beta}{2}\right)\right]^3\ .
\end{equation}
 
\noindent Making $u\!=\!R_{\rm e}/R_{\rm p}\!=\!1+\eta_o(1+\beta)/2$, relation
(\ref{minbet}) becomes $u^3/(u-1)\!\geq\!2(1+\beta)$. Since it is $[u^3/(u-1)]_{\rm
min}$ for $u\!=\!1.5$, condition (\ref{minbet}) requires that $\beta\!\geq\!2.375$ independently of the value of $\eta_o$.\par
  Since the simplest equation to represent the rotationally deformed surfaces shown in Fig.~\ref{fbetlaw} is

\begin{equation}
\label{simplr}
\frac{R_{\rm s}(\theta)}{R_{\rm e}} = 1-\left(1-\frac{R_{\rm p}}{R_{\rm e}}\right)\cos\theta \ ,
\end{equation}

\noindent an alternative expression to relation (\ref{sol3}) for the surface angular velocities accelerated from the equator to the pole could be

\begin{equation}
\label{alter}
\Omega_{\rm s}(\theta) = \frac{\Omega_o}{1+\beta\left[1-\left(1-\frac{R_{\rm p}}{R_{\rm e}}
\right)\cos\theta\right]^2\sin^2\theta}\ ,
\end{equation}

\noindent would require us to infer the ratio $R_{\rm p}/R_{\rm e}$ from interferometry, which according to Eq.~(\ref{rerpbetlaw}) is related to $\eta_o$ and $\beta$, and $\beta$ derived from spectroscopy. {\it However, before using expressions of the type given by either Eq.~(\ref{betlaw}) or (\ref{alter}), it would perhaps be preferable to await ob\-ser\-va\-tio\-nal confirmation that stellar shapes similar to those depicted in Fig.~\ref{fbetlaw} actually exist}.\par

\begin{table}[]
\centering
\caption[]{Models of stars with rigid rotation}
\scriptsize
\label{medk2jt}
\begin{tabular}{cccccccc}
\hline
ZAMS & \multicolumn{7}{c}{$M = 5M_{\odot}$, \ \ \  $t/t_{\rm MS}=0.011$,
\ \ \ $\Omega_{\rm cr}=1.92\times10^{-4}$}\\
\hline
\noalign{\smallskip}
 $\Omega/\Omega_{\rm cr}$ & $\rho_{\rm c}$ & $R_{\rm p}/R_{\odot}$ & 
 $R_{\rm e}/R_{\rm p}$ & $V_{\rm eq}$ & $J/M$ & $\eta_o$ & $K/|W|$ \\
 & & & & & $\times10^{-17}$ &  & $\times10^2$ \\
\noalign{\smallskip}
\hline
\noalign{\smallskip}
  0.00 & 19.498 & 2.644 & 1.000 &   0 & 0.000 & 0.000 & 0.000 \\    
  0.50 & 19.688 & 2.611 & 1.050 & 183 & 2.091 & 0.096 & 0.172 \\
  0.60 & 19.773 & 2.598 & 1.076 & 224 & 2.515 & 0.147 & 0.248 \\
  0.70 & 19.875 & 2.583 & 1.111 & 268 & 2.944 & 0.216 & 0.339 \\
  0.80 & 19.996 & 2.560 & 1.160 & 316 & 3.378 & 0.311 & 0.444 \\
  0.90 & 20.136 & 2.539 & 1.239 & 377 & 3.819 & 0.466 & 0.564 \\
  0.95 & 20.213 & 2.531 & 1.311 & 418 & 4.043 & 0.606 & 0.630 \\
  0.98 & 20.262 & 2.526 & 1.387 & 455 & 4.179 & 0.755 & 0.671 \\
  0.99 & 20.279 & 2.521 & 1.422 & 470 & 4.225 & 0.824 & 0.686 \\
  1.00 & 20.295 & 2.517 & 1.512 & 503 & 4.270 & 0.996 & 0.700 \\
\noalign{\smallskip}
\hline
\noalign{\smallskip}
TAMS & \multicolumn{7}{c}{$M = 5M_{\odot}$, \ \ \ $t/t_{\rm MS}=1.000$,
\ \ \ $\Omega_{\rm cr}=4.79\times10^{-5}$}\\
\noalign{\smallskip}
\hline
\noalign{\smallskip}
  0.00 & 27.479 & 6.454 & 1.000 &   0 & 0.000 & 0.000 & 0.000 \\    
  0.50 & 27.527 & 6.432 & 1.044 & 112 & 0.952 & 0.088 & 0.024 \\
  0.60 & 27.548 & 6.424 & 1.068 & 137 & 1.145 & 0.135 & 0.034 \\
  0.70 & 27.573 & 6.411 & 1.101 & 165 & 1.340 & 0.200 & 0.046 \\
  0.80 & 27.672 & 6.397 & 1.149 & 196 & 1.536 & 0.294 & 0.061 \\
  0.90 & 27.636 & 6.382 & 1.228 & 234 & 1.735 & 0.448 & 0.077 \\
  0.95 & 27.654 & 6.375 & 1.296 & 260 & 1.836 & 0.582 & 0.086 \\
  0.98 & 27.666 & 6.370 & 1.369 & 282 & 1.896 & 0.723 & 0.092 \\
  0.99 & 27.670 & 6.369 & 1.410 & 293 & 1.917 & 0.804 & 0.094 \\
  1.00 & 27.682 & 6.367 & 1.510 & 317 & 1.936 & 0.996 & 0.096 \\
\noalign{\smallskip}
\hline
\noalign{\smallskip}
ZAMS & \multicolumn{7}{c}{$M = 15M_{\odot}$, \ \ \  $t/t_{\rm MS}=0.013$,
\ \ \ $\Omega_{\rm cr}=1.34\times10^{-4}$}\\
\noalign{\smallskip}
\hline
\noalign{\smallskip}
  0.00 & 5.781 & 4.895 & 1.000 &   0 &  0.000 & 0.000 & 0.000 \\    
  0.50 & 5.847 & 4.825 & 1.052 & 237 &  6.510 & 0.099 & 0.268 \\
  0.60 & 5.877 & 4.785 & 1.079 & 290 &  7.839 & 0.151 & 0.386 \\
  0.70 & 5.914 & 4.742 & 1.115 & 346 &  9.184 & 0.220 & 0.528 \\
  0.80 & 5.956 & 4.698 & 1.166 & 409 & 10.551 & 0.319 & 0.692 \\
  0.90 & 6.006 & 4.647 & 1.246 & 485 & 11.947 & 0.475 & 0.880 \\
  0.95 & 6.034 & 4.629 & 1.321 & 540 & 12.660 & 0.619 & 0.984 \\
  0.98 & 6.051 & 4.603 & 1.385 & 580 & 13.095 & 0.743 & 1.050 \\
  0.99 & 6.058 & 4.596 & 1.421 & 600 & 13.240 & 0.814 & 1.072 \\
  1.00 & 6.064 & 4.591 & 1.517 & 644 & 13.387 & 0.994 & 1.095 \\
\noalign{\smallskip}
\hline
\noalign{\smallskip}
TAMS & \multicolumn{7}{c}{$M = 15M_{\odot}$, \ \ \  $t/t_{\rm MS}=1.000$,
\ \ \ $\Omega_{\rm cr}=2.76\times10^{-5}$}\\
\noalign{\smallskip}
\hline
\noalign{\smallskip}
  0.00 & 9.226 & 13.346 & 1.000 &   0 & 0.000 & 0.000 & 0.000 \\    
  0.50 & 9.236 & 13.306 & 1.045 & 133 & 2.268 & 0.088 & 0.022 \\
  0.60 & 9.240 & 13.289 & 1.068 & 163 & 2.728 & 0.135 & 0.032 \\
  0.70 & 9.245 & 13.268 & 1.101 & 196 & 3.192 & 0.200 & 0.044 \\
  0.80 & 9.251 & 13.241 & 1.149 & 233 & 3.661 & 0.294 & 0.057 \\
  0.90 & 9.259 & 13.212 & 1.228 & 279 & 4.136 & 0.447 & 0.072 \\
  0.95 & 9.263 & 13.197 & 1.297 & 310 & 4.377 & 0.581 & 0.081 \\
  0.98 & 9.265 & 13.189 & 1.369 & 336 & 4.522 & 0.722 & 0.087 \\
  0.99 & 9.266 & 13.186 & 1.410 & 350 & 4.571 & 0.802 & 0.088 \\
  1.00 & 9.267 & 13.183 & 1.511 & 378 & 4.619 & 0.996 & 0.090 \\
\noalign{\smallskip}
\hline
\noalign{\smallskip}
\multicolumn{8}{l}{Note: $\rho_{\rm c}$ is given in g\,cm$^{-3}$; 
$\Omega_{cr}$ is given in s$^{-1}$; $V_{\rm eq}$ is given in km\,s$^{-1}$; 
the units of $J/M$}\\
\multicolumn{8}{l}{are cm$^2$s$^{-1}$; $t$ is age and $t_{\rm MS}$ is the MS
life span}\\
\noalign{\smallskip}
\hline
\end{tabular}
\end{table}

\section{The gravity darkening effect in stars with non-conservative rotation laws}
\label{tdss}

  Owing to the radiative equilibrium that prevails in the atmosphere of massive and intermediate-mass stars, the emerging bolometric radiative flux $F$ is proportional to the temperature gradient $\nabla T$ in the outer layers of the stellar envelope. Since these layers are also in hydrostatic equilibrium, their pressure gradient $\nabla P\propto\nabla T$ is balanced by the surface effective gravity ${\bf g_{\rm eff}}$. We have then the following phenomenological relation that describes the principle of von Zeipel's relation

\begin{equation}
\label{vzpp}
F \propto \nabla T \equiv \nabla P \propto {\rm\bf g_{\rm eff}}\ .
\end{equation}

 As $F\propto T^4_{\rm eff}$, where $T_{\rm eff}$ is the effective temperature at a given point $\overline r$ in the stellar surface, it follows the known von Zeipel's
relation \citep{vzei24}

\begin{equation}
\label{vzpr1}
T_{\rm eff} = \mathscr{C}(\overline r)\times{\rm\bf g_{\rm eff}}^{1/4}\ .
\end{equation}

 However, according to Poincar\'e-Wavre's theorem \citep{tass78} only when the rotational law in the external layers is conservative do the constant equipotential, density, and temperature surfaces coincide to consider that $\mathscr{C}(\overline r)= \overline{\mathscr{C}}=$ constant. Otherwise, as for rotational laws discussed in Sect.~\ref{tbbr}, $\mathscr{C}(\overline r)$ is not constant with the colatitude $\theta$. Its detailed expression for the non-conservative case with shellular rotation was obtained by \citet{mae99}.\par
 An equivalent relation to Eq.~(\ref{vzpr1}) for spectroscopic and interferometric analysis can be given the form

\begin{equation}
\label{vzpr}
T_{\rm eff} = \overline{\mathscr{C}}\times{\rm\bf g_{\rm eff}}^{\betaup_{\rm GD}}\ ,
\end{equation}

\noindent where $\betaup_{\rm GD}=0.25+\delta$, so that $\delta$ masks the variation in $\mathscr{C}(\overline r)$ over the stellar surface. The particular notation $\betaup_{\rm GD}$ used for the gravity darkening power in relation (\ref{vzpr}) is introduced to avoid confusions with the $\beta$ parameter that appears in the internal rotation law (\ref{betlaw}). Two-dimensional (2D) models of rotating stars and some observations suggest that $\delta\lesssim0$ [c.f. \citet{lucy67,lovek06,vanb06,monn07,zhao09}]. We insist on $\delta$ not being in principle the function of taking into account the change in the stellar surface from radiative to possible convective equilibrium-dominated layers because of the strong change in $T_{\rm eff}$ with $\theta$, as seems to be understood in \citet{auf06} and \citet{zhao09}, but mainly the non-conservative nature of the rotational law in the external layers.\par

\section{Models of rotating stars}
\label{mrots}

 To study the effect of the centrifugally distorted internal mass distribution on the gravitational potential in the surface of a star, to test the validity of the Roche approximation, it is enough to calculate models of stellar structure where only the global dynamical aspects induced by the rotation are considered. To this end, we differentiate the primary effects produced by the rotation from those induced by the evolution. The primary thermodynamic effects due to the stellar evolution are taken into account by making use of the relations between the pressure and density calculated with one-dimensional models of stellar evolution without rotation. We assume thus that the changes produced on the barotropic relation $P=P(\rho)$ at a given evolutionary stage of a star by the several instabilities and the diffusion of chemical elements unleashed through the stellar evolution by the rotation, have second order effects on the establishment of the dynamical equilibrium of the rotating star. In principle, nothing prevents us from using the barotropic relations derived with models of stellar evolution with shellular rotation, but the results will probably not be more reliable. They all are calculated for low energy ratios $\tau=K/|W|$, which do not correspond to the model inputs regarding the rotational law and energy tested in the present attempt.\par
 Since we are not interested in the precise description of all non-linear time-dependent phenomena associated with the viscosity and internal flows in rotating stars, and because the total energy carried by the meridional circulation is small, our models are axisymmetric, steady state, and circulation free. Owing to these assumptions and Poincar\'e-Wavre's theorem \citep{tass78}, our model stars behave as barotropes. We then adopt  internal rotational laws of conservative form, $\Omega\!=\!\Omega(\varpi)$, where $\varpi$ is the distance to the rotation axis. According to the discussion in Sect.\ref{shr}, the conservative laws are expected to produce stronger stellar geometrical deformations for a given amount of rotational kinetic energy than the homologous non-conservative ones. \par
 For conservative rotation laws, the gravitational potential $\Phi(\varpi,z)$ and the density  distribution $\rho(\varpi,z)$ in the rotating star are simultaneous solutions to the hydrostatic equilibrium equation

\begin{equation}
\rho^{-1}\nabla P = \nabla\Phi+j^2\varpi^{-3}\hat{e}_{\varpi}
\label{hydeq}
\end{equation}

\noindent and the Poisson equation

\begin{equation}
\Delta\Phi = 4\pi G\rho,
\label{poiss}
\end{equation}

\noindent where ($\varpi,\phi,z$) are the cylindrical coordinates with $z$ containing the rotation axis, $\hat{e}_{\varpi}$ is the unit vector perpendicular to the $z$-axis, $P$ is the pressure, and $j=\Omega\varpi^2$ is the specific angular momentum.\par
  Equations (\ref{hydeq}) and (\ref{poiss}) are solved with the adopted complementary barotropic relation

\begin{equation}
P = P(\rho) \ .
\label{barot}
\end{equation}

\noindent The $P\!=\!P(\rho)$ relations used in this work have a two-component polytropic character

\begin{equation}
P = a\rho^{\gamma_a}+b\rho^{\gamma_b}\ ,
\label{twobarot}
\end{equation}

\noindent where the constants $a$, $b$, $\gamma_a$, and $\gamma_b$ were adjusted so as to: a) reproduce the pressure $P_{\rm c}$ and the density $\rho_{\rm c}$ in the center of the non-rotating star of given mass and evolutionary stage; b) ensure a continuous distribution of the pressure-density relation at the radius of the stellar core; c) obtain the correct stellar mass at the stellar radius as tabulated by \citet{schal92} for 1-D evolutionary models for the initial metallicity $Z=0.02$. The function (\ref{twobarot}) is continued in the stellar atmosphere by another pressure-density relation calculated by \citet{caskur03} for stellar atmospheres as a function of the parameters ($T_{\rm eff},\log g$).\par
 The first order effects due to the stellar evolution are thus accounted for by the pressure-density relations at the center of the star and the $\partial P/\partial\rho$ gradients. An additional term in relation (\ref{twobarot}) could also in principle take into account the presence of the convective regions in the envelope induced by the rapid rotation, but this was not done in the present approach. The only rotational effect on the $P\!=\!P(\rho)$ relation considered here is by means of the mass-compensation effect \citep{sack70}, which increases the density $\rho_{\rm c}$ at the center of the star. To this end, we iterated $\rho_{\rm c}$ until the nominal stellar mass $M$ was attained. This iteration also implied that the central pressure $P_{\rm c}$ was changed in accordance.\par
  The gravitational potential $\Phi(\varpi,z)$ is obtained by solving the Poisson equation given in Eq.~(\ref{poiss}) with the cell-method adapted by \citet{clem74} to stellar structure calculations. The density distribution $\rho(\varpi,z)$ is derived from the integrated form of Eq.~(\ref{hydeq})

\begin{equation}
\int_{\rho_{\rm c}}^{\rho}\frac{dP}{\rho} = \Phi(\varpi,z)-\Phi_{\rm c}+\int_0^{\varpi}
\Omega^2(\varpi)\varpi d\varpi\ ,
\label{bernou}
\end{equation}

\noindent where $\Phi_{\rm c}$ is the gravitational potential at the stellar center. Given a rotational law $\Omega(\varpi)$ and a barotropic relation in Eq.~(\ref{barot}), the common solution to equations (\ref{hydeq}) and (\ref{poiss}) is performed over the entire space. The iteration of $\Phi$ and $\rho$ is stopped when the highest density difference in the ($\varpi,z$)-space is $max(\delta\rho/\rho)\!\la\!10^{-6}$. In our iterations, the virial relation $\delta=[2(K+U)-W]/|W|=0$, where $K$ = kinetic energy, $U$ = internal energy, and $W$ = total gravitational potential energy, is verified to better than $\delta\approx2\times10^{-4}$ in the ZAMS models, and $\delta\approx6\times10^{-3}$ by the TAMS models. Since in the framework of conservative rotational laws, the surfaces of constant pressure, density, and total potential are parallel, the rotationally distorted shape of our models is defined by the total equipotential surface that contains the polar `photospheric' radius $R_{\rm p}$. This radius is identified by the layer whose density satisfies the model-atmosphere relation $\tau_{\rm Ross}(\rho)\!=\!2/3$  in the stellar atmosphere models of \citet{caskur03}. The local effective temperature in the pole needs to satisfy also the gravity darkening effect. We modified accordingly the effective temperature given by \citet{schal92} for the given mass $M$ using von Zeipel's approximation \citep{vzei24}. The transformation to the rotation-dependent effective temperature was performed following the procedure given in \citet{frem05}.\par 

\begin{figure*}[ht!]
\centerline{\psfig{file=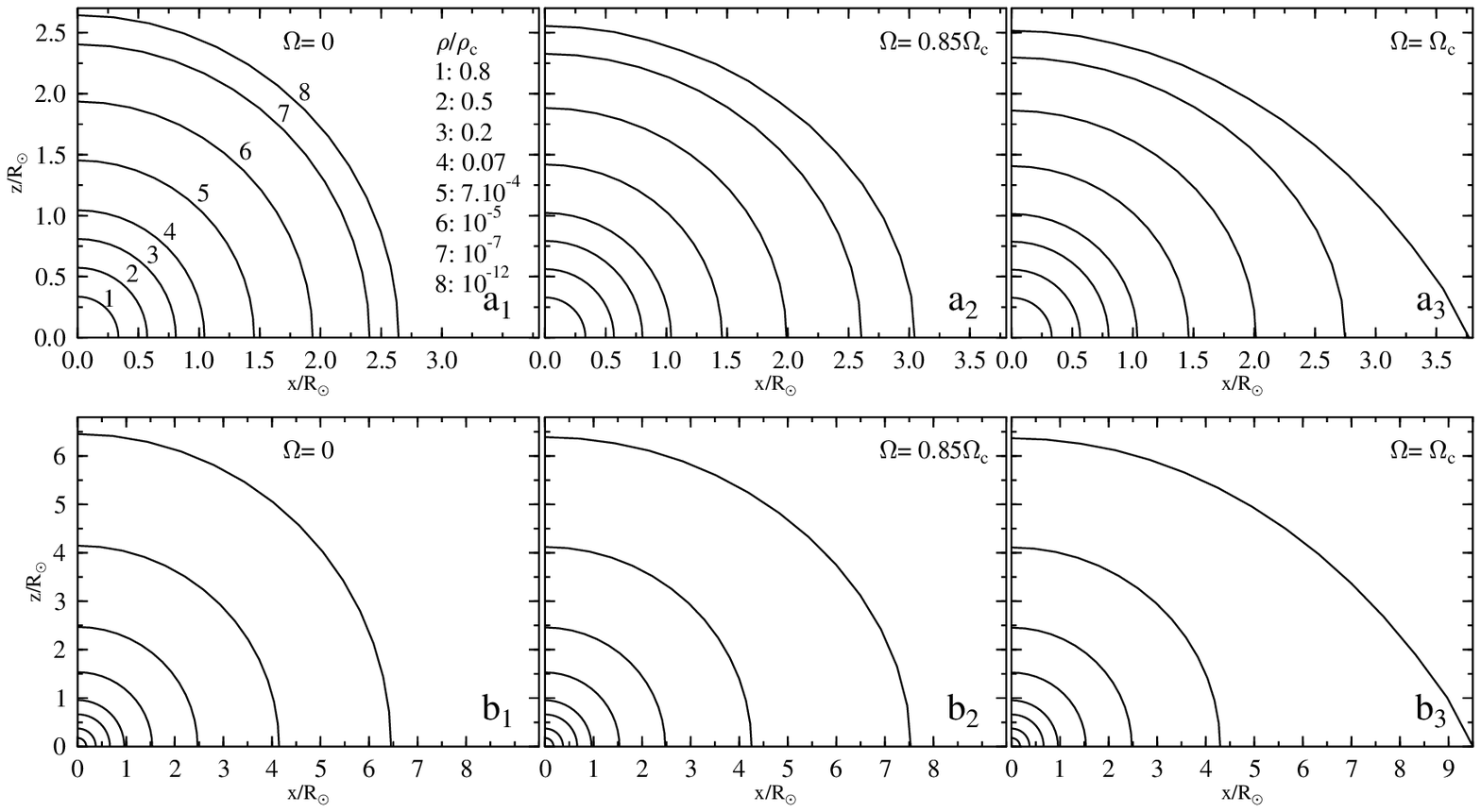,width=18truecm,height=9.8truecm}}
\caption[]{\label{kw-f} \underline{Rigid rotators}. {\bf a}: Iso-density surfaces in model stars of $M=5M_{\odot}$ in the ZAMS rotating at (a$_1$) $\Omega=0$; (a$_2$) $\Omega=0.85\Omega_{\rm c}$; (a$_3$) $\Omega=\Omega_{\rm c}=1.92\times10^{-4}$ s$^{-1}$; {\bf b}: Iso-density surfaces in stars of $M=5M_{\odot}$ in the TAMS rotating at (b$_1$) $\Omega=0$; (b$_2$) $\Omega=0.85\Omega_{\rm c}$; (b$_3$) $\Omega =\Omega_{\rm c}=4.79\times10^{-5}$ s$^{-1}$.The iso-density surfaces are labeled with the corresponding density ratios $\rho/\rho_{\rm c}$, which are the same in all panels of the figure.}
\centerline{\psfig{file=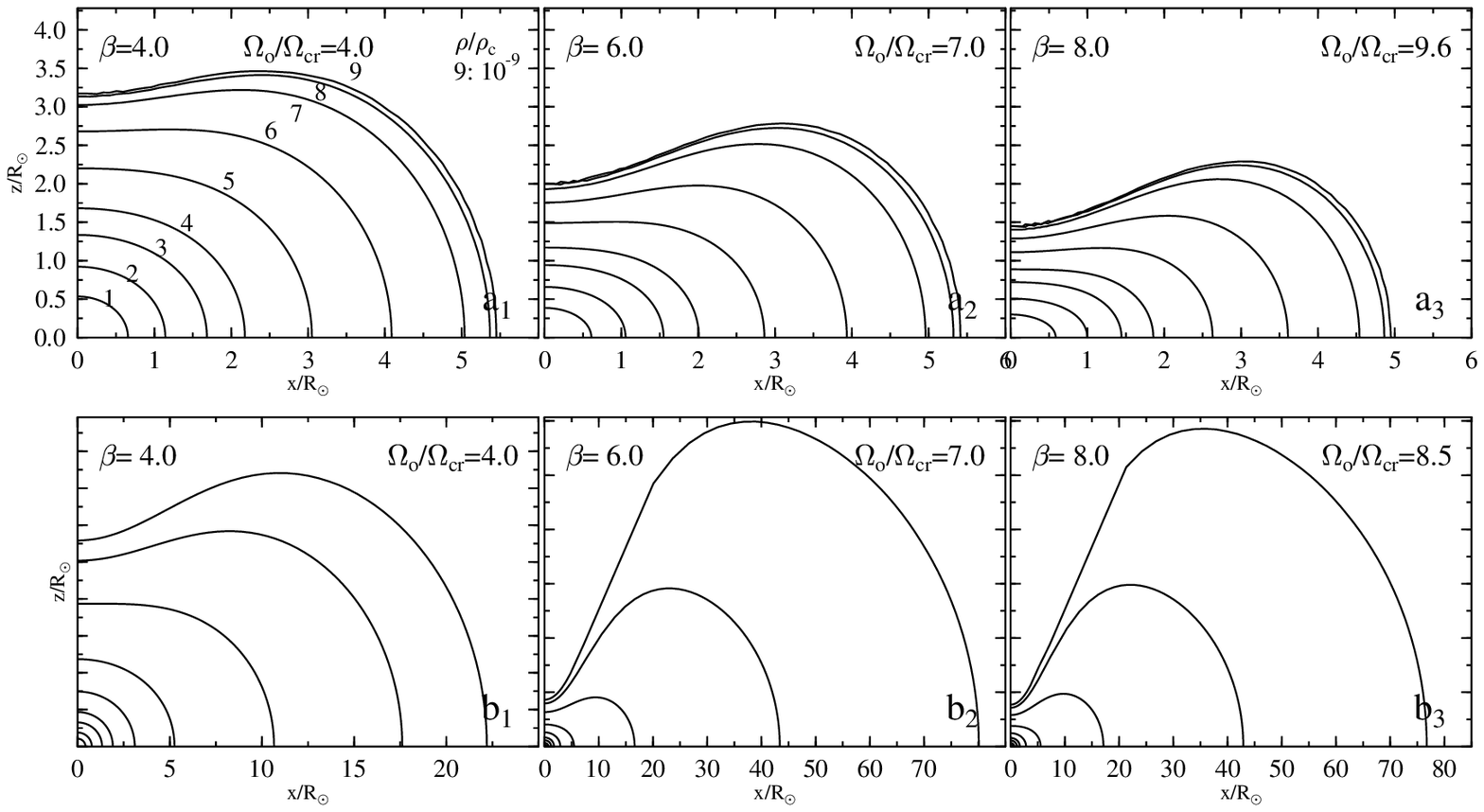,width=18truecm,height=9.8truecm}}
\caption[]{\label{dkw-f} \underline{Differential rotators}. {\bf a}: Iso-density surfaces in model stars of $M=15M_{\odot}$ in the ZAMS having the internal rotation law (\ref{bo}), whose parameters $\Omega_o/\Omega_{cr}$ and $\beta$ are indicated. $\Omega_{cr}$ is for the critical rigid rotation. The iso-density surfaces are labeled with the corresponding density ratios $\rho/\rho_c$, which are the same as in Fig.~\ref{kw-f} and for all panels of this figure. {\bf b}: Iso-density surfaces in model stars of $M=15M_{\odot}$ in the TAMS having the internal rotation law (\ref{bo}), whose parameters $\Omega_o/\Omega_{cr}$ and $\beta$ are indicated. In each panel, the ordinates are in the same scale as the abscissas, but they differ from one to the other panel.}
\end{figure*}

\begin{table*}[htpb]
\centering
\caption[]{Models of stars with internal differential rotation}
\scriptsize
\label{modkw}
\begin{tabular}{cccccccc|cccccccc}
\hline
ZAMS & \multicolumn{7}{c|}{$M = 5M_{\odot}$,\ \ \ $\Omega_{\rm cr}=1.92\times10^{-4}$} & ZAMS & 
\multicolumn{7}{c}{$M = 15M_{\odot}$,\ \ \ $\Omega_{\rm cr}=1.34\times10^{-4}$}\\
\hline
\noalign{\smallskip}
\smallskip
$\Omega_o/\Omega_{\rm cr}$ & $\rho_{\rm c}$ & $R_{\rm av}/R_{\odot}$ & $R_{\rm e}/R_{\odot}$ & $R_{\rm e}/R_{\rm p}$& $V_{\rm eq}$ & $J/M$            & $K/|W|$ &
$\Omega_o/\Omega_{\rm cr}$ & $\rho_{\rm c}$ & $R_{\rm av}/R_{\odot}$ & $R_{\rm e}/R_{\odot}$ & $R_{\rm e}/R_{\rm p}$& $V_{\rm eq}$ & $J/M$            & $K/|W|$ \\
                           &                &                        &                       &                      &         & $\times10^{-17}$ &         &
                           &                &                        &                       &                      &         & $\times10^{-17}$ &         \\
\noalign{\smallskip}
\hline
\noalign{\smallskip}
\multicolumn{8}{c|}{$\beta=2$} & \multicolumn{8}{c}{$\beta=2$} \\
  1.0 & 19.996 &  2.671 &  2.711 & 1.052 & 117 &  3.326 & 0.0044  &  1.0 &  5.945 &  4.915 &  4.986 & 1.047 & 151 &  9.920 & 0.0064  \\
  2.0 & 21.593 &  2.821 &  2.967 & 1.269 & 225 &  6.723 & 0.0175  &  2.0 &  6.482 &  5.092 &  5.383 & 1.265 & 294 & 20.091 & 0.0250 \\
  3.0 & 24.741 &  3.451 &  3.938 & 1.947 & 290 & 10.437 & 0.0389  &  3.0 &  7.580 &  5.828 &  6.666 & 1.878 & 396 & 31.115 & 0.0553 \\
  3.5 & 27.451 &  4.813 &  5.799 & 3.143 & 255 & 12.898 & 0.0532  &  3.5 &  8.529 &  7.330 &  8.938 & 2.809 & 381 & 38.020 & 0.0752 \\
  3.9 & 32.121 &  9.933 & 11.652 & 6.998 & 152 & 17.775 & 0.0683  &  4.0 & 10.930 & 16.655 & 20.092 & 7.455 & 216 & 58.322 & 0.1035 \\
\multicolumn{8}{c|}{$\beta=4$} & \multicolumn{8}{c}{$\beta=4$} \\
  1.0 & 19.870 &  2.639 &  2.656 & 1.023 &  70 &  2.824 & 0.0033  &  1.0 &  5.899 &  4.875 &  4.915 & 1.025 &  91 &  8.220 & 0.0046 \\
  2.0 & 21.022 &  2.667 &  2.729 & 1.138 & 139 &  5.642 & 0.0130  &  2.0 &  6.273 &  4.861 &  4.981 & 1.133 & 181 & 16.443 & 0.0179 \\
  3.0 & 23.095 &  2.731 &  2.867 & 1.353 & 202 &  8.458 & 0.0285  &  4.0 &  8.085 &  4.955 &  5.455 & 1.697 & 341 & 32.810 & 0.0663 \\
  4.0 & 26.378 &  2.906 &  3.185 & 1.728 & 250 & 11.318 & 0.0489  &  5.5 & 11.219 &  5.774 &  6.887 & 2.843 & 396 & 45.878 & 0.1163 \\
  5.0 & 31.538 &  3.400 &  3.899 & 2.489 & 269 & 14.450 & 0.0736  &  6.2 & 14.319 &  7.599 &  9.291 & 4.558 & 349 & 56.026 & 0.1457 \\
  6.2 & 46.905 &  6.692 &  7.790 & 6.456 & 181 & 22.434 & 0.1128  &  6.4 & 16.610 &  9.221 & 11.247 & 5.958 & 304 & 64.434 & 0.1588 \\
\multicolumn{8}{c|}{$\beta=6$} & \multicolumn{8}{c}{$\beta=6$} \\
  1.0  & 19.796 &  2.634 &  2.652 & 1.022 &  50 &  2.482 & 0.0027  &  1.0  &  5.874 &  4.856 &  4.889 & 1.010 &  65 &  7.103 & 0.0035 \\
  2.0  & 21.022 &  2.626 &  2.670 & 1.095 & 100 &  4.946 & 0.0104  &  3.0  &  6.681 &  4.726 &  4.869 & 1.221 & 196 & 21.149 & 0.0303 \\
  4.0  & 23.095 &  2.631 &  2.768 & 1.426 & 195 &  9.761 & 0.0392  &  5.0  &  8.684 &  4.544 &  4.941 & 1.693 & 324 & 34.451 & 0.0770 \\
  5.0  & 28.193 &  2.677 &  2.907 & 1.727 & 235 & 12.103 & 0.0586  &  7.0  & 12.871 &  4.618 &  5.417 & 2.664 & 424 & 46.715 & 0.1336 \\
  7.0  & 40.196 &  3.160 &  3.641 & 2.969 & 275 & 17.048 & 0.1047  &  8.0  & 16.731 &  5.171 &  6.299 & 3.689 & 430 & 53.793 & 0.1648 \\
  8.35 & 67.403 &  5.514 &  6.446 & 6.938 & 196 & 25.828 & 0.1483  &  8.4  & 19.300 &  5.767 &  7.106 & 4.564 & 408 & 58.318 & 0.1796 \\
\multicolumn{8}{c|}{$\beta=8$} & \multicolumn{8}{c}{$\beta=8$} \\
  1.0  & 19.746 &  2.626 &  2.640 & 1.009 &  39 &  2.227 & 0.0022  &  1.0  &  5.857 &  4.857 &  4.884 & 1.011 &  51 &  6.293 & 0.0029 \\  
  3.0  & 21.808 &  2.580 &  2.630 & 1.164 & 118 &  6.610 & 0.0190  &  2.0  &  6.092 &  4.790 &  4.837 & 1.066 & 102 & 12.548 & 0.0112 \\ 
  5.0  & 26.464 &  2.519 &  2.669 & 1.495 & 195 & 10.776 & 0.0492  &  4.0  &  7.143 &  4.551 &  4.722 & 1.304 & 209 & 24.704 & 0.0425 \\
  7.0  & 34.953 &  2.529 &  2.788 & 2.070 & 264 & 14.665 & 0.0878  &  6.0  &  9.316 &  4.229 &  4.610 & 1.733 & 319 & 35.796 & 0.0869 \\ 
  9.0  & 50.578 &  2.873 &  3.343 & 3.305 & 292 & 18.741 & 0.1313  &  8.0  & 13.338 &  3.980 &  4.573 & 2.373 & 428 & 45.301 & 0.1363 \\
 10.0  & 66.836 &  3.572 &  4.219 & 4.962 & 264 & 22.199 & 0.1575  &  9.6  & 18.901 &  4.078 &  4.940 & 3.327 & 484 & 52.380 & 0.1764 \\ 
\noalign{\smallskip}
\hline
\noalign{\smallskip}
TAMS & \multicolumn{7}{c|}{$M = 5M_{\odot}$,\ \ \ $\Omega_{\rm cr}=4.79\times10^{-4}$} & TAMS & \multicolumn{7}{c}{$M = 15M_{\odot}$,\ \ \ $\Omega_{\rm cr}=2.76\times10^{-4}$}\\
\noalign{\smallskip}
\hline
\noalign{\smallskip}
\multicolumn{8}{c|}{$\beta=2$} & \multicolumn{8}{c}{$\beta=2$} \\
  1.0 & 27.626 &  6.639 &  7.708 & 1.051 &  71 &  1.692 & 0.0008  &  1.0 & 9.255 & 13.564 & 13.777 & 1.049 &  86 &  4.119 & 0.0007 \\
  2.0 & 28.085 &  7.556 &  7.915 & 1.280 & 132 &  3.482 & 0.0031  &  2.0 & 9.357 & 15.546 & 16.370 & 1.285 & 160 &  8.521 & 0.0030 \\
  3.0 & 29.015 & 12.077 & 13.648 & 2.334 & 137 &  5.633 & 0.0072  &  3.0 & 9.572 & 25.858 & 25.259 & 2.411 & 162 & 14.021 & 0.0073 \\
  3.5 & 29.954 & 27.310 & 30.655 & 5.467 &  78 &  7.475 & 0.0104  &  3.5 & 9.982 & 66.541 & 73.661 & 6.333 &  81 & 19.758 & 0.0107 \\
\multicolumn{8}{c|}{$\beta=4$} & \multicolumn{8}{c}{$\beta=4$} \\
  1.0  & 27.601 &  6.535 &  6.564 &  1.028 &  43 &  1.541 & 0.0006  & 1.0  &  9.249 & 13.363 & 13.483 &  1.026 &  52 &  3.753 & 0.0006 \\
  3.0  & 28.632 &  7.844 &  8.169 &  1.384 & 110 &  4.829 & 0.0059  & 2.0  &  9.331 & 14.200 & 14.476 &  1.128 &  99 &  7.644 & 0.0025 \\
  4.0  & 29.723 & 10.200 & 11.025 &  1.984 & 116 &  6.805 & 0.0107  & 3.0  &  9.480 & 16.212 & 16.970 &  1.387 & 133 & 11.882 & 0.0058 \\
  5.0  & 31.729 & 19.386 & 20.867 &  4.110 &  81 &  9.748 & 0.0177  & 4.0  &  9.732 & 21.574 & 23.300 &  2.021 & 138 & 16.979 & 0.0108 \\
  5.5  & 34.006 & 42.148 & 44.030 &  9.199 &  43 & 13.124 & 0.0221  & 5.0  & 10.236 & 44.869 & 48.918 &  6.602 &  87 & 25.481 & 0.0183 \\
  5.7  & 36.217 & 85.056 & 88.533 & 19.108 &  22 & 16.649 & 0.0246  & 5.5  & 11.028 & 129.78 & 136.36 &  9.820 &  35 & 39.665 & 0.0240 \\
\multicolumn{8}{c|}{$\beta=6$} & \multicolumn{8}{c}{$\beta=6$} \\
  1.0  & 27.584 &  6.497 &  6.515 &  1.021 &  31 &  1.428 & 0.0006  & 1.0  &  9.246 & 13.246 & 13.307 &  1.013 &  37 &  3.479 & 0.0006 \\
  3.0  & 28.440 &  7.200 &  7.383 &  1.232 &  83 &  4.396 & 0.0051  & 3.0  &  9.437 & 14.815 & 15.268 &  1.238 &  72 & 10.791 & 0.0050 \\
  4.0  & 29.275 &  8.086 &  8.447 &  1.494 & 100 &  6.025 & 0.0091  & 4.0  &  9.626 & 16.812 & 17.571 &  1.498 & 121 & 14.905 & 0.0091 \\
  5.0  & 30.516 &  9.898 & 10.573 &  2.007 & 103 &  7.883 & 0.0145  & 5.0  &  9.914 & 20.975 & 22.325 &  2.043 & 123 & 19.763 & 0.0148 \\
  6.0  & 32.471 & 14.334 & 15.413 &  3.190 &  88 & 10.297 & 0.0216  & 6.0  & 10.393 & 31.844 & 34.264 &  3.419 &  99 & 26.539 & 0.0223 \\
  7.0  & 36.522 & 32.588 & 33.300 &  7.737 &  48 & 14.903 & 0.0308  & 7.0  & 11.634 & 91.051 & 95.665 & 10.725 &  42 & 43.227 & 0.0334 \\
\multicolumn{8}{c|}{$\beta=8$} & \multicolumn{8}{c}{$\beta=8$} \\
  1.0  & 27.571 &  6.474 &  6.496 &  1.018 &  24 &  1.337 & 0.0005  & 1.0  &  9.243 & 13.239 & 13.327 &  1.006 &  29 &  3.260 & 0.0005 \\
  3.0  & 28.311 &  6.933 &  7.037 &  1.173 &  67 &  4.082 & 0.0045  & 3.0  &  9.408 & 14.238 & 14.472 &  1.164 &  82 & 10.008 & 0.0049 \\
  5.0  & 29.992 &  8.280 &  8.640 &  1.610 &  94 &  7.099 & 0.0127  & 5.0  &  9.790 & 17.301 & 18.070 &  1.624 & 113 & 17.656 & 0.0128 \\
  6.0  & 31.368 &  9.791 & 10.361 &  2.081 &  96 &  8.858 & 0.0185  & 6.0  & 10.112 & 20.816 & 22.169 &  2.148 & 113 & 22.314 & 0.0189 \\
  8.0  & 36.726 & 20.408 & 21.171 &  5.150 &  65 & 14.213 & 0.0345  & 7.0  & 10.602 & 28.159 & 30.082 &  3.159 &  99 & 28.363 & 0.0267 \\
  9.0  & 47.921 & 88.304 & 88.136 & 24.993 &  18 & 24.175 & 0.0453  & 8.5  & 12.570 & 86.558 & 90.257 & 11.410 &  41 & 49.916 & 0.0438 \\
\noalign{\smallskip}
\hline
\noalign{\smallskip}
\multicolumn{16}{l}{Note: $\rho_{\rm c}$ is given in g\,cm$^{-3}$; $\Omega_{cr}$ is given in s$^{-1}$; $V_{\rm eq}$ is given in km\,s$^{-1}$; the units of $J/M$ are cm$^2$s$^{-1}$; $R_{\rm av}/R_{\odot}$ is the radius of a sphere having the same volume as the rotatio-}\\
\multicolumn{16}{l}{nally deformed object}\\
\noalign{\smallskip}
\hline
\end{tabular}
\end{table*}

\subsection{Rigid rotation}
\label{rr}
 
 The simplest conservative rotational law is that of rigid rotation. The models of stars with rigid rotation were calculated assuming that at each evolutionary phase they experience instantaneous total redistribution of their internal angular momentum. The characteristics of these models are given in Table~\ref{medk2jt}. In this table, $t/t_{\rm MS}$ is the fractional age of the star, where $t_{\rm MS}$ is the time that a non-rotating star of mass $M$ spends on the main sequence, $\Omega_{\rm cr}$ is the critical angular velocity, $\Omega/\Omega_{\rm cr}$ represents the angular velocity for which the model was calculated, $\rho_{\rm c}$ is the core density of the rotating object, $R_{\rm e}/R_{\odot}$ is the equatorial radius of the model star in solar units and $R_{\rm e}/R_{\rm p}$ the equatorial-to-polar radii ratio, $V_{\rm eq}$ is the equatorial linear velocity in km\,s$^{-1}$, $J/M$ is the total specific angular momentum, $\eta\!=\!\Omega^2R_{\rm e}^3/GM$ is the ratio of centrifugal to the gravitational acceleration in the equator, and $K/|W|$ is the ratio of the kinetic rotational energy ($K$) to the absolute value of the gravitational potential energy ($W$). Some of these models are shown in Fig.~\ref{kw-f}, where in all cases the iso-density surfaces are for the same $\rho/\rho_{\rm c}$ density ratios. Similar models were also calculated for other masses and age ratios $t/t_{\rm MS}$. In spite of the simple approach used to calculate them, the radius ratios compare very closely with those calculated by \citet{ekst08} for the same $K/|W|$ energy ratios. We note, however, that the models obtained by these authors cannot be compared directly with ours because the distribution of the internal rotational velocity is not the same. Starting from a  quasi-rigid rotation in the ZAMS, \citet{ekst08} accounted for a consistent evolution of the angular momentum distribution inside the star throughout the calculated stellar evolution span.\par

\subsection{Shellular differential rotation}
\label{shr}

\begin{figure}[Ht!]
\centerline{\psfig{file=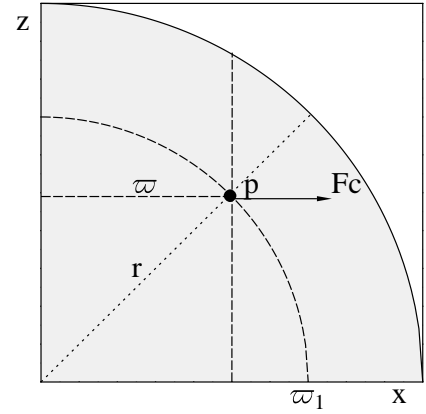,width=7.0truecm,height=7.0truecm}}
\caption[]{\label{shcyl}Schematic comparison of the centrifugal force produced by a shellular and a cylindrical rotation law when both are described by the same analytical function}
\end{figure}

 \citet{zahn10} studied the shapes of stars with internal shellular differential rotation and concluded that in $7M_{\odot}$ stars the radius ratio $R_{\rm e}/R_{\rm p}$ at critical equatorial rotation can be enlarged from 1\% to 4\%, depending on the internal rotational energy content and the evolutionary phase. However, it can easily be seen that when two rotation laws are described with the same function and both imply the same central to surface angular velocity ratios $\Omega_{\rm center}/\Omega_{\rm surface}$, but one is shellular and the other cylindrical, the law that is shellular may have a weaker effect on the global internal density distribution than the cylindrical one. For simplicity, we change the independent variable P (pressure) in the shellular law by the radius $r$. The comparison is depicted in Fig.~\ref{shcyl}: the specific centrifugal force acting at a point $p$ in the stellar interior is

\begin{eqnarray}
F_{\rm c}^{\rm shellular}   & = & \Omega^2(r)\varpi \nonumber \\ 
F_{\rm c}^{\rm cylindrical} & = & \Omega^2(\varpi,\theta)\varpi
\label{ctrfor1}
\end{eqnarray}

\noindent for shellular and cylindrical angular velocity distribution laws, respectively ($\varpi\!\!=r\sin\theta$ is the distance to the rotation axis). At the equator, if we assume that $\Omega^{\rm cylindrical}(\varpi,\pi/2)\!=\!\Omega^{\rm shellular}(r,\pi/2)$, for a given point $p(z,\varpi)$, it will be

\begin{eqnarray}
\Omega(p)^{\rm cylindrical} & > & \Omega^{\rm shellular}(p)\ , \nonumber \\ 
F_{\rm c}^{\rm cylindrical}(p) & > & F_{\rm c}^{\rm shellular}(p)\ .
\label{ctrfor2}
\end{eqnarray}

\noindent This holds in particular for the points in the stellar surface. However, since it has been shown in Sect.~\ref{form} that the external geometrical deformation of the star is mainly a function of the angular velocity law on the surface, cylindrical rotation laws will carry stronger geometrical deformations than shellular ones for a similar total kinetic energy. Moreover, in Sect.~\ref{tbbr} we have anticipated that strong cylindrical components of the angular velocity distributions in the external layers probably exist. It is then important to determine the order of magnitude of the quadrupole factor $\gamma$ in rotating stars using conservative internal rotational laws. To this end, in the next section we calculate two-dimensional models of rotating stars with conservative internal rotational laws.\par

\subsection{Cylindrical differential rotation}
\label{dr}

 The models with differential rotation were calculated using the same barotropic relations $P\!=\!P(\rho)$ as for the solid-body rotation. We adopted the internal angular velocity distributed as

\begin{equation}
\Omega(\varpi) = \frac{\Omega_o}{1+\beta(\varpi/R_{\rm e})^n}\ ,
\label{bo}
\end{equation}

\noindent where $\beta>0$ is a free parameter and $\Omega_o$ is the angular velocity at the axis of rotation. For simplicity, we considered $n\!=\!2$. Nevertheless, it can be shown that all models with $n\!\leq\!2$ obey the Solberg-H\o iland stability criterion, and that at whatever distance $\varpi$ the energy ratio $\tau(\varpi)$ is $K(\varpi)/|W(\varpi)|\!\lesssim\!0.138$. This ensures that no region in the stellar interior is unstable to the secular instability that can carry it to a three-axial Jacoby ellipsoidal configuration. \par 
  In Table~\ref{modkw}, we indicate the characteristics of the calculated models. The entries have the same meaning as in Table~\ref{medk2jt}. The higher $K/|W|$ values given in Table~\ref{modkw} are about the highest we could obtain for the given masses, ages, and $\beta$ parameters. Figure~\ref{dkw-f} shows some of these calculated models.\par
  Using the same algorithm, we can obtain models that approach more closely the limit of dynamical stability $K/|W|\sim1/4$ if the objects are on the ZAMS and have masses higher than $M=15M_{\odot}$. However, \citet{clem79} in his analysis of the secular stability of differentially rotating objects already pointed out the unstable character of model stars with $K/|W|\gtrsim0.1$.\par 

\subsection{The Roche approximation}
\label{rocheapp}

\label{totau}
\begin{table}[]
\centering
\caption[]{\label{quadf}Quadrupole factors in stars with rigid rotation at different angular velocity ratios $\Omega/\Omega_c$}
\tabcolsep 4.0pt
\begin{tabular}{rcccccc}
\hline
 & $\Omega/\Omega_c=$ & 0.5 & 0.7 & 0.9 & 0.95 & 1.0 \\
$M/M_{\odot}$ & Phase & \multicolumn{5}{c}{$\gamma$} \\
\hline
\noalign{\smallskip}
 5.0 & ZAMS & 0.0012 & 0.0018 & 0.0026 & 0.0028 & 0.0030 \\
 5.0 & TAMS & 0.0003 & 0.0007 & 0.0016 & 0.0020 & 0.0025 \\
15.0 & ZAMS & 0.0015 & 0.0025 & 0.0037 & 0.0040 & 0.0043 \\
15.0 & TAMS & 0.0004 & 0.0007 & 0.0018 & 0.0023 & 0.0028 \\
\noalign{\smallskip}
\hline
\end{tabular}
\end{table}

  With the models thus calculated, which give us the shape of a rotationally distorted star, we can test to what degree the gravitational potential $\Phi_{\rm G}$ in the stellar surface may deviate from the simple central-field expression used in the Roche approximation.\par
  \citet{hubb75} showed that the gravitational potential of rotating centrally condensed objects can be given by a multipol expansion

\begin{equation}
\label{hubb1}
\Phi_{\rm G} = -\frac{GM}{R(\theta)}\left[1-\sum_{n=1}^{\infty}\left(\frac{R_o}{R(\theta)}
\right)^{2n}J_{2n}P_{2n}(\cos\theta)\right]\ ,
\end{equation}

\noindent where $R_o$ is the radius of the rotationally undistorted star, $J_{2n}$ are the zonal harmonic coefficients, and $P_{2n}(\cos\theta)$ are the Legendre polynomials. To derive relation (\ref{hubb1}), \citet{hubb75} assumed that the density can be expanded in Legendre polynomials. The $J_{2n}$ coefficients are then obtained as integrals over the volume of the axisymmetric object which generally produce values so that $J_{2}>0$, $J_{4}<0$, $J_{6}>0$, $J_{8}<0$, and so on. Since we are not interested in developing a detailed theory of the gravitational potential of rotating stars, but only to test the validity of the Roche approximation, we do not calculate the volume integrals over the rotationally deformed star, but use the harmonic coefficients coefficients as mere parameters giving a quantitative indication of the deviation from the central-field form of the gravitational potential. We determine the quantities $J_{2n}$ in expression (\ref{hubb1}) by simply fitting the gravitation potential predicted by the 2D models of rotating stars calculated in Sects.~\ref{rr} and \ref{dr} using a least squares method. In this algebraic estimate of $J_{2n}$, their order of magnitude are preserved, but not always the expected 
sign.\par
  In relation (\ref{hubb1}), we retain only the quadrupole moment to estimate the effects in rigid rotators

\begin{equation}
\label{hubb2}
\Phi_{\rm G} = -\frac{GM}{R(\theta)}\left\{1-\left[\frac{R_o}{R(\theta)}\right]^2J_2P_2(\cos\theta)\right\}\ .
\end{equation}

 The shape of the surface thus becomes

\begin{equation}
\label{hubb2}
\frac{R_{\rm s}(\theta)}{R_{\rm e}} = 
\frac{1-\gamma\left(\frac{R_{\rm s}(\theta)}{R_{\rm e}}\right)^2P_2(\cos\theta)}{\left(1+\frac{1}{2}\gamma\right)+\frac{1}{2}\eta_o\left[1-\left(\frac{R_{\rm s}(\theta)}{R_{\rm e}}\right)^2\sin^2\theta\right]}\ ,
\end{equation}

\noindent where we have introduced the notation

\begin{equation}
\gamma = J_2\left(\frac{R_o}{R_{\rm e}}\right)^2\ .
\label{hubb4}
\end{equation}

 The largest difference between the estimate given by Eq.~(\ref{hubb2}) to the ratio $R_{\rm s}(\theta)/R{\rm e}$ and that produced by the Roche approximation in Eq.~(\ref{rerpr}), is expected for $\theta\!=\!0$. Thus according to (\ref{hubb2}) we have

\begin{equation}
\label{hubb3}
\frac{R_{\rm e}}{R_{\rm p}} = \frac{1+\frac{1}{2}(\eta_o+\gamma)}{1-\gamma\left(\frac{R_{\rm e}}{R_{\rm p}}\right)^2}\ .
\end{equation}

\noindent Since it is always true that $R_{\rm e}/R_{\rm p}\geq1$, relation (\ref{hubb3}) implies that {\it for whatever $0\leq\eta_o\leq1$ and $\gamma\neq0$ the equator to polar radii ratio is slightly larger than obtained from the sheer Roche approximation}.\par
  In Table~\ref{quadf}, we indicate the quadrupole factors $\gamma$ for 5$M_{\odot}$ and 15$M_{\odot}$ stars in the ZAMS and TAMS evolutionary phases obtained by fitting relation (\ref{hubb2}) with surface gravitational potential obtained with the model calculation. In this table, we see that the smaller $J_2$ the more evolved is the star, simply because the star is more centrally condensed such that the central-field approximation for $\Phi_{\rm G}$ holds better. \citet{zahn10} calculated the same factors for rigid and shellular differential rotators. Reducing their $J_2$ estimates by $(R_{\rm p}/R_{\rm e})^2$ to be able to approach the ratio $(R_{\rm o}/R_{\rm e})^2$ more closely than using the average stellar radius for $R_{\rm o}$, we see that for the ZAMS and TAMS epochs of rigid critical rotators we obtain the same values for the harmonic coefficient $J_2$ than \citet{zahn10}. Having then $\gamma\lesssim0.004$, from (\ref{rerpr})  we see that neglecting the quadrupole term in (\ref{hubb2}), in rigid rotators at critical rotation the $R_{\rm e}/R_{\rm p}$ can be underestimated by less than 2\%. For shellular rotators, \citet{zahn10} found that at critical equatorial rotation the flattening changes by 4\% because $\Omega_{\rm center}/\Omega_{\rm surface}=4$ in the ZAMS and is smaller near the TAMS.\par
  Finally, we note that within approximation (\ref{hubb2}), the actual ratio of the centrifugal to gravity acceleration at the equator should be

\begin{equation}
\label{accgr}
\eta_J = \frac{\eta_o}{1+\frac{3}{2}\gamma}\ .
\end{equation}

\noindent In stars with conservative differential rotation laws of the type given in Eq.~(\ref{bo}) that imply rotational energies $\tau\!=\!K/|W|$ $>$ $\tau_{cr}$, the deformations of the internal mass distribution and the deviations to the central-field gravitational potential are expected to be stronger than in stars with  rigid rotation. To test the use of the Roche approximation in these cases, we calculated the zonal harmonic coefficients $\gamma_{2n}\!=\!J_{2n}(R_o/R_{\rm e})^2$ from $n\!=\!1$ to 4. Table~\ref{modkwequi} lists the zonal multipolar coefficients $\gamma_{2n}$ obtained by fitting relation (\ref{hubb1}) with the calculated gravitational potential for several values of the coefficient $\beta$ and angular velocity ratio $\Omega_o/\Omega_{cr}$, where $\Omega_o$ is the angular velocity at the rotation axis, and $\Omega_{cr}$ is the critical angular velocity of an homologous star (same mass and evolutionary stage) behaving as a rigid rotator. We insist on the algebraic nature of the $\gamma_{2n}$ coefficients, since they were derived as fitting parameters to model-calculated gravitational potential. In this Table, we also give the fractional deviations of the radius vectors $\Delta R(\theta)/R(\theta)=[R(\theta)_{\rm model}-R(\theta)_{\gamma_{2n}}]/R(\theta)_{\rm model}\times10^4$ for different colatitude angles $\theta$, where $R(\theta)_{\rm model}$ is obtained with the models calculated in Sect.~\ref{dr}, while the $R(\theta)_{\gamma_{2n}}$ values refer to those estimated with the total potential that includes the gravitational potential given by Eq.~(\ref{hubb1}). A quick inspection of Table~\ref{modkwequi} shows that in most cases a fairly precise description of the stellar surface can be obtained by considering only $\gamma_2$ and $\gamma_4$ in the gravitational potential given by Eq.~(\ref{hubb1}). Nevertheless, in all cases the Roche approximation can be valid to better than 5\% for ratios $\Omega_{\rm center}/\Omega_{\rm surface}\simeq4$. The approximation works better the more evolved the star is and the higher the rotational energy parameter $\tau$.\par

\section{Attainable stellar parameters}
\label{attain}

  According to Vogt-Russell's theorem, ``the complete structure of a star in hydrostatic and thermal equilibrium is uniquely determined by the total mass $M$ and the run of chemical composition throughout the star, provided the structure equations are function of local parameters" \citep{cox68}. Then, for a given chemical composition, denoted here in a generic way by $Q$,  it follows that parameters that conventionally specify the physical properties of a star are the luminosity $L$, the mass $M$, and the radius $R$ (or the stellar age $t$).\par
  An object in permanent rotation can still be considered to be in hydrostatic and thermal equilibrium. However, owing to the numerous effects the rotation induces, the number of parameters needed to characterize an object can be high: mass, total luminosity, age, total angular momentum, internal distribution of the angular momentum, dimensions or stellar geometry, inclination angle, degree of non-uniform distribution of the temperature in the surface, mixing of chemical elements, and degree of their non-uniform internal and surface distribution, etc. This number is certainly higher than we can tackle and/or determine from observations. We consider here only parameters we can actually determine with the diagnostic tools we presently have at our disposal. However, since many of these quantities are model-dependent, the resulting characterization of rotating stars would probably be  limited or incomplete.\par
 We may classify the accessible stellar parameters into three classes: 1) those related to the stellar internal structure; 2) parameters relevant to the structure of the atmosphere over the observed stellar hemisphere; 3) quantities related to the apparent stellar geometry.\par
\medskip
  \underbar{\it Models of stellar structure}: The existing models of stars with rotation start their MS evolution with low rotational energies in the ZAMS, i.e.

\begin{equation}
\label{taums1}
\tau = K/|W| \leq \tau_{cr}^{rig}(M)\ , 
\end{equation}

\noindent where $K$ is the rotational kinetic energy, $W$ is the gravitational potential energy, and $\tau_{cr}^{rig}$ is the energy ratio for a rigid rotator at a critical equatorial rotation in the ZAMS. For stars with masses ranging from $3M_{\odot}$ to some $60M_{\odot}$, it is on average $\tau_{cr}^{rig}\lesssim0.015$ \citep{zo88a}. Fom Table~\ref{medk2jt}, we see that $\tau_{\rm ZAMS}$ is a function of $M$ and $V_{\rm eq}$. For rigid rotators in the ZAMS, $V_{\rm eq}$ and $\Omega/\Omega_{\rm c}$ are equivalent parameters. \par
  Owing to mass and angular momentum losses, the energy ratio $\tau$ becomes a function of the time, so that $\tau(M,t)<\tau_{\rm ZAMS}^{rig}(M)$. On account of possible redistribution phenomena of angular momentum that could take place in the pre-MS evolutionary phases, we should characterize the stellar initial rotational law with at least one single quantity. This single parameter should also correspond to the expected value of the kinetic energy ratio when $\tau_{\rm ZAMS}>\tau_{\rm ZAMS;cr}^{rig}$. However, existing models of rotating stars that provide us today with evolutionary paths to infer the stellar masses and ages, are calculated with initial rigid rotations in the ZAMS. In this case, $\tau$, $\Omega/\Omega_{\rm c}$, and $V_{\rm eq}$ can be assumed to carry the same information. We must keep in mind that the masses and ages we infer from these models are depend on the assumption given in Eq.~(\ref{taums1}). The models provide us with quantities averaged over the stellar surface (cf. \citet{meynet2000,ekst08}), namely surface-averaged bolometric luminosity $\langle{L}\rangle$, effective temperature $\langle{T_{\rm eff}}\rangle$, surface effective gravity $\langle{\log g_{\rm eff}}\rangle$ (occasionally), and equatorial linear rotational velocity $V_{\rm eq}$

\begin{eqnarray}
\label{paramm}
\langle{L}\rangle & = & \langle L(M,t,Q,\tau_{ZAMS})\rangle\ ,  \nonumber \\
\langle{T_{\rm eff}}\rangle & = & \langle{T_{\rm eff}}(M,t,Q,\tau_{ZAMS})\rangle\ , \nonumber \\
\langle{\log g_{\rm eff}}\rangle & = & \langle{\log g_{\rm eff}}(M,t,Q,\tau_{ZAMS})\rangle\ , \nonumber \\
V_{\rm eq} & = & V_{\rm eq}(M,t,Q,\tau_{ZAMS})\ . 
\end{eqnarray}

  From \citet{maeder00}, we can see that because of rotation, stars behave as though they
have lower effective mass than in reality, which mainly affects the emitted bolometric luminosity. According to this mass-compensation effect \citep{sack70} and the core fuelling with hydrogen by the meridional circulation, rotating stars have longer MS life-spans. Owing
to the change in the opacity of the envelope by the diffusion of chemical elements from the core, mainly He, stars become more luminous than their mass-homologous rotationless objects. Evolutionary timescales in the post-MS phases are also sensitively changed. If stars were actually {\it neat} differential rotators, i.e. with rotational energies $\tau_{\rm ZAMS}>\tau_{cr}^{rig}$ \citep{boden71,clem79}, the evolutionary tracks needed to interpolate stellar masses would also differ from those existing today.\par
\medskip
 \underbar{\it Models of stellar atmospheres}: High resolution spectra and spectral energy  distributions of rapidly rotating stars are both $i-$dependent ($i$ is the inclination angle). They can be modeled by taking into account the rotationally-induced stellar geometrical deformation, but mainly the concomitant latitude-dependent surface gravity and temperature distribution: gravity darkening \citep{frem05,lovek06,gill08}.\par
  Since a priori we do not have any indication on what may be the function describing the surface angular velocity of a star, in Sect.~\ref{avdss} we adopted a simple Maunder-like expression given by Eq.~\ref{sol3}, i.e., $\Omega(\theta)=\Omega(\Omega_o,\alpha,\theta)$, where $\Omega_o$ is the angular velocity of the equator, $\alpha$ is the surface differential rotation parameter, and $\theta$ is the colatitude angle. Instead of using $\Omega_o$ we prefer to use the force ratio parameter $\eta_o$ defined in Eq.~(\ref{etao}). Since $\alpha\neq0$, the iso-radial velocity lines responsible for the Doppler shifts and
causing the rotationally broadening of spectral lines are no longer straight lines but curves, so that  the true velocity in $V\sin i$ does not necessarily correspond to the equatorial linear velocity \citep{zo88b,zor04}. To this adds the uncertainty in the $V\sin i$ parameter related to the less effective contribution to the line broadening by the equatorial regions affected by the gravitational darkening \citep{stoe68,town04,frem05}. \par
  For the colatitude-dependent distribution of the effective temperature, we concluded in Sect.~\ref{tdss} that $T_{\rm eff}(\theta)=\overline{\mathscr{C}}\times{\rm\bf g_{\rm eff}}^{\betaup_{\rm GD}}$, where $\overline{\mathscr{C}}$ can be assumed to be caused by the average stellar geometrical deformation and where $\betaup_{\rm GD}=0.25+\delta$ has a free parameter $\delta$ that should be $\delta\leq0$ according to what is suggested in the literature [cf. \citet{lucy67,lovek06,vanb06,monn07,zhao09}]. For models that begin evolving in the ZAMS as rigid rotators, $\tau$ and $\eta_o$ are also synonymous. The apparent, hemisphere-dependent bolometric luminosity, effective temperature, effective gravity and rotation parameter are then functions of at least the seven unknowns of mass $M$, age $t$, initial chemical composition $Q$, equatorial rotation parameter $\eta_o$, surface differential parameter $\alpha$, power $\betaup_{\rm GD}$ in the von Zeipel relation, and the inclination angle $i$

\begin{eqnarray}
\label{ltgap} 
L^{\rm app} & = & L(M,t,Q,\eta_o,\alpha,\betaup_{\rm GD},i)\ , \nonumber \\
T_{\rm eff}^{\rm app} & = & T_{\rm eff}(M,t,Q,\eta_o,\alpha,\betaup_{\rm GD},i)\ , \nonumber \\
\log g^{\rm app} & = & \log g(M,t,Q,\eta_o,\alpha,\betaup_{\rm GD},i)\ , \nonumber \\
(V\!\sin i)^{\rm app} & = & V(M,t,Q,\eta_o,\alpha,\betaup_{\rm GD})\times\sin i\ .
\end{eqnarray}

 We also include the dependence on $\betaup_{\rm GD}$ in $\log g^{\rm app}$ and $(V\!\sin i)^{\rm app}$, since their prediction is based on spectral lines produced by gravity-darkened models. The above seven-fold dependence is valid also for the equivalent widths ($W_{\lambda}$) of spectral lines, the line FWHM (full width at half maximum), the zeroes $q_n$ of the Fourier transforms of spectral lines, as well as the energy distributions and spectrophotometric parameters such as ($\lambda_1,D$) of the BCD system, which describe the Balmer discontinuity and are highly sensitive to ($\log g^{\rm app},T_{\rm eff}^{\rm app}$) \citep{zor09}. Referring to all these observable quantities with a generic letter $\mathscr{O}_{\lambda}$, they are formally

\begin{equation}
\label{wfqld} 
\mathscr{O}_{\lambda} = \mathscr{O}_{\lambda}(M,t,Q,\eta_o,\alpha,\betaup_{\rm GD},i)\ .
\end{equation}
 
  We note that for stars with a given initial chemical composition $Q$ and a prescription for their internal rotation, the parametric couples ($\langle{L}\rangle,\langle{T_{\rm eff}}\rangle$), and ($M,t$) are equivalent quantities. From observations, we derive $L^{\rm app}$, $T_{\rm eff}^{\rm app}$, and $\log g^{\rm app}$, which are $i-$dependent, while the models of stellar evolution provide us with $\langle{L}\rangle$, $\langle{T_{\rm eff}}\rangle$, and sometimes $\langle\log g\rangle$. This means that $M$ and $t$ can be estimated by iteration, where we also iterate simultaneously the inclination angle $i$ \citep{zor05}.\par  
\medskip
 \underbar{\it Stellar geometry}: Since it is possible to use the Roche approximation to describe the rotationally induced stellar deformation, from Eq.~(\ref{nonc3}) it comes that the inclination angle-dependent shape of rapidly rotating star depends only on $\eta_o$, $\alpha$ and $i$. If we are not interested in carrying out imaging of fast rotating objects, interferometry can provide us with data that to a first approximation do not depend on $M$, $t$, $Q$, and $\betaup_{\rm GD}$, i.e. the apparent shape of the star $(y,x)$, where $x$ is the coordinate measured along the equatorial line projected onto the sky, and $y$ is:

\begin{equation}
\label{rprei} 
y(x) = \frac{R(\eta_o,\alpha,i,x)}{R_{\rm e}}\ , 
\end{equation}

\noindent so that the apparent polar-to-equatorial radius ratios are $y(x=0)=R_{\rm p}(i)/R_{\rm e}$ and $y(x=1)=0.0$. We examine this in greater detail in Sect.~\ref{ifi}.\par
\medskip
  If models are calculated for a given metallicity $Z$, which stands for the initial generic quantity $Q$ used above, we conclude that a massive and intermediate-mass fast rotating star with surface differential rotation can be described as a function of six parameters 

\begin{eqnarray}
\label{lmetabi}
\ \ \ \ \ \ \ \ \ \ \ \ \ \ \ & M,\ \  \eta_o,\ \  t,\ \  \alpha,\ \  \betaup_{\rm GD}, \ \ {\rm and}\ \  i\ , &
\end{eqnarray}

\noindent which we should be able to derive by interpreting the observed data. We briefly review some of the data and techniques that can help us to determine some of these parameters, in particular $\eta_o$, $\alpha$, and
$i$.\par

\subsection{Information on the differential rotation from spectral lines}
\label{ifsl}

\begin{figure*}[Ht!]
\centerline{\psfig{file=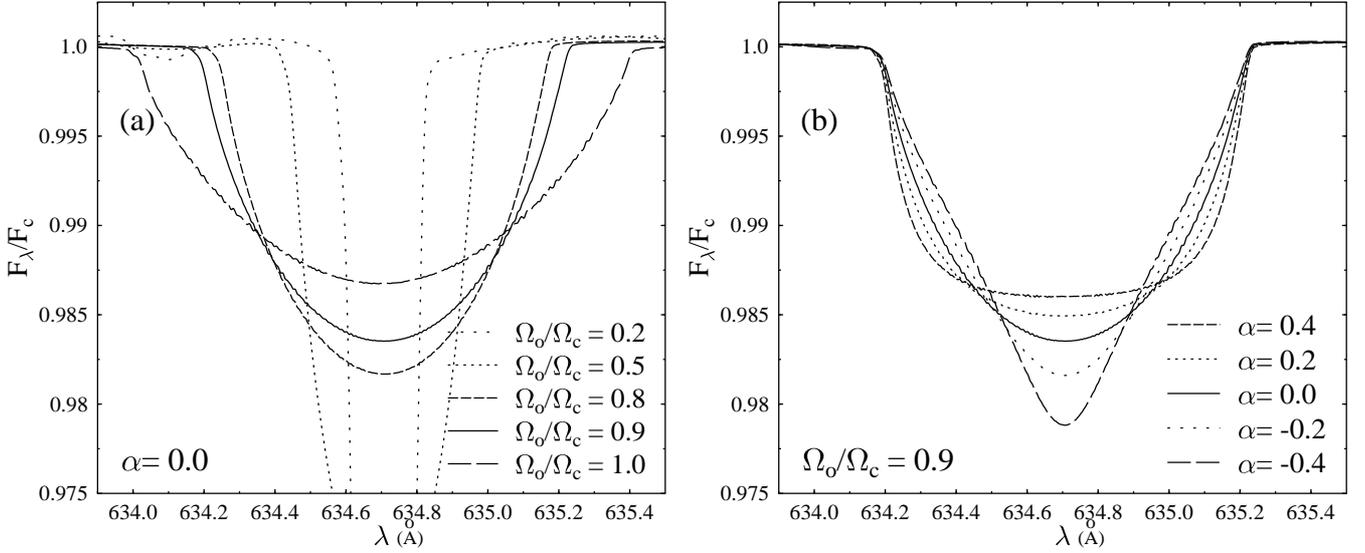,width=18truecm,height=7.7truecm}}
\caption[]{\label{w_w} Rotationally broadened SiII 6347 model line profile in a main sequence B2-type star (rest $T_{\rm eff}\!=\!19000$K, $\log g\!=\!4.0$) seen at the inclination angle $i\!=\!45^o$. (a): Line profiles affected by gravitational darkening in stars rotating at different surface $\Omega/\Omega_c$ rates and differential rotation parameter $\alpha\!=\!0.0$; (b): Line profiles with gravitational darkening effect in stars rotating at $\Omega_o/\Omega_c\!=\!0.9$ and for several values of the differential rotation parameter $\alpha$. To calculate the line profiles, we used $\betaup_{\rm GD}=0.25$.}
\end{figure*}

  As stated in Sect.~\ref{tdss}, the distribution of the effective temperature in the stellar surface deviates from von Zeipel's prescription, i.e. $T_{\rm eff}\!\propto\!g^{\betaup_{\rm GD}}_{\rm eff}$ with $\betaup_{\rm GD}\neq1/4$ if $\alpha\!\neq\!0$. However, in what follows of the present discussion we shall assume the validity of the original von Zeipel relation, but use a modified expression of the effective gravity $g_{\rm eff}$ according to the latitudinal dependence of the surface differential angular velocity $\Omega(\eta,\alpha,\theta)$ given by Eq.~(\ref{sol3}).\par
  In this section, we report first estimates of the effects produced by surface rotational laws $\Omega(\eta,\alpha,\theta)$ on some spectral lines and envision with them the way we can obtain estimates of $\alpha$ and the inclination angle $i$ from spectroscopy .\par
  We modified our FASTROT code \citep{frem05}, initially written for a uniform angular velocity on the stellar surface, to take into account an external Maunder-like rotational profile and the related changes in the gravity and the effective temperature as a function of the colatitude $\theta$ in the calculation of spectral line profiles. In 
Fig.~\ref{w_w}a, we show the \ion{Si}{ii}~6347 line profiles obtained with the new FASTROT program, which are produced by MS stars with $T_{\rm eff}\!=\!19500$ K and $\log g\!=\!4.0$ at rest, whose rotation is characterized by several ratios of the angular velocity in the equator $\Omega_o/\Omega_{\rm c}$ without surface differential rotation ($\alpha=0$). In Fig.~\ref{w_w}b, we show the \ion{Si}{ii}~6347 line profiles for stars with the same ($T_{\rm eff},\log g$) rest  parameters as before, but now for $\Omega_o/\Omega_{\rm c}=0.9$ and several degrees of differential rotation ($\alpha\neq0$). All line profiles in Figs.~\ref{w_w}a and \ref{w_w}b were obtained with gravity-darkened models seen at the inclination angle is $i\!=\!45^o$. We note that the kinematic effect on the line broadening with surface differential rotation is determined by curves of constant Doppler shift, which are not straight lines as in the solid body rotation, but curves whose shapes depend on the value of $\alpha$ and the inclination angle $i$ \citep{arm04a}. Other glimpses into the sensitivity of different spectral lines in early-type rapid rotators to the effects carried by a surface rotation law $\Omega(\eta,\alpha,\theta)$ can be seen  in \citet{zor07}.\par
\begin{figure}[]
\centerline{\psfig{file=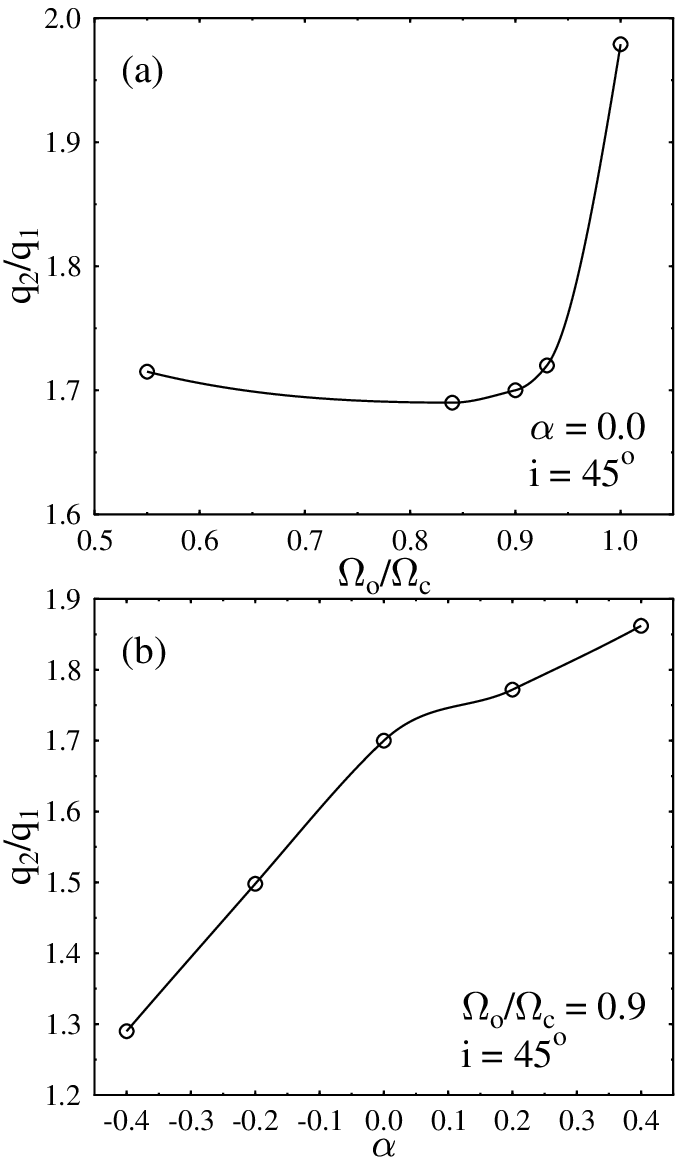}}
\caption[]{\label{q2q1} Ratio $q_2/q_1$ of the second to the first zero of the Fourier transform of the spectral lines shown in Fig.~\ref{w_w}. (a) Ratios for line profiles that correspond to stars rotating at different $\Omega_o/\Omega_c$, which are affected by gravitational darkening, but without surface differential rotation ($\alpha=0.0$). (b) Ratios affected by differential rotation and gravitational darkening, as a function of the differential rotation parameter $\alpha$ in stars where $\Omega_o/\Omega_c=0.9$. All relations shown in this figure are for the inclination angle $i\!=\!45^o$.}
\end{figure}

  The Fourier transform can give a quantitative description of the effects induced on the spectral lines by the latitudinal differential rotation and perhaps on the associated gravity darkening effect. This information can be carried by the ratio $q_2/q_1$ of the second ($q_2$) to the first zero ($q_1$) of the Fourier lobes, as suggested by 
\citet[][\ \ and references therein]{rei04}. In Fig.~\ref{q2q1}a, we see the ratio $q_2/q_1$ determined by the Fourier analysis of the lines shown in Fig.~\ref{w_w}b, as a function of the ratio $\Omega_o/\Omega_{\rm c}$ for stars without surface differential rotation ($\alpha=0$). We see in this figure that the ratio $q_2/q_1$ becomes sensitive to the gravity darkening only when $\Omega_o/\Omega_{\rm c}$ approaches a critical value. In Fig.~\ref{q2q1}b, we show the ratios $q_2/q_1$ for the profiles in Fig.~\ref{w_w}b, i.e. $\Omega_o/\Omega_{\rm c}=0.9$ and $\alpha\neq0$, where the corresponding gravity darkening is also taken into account. As in Fig.~\ref{w_w}, the inclination in Fig.~\ref{q2q1} is $i\!=\!45^o$. From the results shown in Fig.~\ref{q2q1}b, we see that the differential rotation in fast rotating early-type star produces fairly large effects on the ratio $q_2/q_1$, which are then expected to be measurable. It is also expected that a series of relations of the type

\begin{equation}
\label{q2q1l}
q_2/q_1 = f(\alpha,i)
\end{equation}

\noindent can be obtained, whose functional dependence on  $\alpha$ and $i$ differ from line to line. This is because the dominant excitation mechanism of spectral lines is not the same for all of them, and that they can react more or less in different ways, or be more or less sensitive to the non-uniformity of the surface effective temperature and gravity. This precludes our being able to define a prototype line-broadening function that could be used for all spectral lines analyzed in a spectrum, which will be studied in detail in a following paper.\par
\begin{figure*}[ht!]
\centerline{\psfig{file=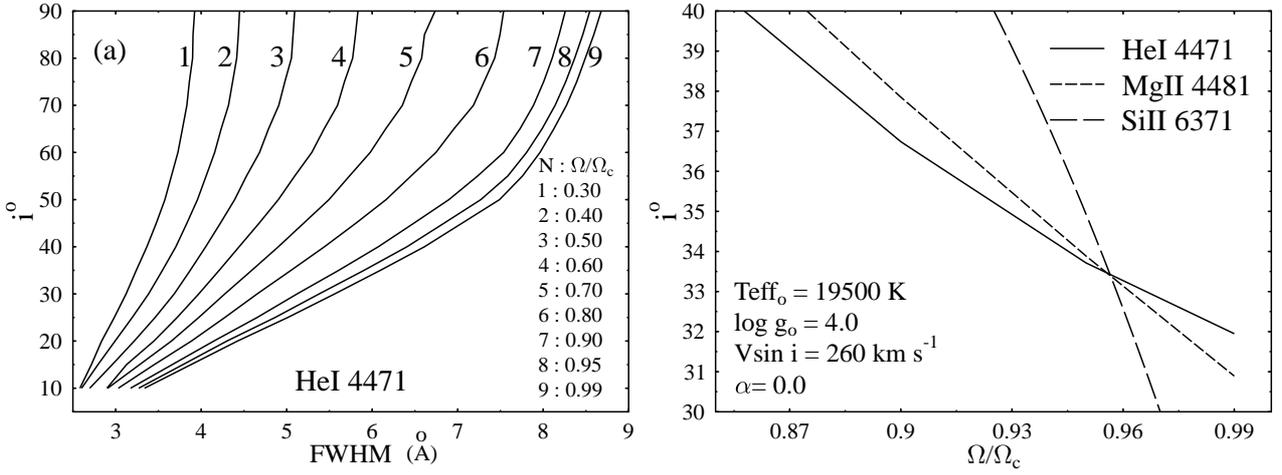,width=17.4truecm,height=6.4truecm}}
\caption[]{\label{w_i} (a): Inclination angle as a function of the equivalent half-intensity width of the HeI 4471 line for gravitationally darkened stars rotating at the 
indicated $\Omega_o/\Omega_c$ values, where the differential rotation parameter is set to $\alpha\!=\!0.0$. ; (b): Intersection of constant FWHM(HeI 4471)- and FWHM(MgII 4481)- curves of an early-type rapid rotator seen nearly pole-on.}
\end{figure*}

  We note that for this type of study, only spectral lines that do not undergo strong Stark broadening can be used. Accordingly, in early-type stars the \ion{He}{i}~4471 and \ion{Mg}{ii}~4481 lines are more frequently used to determine the $V\!\sin i$ parameter (c.f. \citet{slet82,chau01,frem05}). To test the sensitivity of these lines to the effects related to the gravitational darkening, we calculated the line profiles of \ion{He}{i}~4471, \ion{Mg}{ii}~4481, and \ion{Si}{ii}~6371, and obtained the diagrams shown in Fig.~\ref{w_i}a, where for each studied line the inclination of the rotation axis is presented against the equivalent width and the FWHM (full width at half maximum). For the calculations shown in Fig.~\ref{w_i}, we set the differential rotation parameter to $\alpha\!=\!0$. However, similar behaviors occur when $\alpha\!\neq\!0$. The curves in Fig.~\ref{w_i} show the differentiated sensitivity of the \ion{He}{i}, \ion{Mg}{ii}, and \ion{Si}{ii} lines to the effects carried by the rotation.\par
 Using both the measured equivalent widths and FWHM of lines, and the diagrams of the type shown in Fig.~\ref{w_i}a, we obtain intersections as shown in Fig.~\ref{w_i}b that determine estimates of $\Omega/\Omega_{\rm c}$ and $i$. From Eq.~(\ref{q2q1l}), we can then infer a value of $\alpha$ and use it to obtain a new series of curves similar to those in Fig.~\ref{w_i}a. This iteration can be followed until convergence for $\Omega/\Omega_{\rm c}$, $i$, and $\alpha$ is attained. To infer a first estimate of uncertainties related to this method, we note that the curves of Fig.~\ref{w_i} can be reasonably fitted with relations such as

\begin{equation}
\label{fitiw}
 i = A(\omega)W^{B(\omega)}+C(\omega) \ ,
\end{equation}

\noindent where $\omega\!=\!\Omega/\Omega_{\rm c}$. It then follows that uncertainties $\delta W/W$ in the equivalent widths carry errors in the inclination angle $i$ which are on the order of:

\begin{equation}
\label{erriw}
\frac{\delta i}{i} \simeq B(\omega)\left[1-\frac{C(\omega)}{i}\right]\frac{\delta W}{W}\ .
\end{equation}

\noindent For an order of magnitude estimate, we consider Achernar, for which measurements of the apparent radii ratio $R_{\rm e}/R_{\rm p}$ were obtained \citep{arm03,kerv06} and another analysis of the inclination angle was performed \citep{vinic06}. Assuming for this star that $\eta_o\!\sim\!0.8-0.9$ and $i\!\lesssim\!75^o$, from the \ion{He}{i}4471 and \ion{Mg}{ii}4481 we can infer, respectively, that $B_{4471}\!=\!3.3$ and $C_{4471}\!=\!11.6$, $B_{4481}\!=\!2.5$ and $C_{4481}\!=\!13.7$, and $\delta W/W\!\sim\!0.03$ that lead to $\delta i\!\sim\!5^o$, which is an upper limit, since as seen in Fig.~\ref{w_i}b, the \ion{Si}{ii} lines reduce this uncertainty. \par
  A simple polynomial fit of $\alpha$ as a function of the ratio $R_{\rm q}\!=\!q_2/q_1$ shown in Fig.~\ref{q2q1} leads to uncertainties of $\delta\alpha\!\simeq\!0.15$ for $R_{\rm q}\!=\!1.65$ and $\delta R_{\rm q}\!\simeq\!0.1$, and smaller values of $\delta\alpha$ for other ratios $R_{\rm q}$. However, the determination of the parameters $\Omega/\Omega_{\rm c}$ (or $\eta_o$) and $i$, and then $\alpha$, can be greatly improved using curves similar to Fig.~\ref{w_i} calculated for many other spectral lines of different excitation character, i.e, collision-dominated or photoionization-dominated, which enable then to infer the inclination angle factor $\sin i$ more precisely \citep{husto77,ruu89}. A future paper will be devoted to the discussion of this method and the related uncertainties.\par

\subsection{Information on the differential rotation from interferometry}
\label{ifi}

 Interferometry can help provide different types of information on rotating stars: a) data related to the geometrical shape of stars; b) independent estimates of the differential rotation parameter $\alpha$ and the inclination $i$; c) imaging of the non-uniform hemisphere of a rotationally deformed star. In this introductory paper, we do not address the imaging techniques.\par
\smallskip
 a) \underline{Stellar geometry}. The main goal is to measure the geometrical deformations carried by the stellar rotation, and in particular those added by the surface differential rotation described in Sect.~\ref{mrl}. Using the integrated fluxes in different spectral bands, the geometrical shape of the sources can be determined from the visibility measurements using the widest possible coverage of the plane (u,v) \citep{vbelle01,arm03,
auf06,vanb06,mcalis05,kerv06,monn07,zhao09}. The resulting description of the geometry of the source is however $i$-inclination angle dependent.\par
 As in Eq.~(\ref{rprei}), we write as $y(x,i)$ the coordinates measured perpendicular to the $x$ axis drawn in the direction of the stellar equator projected onto the sky. The $y(x,i)$ ordinates determined from the interferometry are then related to the true shape of the star by

\begin{eqnarray}
\label{interfdat}
y(i,x) & = & [R^2_{\rm s}(\theta,\eta_o,\alpha)\sin^2\theta-x^2]^{1/2}\cos i+ \nonumber  \\
       &   & R_{\rm s}(\theta,\eta_o,\alpha)\cos\theta\sin i \ ,
\end{eqnarray}

\noindent where $R_{\rm s}(\theta,\eta_o,\alpha)$ is the equation of the stellar surface normalized to $R_{\rm e}$ given by Eq.~(\ref{iter1}) and $\theta$ is the colatitude angle measured in the stellar reference system. Equations (\ref{interfdat}) and (\ref{iter1}) are two independent equations that for a given pair of prameters ($i,\alpha$) enable us to obtain an $x\!=\!x(\eta,\theta)$ relation and determine the sought-after parameter $\eta$. This leads us to the final objective, i.e., to derive the true radii ratio $R_{\rm p}/R_{\rm e}\!=\!R_{\rm s}(\theta\!=\!0)/R_{\rm e}$ of the star from the apparent stellar shape $y(i,x)/R_{\rm e}$. To achieve this purpose, the inclination angle $i$ can be inferred using several iteration and modeling methods: 1) approximate ones such as those tested by \citet{vinic06} and \citet{carch08}; 2) the spectroscopic inferences described in Sect.~\ref{ifsl}; 3) spectrophotometric constraints imposed by the ($\lambda_1,D$) parameters of the Balmer discontinuity \citep{zor05}; 4) far and near-UV spectrophotometry \citep{fre02,frem05}; 5) differential interferometry as described below (see item b).\par
 A first insight into the uncertainties incurred when using the VINCI/VLTI interferometric data can be inferred by assuming that the apparent star corresponds to the projection of an axisymmetric ellipsoid. Equation (\ref{interfdat}) then becomes \citep{kana08}

\begin{equation}
\label{roapv}
\rho_{\rm ap} = \left[1-(1-\rho^2_{\rm tr})\sin^2i\right]^{1/2}\ ,
\end{equation}

\noindent where we have $\rho_{\rm ap}\!=\!(R_{\rm p}/R_{\rm e})_{\rm apparent}$ and $\rho_{\rm tr}\!=\!(R_{\rm p}/R_{\rm e})_{\rm true}$. For the VINCI/VLTI data on Achernar analyzed by \citet{kerv06}, we infer that $\delta\rho_{\rm ap}\!\simeq\!0.02$. Using a Monte Carlo simulation for the propagation of errors, we obtain $\delta\rho_{\rm tr}\!\simeq\!0.03$, which from Table~\ref{rerpdt} implies that the differential rotation parameter $\alpha$ can be estimated with an uncertainty of $\delta\alpha\!\lesssim\!0.15$.\par
 Improved estimates of $\rho_{\rm ap}$ are expected from the VEGA/CHARA instrument \citep{mourard09}. To determine the lowest stellar fattenings that we can reliably measure and the related uncertainties, the important quantity we need to estimate is the visibility ratio

\begin{equation}
\label{dv2v2}
\frac{dV^2}{\langle{V^2}\rangle} = \frac{V_{\rm p}^2-V_{\rm e}^2}{\frac{1}{2}\left(V_{\rm p}^2+V_{\rm e}^2 \right)}\ ,
\end{equation}

\noindent where $V_{\rm p}^2$ is the visibility in the polar direction of the apparent shape of the star and $V_{\rm e}^2$ is the visibility in its equatorial direction. \citet{mourard09} demonstrated that highly reliable measurements with the VEGA/CHARA instrument can be performed of the angular diameter when the relative uncertainty in the squared visibility curve is $\sigma_{V^2}/V^2\lesssim0.02$. We then expect that visibility curves obtained in directions towards the apparent stellar polar and equatorial directions, respectively, be able to be differentiated and thus a difference between $R_{\rm e}$ and $R_{\rm p}$ be found as soon as $dV^2/\langle{V^2}\rangle\gtrsim0.02$. We then tested the ratio in Eq.~(\ref{dv2v2}) for B0, B2, and B8-type objects of luminosity classes V and III with apparent magnitudes $V=5$ mag and $V=4$ mag. The magnitude $V=5$ mag corresponds to an upper conservative limiting magnitude for the VEGA/CHARA interferometer in the 
medium-resolution spectral mode. The tested B sub-spectral types were chosen to be B0 to include massive objects for which rapid rotators are frequent, B2 spectral type which represent the highest frequency of Be stars, and B8 to represent Bn stars, which are rapid rotators without emission lines and most currently found among late B-type stars.\par
 Using the fundamental parameters given in \citet{zorbri91}, \citet{div82}, and \citet{zor09} for the visible absolute magnitude $M_{\rm V}$, the absolute bolometric magnitude $M_{\rm bol}$, and the effective temperature $T_{\rm eff}$, respectively, we inferred the circular angular diameters $\theta_o$ of the tested B sub-spectral types associated with their apparent average radius $R_o=(R_{\rm e}+R_{\rm p})/2$. We calculated the visibilities for two baselines, $B=156$ m and $B=300$ m, attainable with the VEGA/CHARA instrument at the effective wavelength $\lambda=5500~\AA$, assuming that stellar discs have uniform brightness. The visibilities were calculated separately for the polar and equatorial directions and test radius ratios ranging from $R_{\rm e}/R_{\rm p}=1.05$ to $R_{\rm e}/R_{\rm p}=1.25$, to account for the lower deformations of rotators characterized by parameters ($\eta_o,\alpha$) reported in Table~\ref{rerpdt}. The results obtained are displayed in Table~\ref{dv2v2c}. We note that the fundamental parameters given in Table~\ref{dv2v2c} are not for specific stars, but correspond to hypothetical objects having the quoted MK spectral types.\par
 From the results displayed in Table~\ref{dv2v2c} we conclude that provided the objects are observed with large enough baselines, even a modest flattening $R_{\rm e}/R_{\rm p}\sim1.05$ could be detected at the limiting magnitude $V=5$ mag. Brighter objects with slightly larger flattenings are found to have visibility ratios $dV^2/\langle{V^2}\rangle$ that are easily beyond the uncertainty range of $3\sigma_{V^2}/V^2\sim0.06$. Hence, rotational parameters ($\eta_o,\alpha$) can be inferred  with the VEGA/CHARA instrument to much better than 2\% in rotators brighter than $V=5$ mag with rates $\eta_o\gtrsim0.4$. \par
\begin{figure}[]
\centerline{\psfig{file=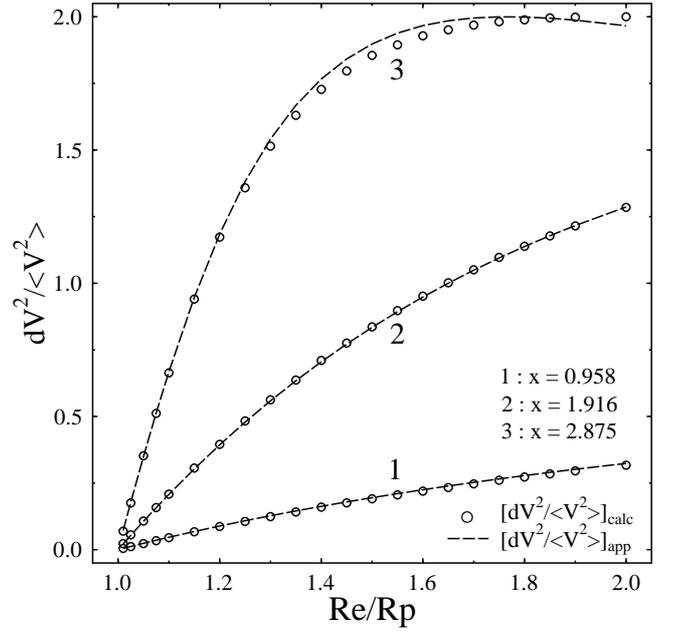}} 
\caption[]{\label{dv2_v2} Calculated (circles) and analytically approximated (dashed line) visibility ratios $dV^2/\langle V^2\rangle$ as a function of the apparent radii ratio $R_{\rm e}/R_{\rm p}$, for different parameters $x = \pi\theta_o(B/\lambda)$.}
\end{figure}

   
 For the sake of completeness, we derive in Appendix~\ref{aar} an expression for the visibility ratio $dV^2/\langle{V^2}\rangle$ as a function of the apparent flattening $R_{\rm e}/R_{\rm p}$ and the argument $x=\pi\theta_o(B/\lambda)$, where $\theta_o=(2/D)[(R_{\rm e}+R_{\rm p})/2$] is the average angular diameter, given by

\begin{equation}
\left.\begin{array}{lcl}
\label{dv2v2app}
\frac{dV^2}{\langle{V^2}\rangle} & = & -4\left\{\frac{\delta_{\rm ep}[\Gamma(x)/V(x)]}{1+\delta_{\rm ep}^2[\Gamma(x)/V(v)]^2}\right\} \nonumber \\
\Gamma(x) & = & dV(x)/d\ln x \nonumber \\
\delta_{\rm ep} & = & \frac{(R_{\rm e}/R_{\rm p})-1}{(R_{\rm e}/R_{\rm p})+1} \nonumber \\
\end{array}
\right\}
\end{equation}

\noindent which gives a clear insight into the dependence of $dV^2/\langle{V^2}\rangle$ on  $R_{\rm e}/R_{\rm p}$, provided that $V_{\rm e}$ and $V_{\rm p}$ are both in the respective first lobes of $J_1(x)$, i.e. $x_{\rm p}\!\lesssim\!3.8327$ and $x_{\rm e}\!\lesssim\!3.8327$. In Fig.~\ref{dv2_v2}, we compare the ratios $dV^2/\langle{V^2}\rangle$ given as a function of the apparent flattening $R_{\rm e}/R_{\rm p}$ and calculated for several average arguments $x$ (circles). In the same figure, we also superimpose the estimates derived from the approximation given in Eq.~(\ref{dv2v2app}) (dashed lines). Figure~\ref{dv2_v2} may help us to sharpen the observation strategy. From the figure, it is clear that when $R_{\rm e}\sim R_{\rm p}$, it is $dV^2/\langle{V^2}\rangle\to0$ and when for the same baseline one of the visibilities is significantly higher than the other the visibility ratio saturates, $dV^2/\langle{V^2}\rangle\to2$, so that the resolution for the determination of high flattenings becomes less obvious.\par 
 Finally, we note that for $\eta_o\!\simeq\!0.4$ and $\alpha\!\thickapprox\!0$ the flattening is $R_{\rm e}/R_{\rm p}\!\simeq\!1.2$, which also implies that $\Omega_o\Omega_{\rm c}\!\simeq\!0.63$ and $V_{\rm eq}/V_{\rm crit}\!\simeq\!0.7$. This translates into $V_{\rm eq}\!\sim\!260[(M/M_{\odot})/R_o/R_{\odot})]$ km~s$^{-1}$ \citep{frem05}, which means that the equatorial velocities range from 300 to 220 km~s$^{-1}$ in dwarf stars with spectral types going from B0 to A0, and from 230 to 180 km~s$^{-1}$ in giant stars of the same spectral type. Searching in the catalog by \citet{gleb00} for emission-less objects hotter than A2, with $V\!\sin i\!\gtrsim\!200$ km~s$^{-1}$, apparent magnitudes $V\! \leq \!5$ mag, and declinations $\delta>-20^o$, we found 63 OB stars that can be observed with VEGA/CHARA in the framework of the present study. Since their average inclination angle is probably $i\gtrsim52^o\!\sim\!\arcsin(\pi/4)$, we may expect that many of them have true equatorial velocities in the range $250~{\rm km~s^{-1}}\!\lesssim V_{\rm eq\!}\lesssim\!V_{\rm crit}$.\par

\setcounter{table}{5}
\begin{table}[]
\centering
\caption[]{\label{dv2v2c}Visibility ratios $dV^2/\langle{V^2}\rangle$ calculated for different B sub-spectral types and luminosity classes, with several equator-to-polar radius ratios $R_{\rm e}/R_{\rm p}$. The visibility ratios are calculated at $\lambda = 5500\ \AA$ for two baselines and apparent visible magnitudes V.}
\tabcolsep 4.5pt
\begin{tabular}{rc|ccc|ccc}
\hline
\noalign{\smallskip}
       &                  & \multicolumn{3}{c|}{B = 156 m} & \multicolumn{3}{c}{B = 300 m}  \\
\cline{3-8}
&&  \multicolumn{3}{c|}{$R_{\rm e}/R_{\rm p}$} & \multicolumn{3}{c}{$R_{\rm e}/R_{\rm p}$}  \\
D (pc) & $\theta_o$ (mas) &  1.05   & 1.10   & 1.25   & 1.05   & 1.10   & 1.25 \\
\cline{3-8}
\noalign{\smallskip}
&& \multicolumn{6}{c}{B0V} \\
\noalign{\smallskip}
316    & 0.203            &  0.036  & 0.070  &  0.143 & 0.151  &  0.274  & 0.550 \\
501    & 0.128            &  0.015  & 0.029  &  0.055 & 0.056  &  0.102  & 0.210 \\
\noalign{\smallskip}
&& \multicolumn{6}{c}{B0III} \\
\noalign{\smallskip}
457    & 0.195            &  0.036  & 0.063  &  0.131 & 0.135  &  0.252  & 0.510 \\
724    & 0.123            &  0.015  & 0.026  &  0.052 & 0.050  &  0.094  & 0.194 \\
\noalign{\smallskip}
&& \multicolumn{6}{c}{B2V} \\
\noalign{\smallskip}
174    & 0.262            &  0.063  & 0.117  &  0.242 & 0.277  &  0.504  & 0.954 \\
275    & 0.165            &  0.024  & 0.044  &  0.093 & 0.096  &  0.176  & 0.362 \\
\noalign{\smallskip}
&& \multicolumn{6}{c}{B2III} \\
\noalign{\smallskip}
302    & 0.250           &  0.056  & 0.107  &  0.218 & 0.248  &  0.453  & 0.868 \\
479    & 0.158           &  0.021  & 0.041  &  0.084 & 0.085  &  0.161  & 0.327 \\
\noalign{\smallskip}
&& \multicolumn{6}{c}{B8V} \\
\noalign{\smallskip}
 57    & 0.433           &  0.036  & 0.067  &  0.139 & 0.145  &  0.269  & 0.536 \\
 91    & 0.273           &  0.015  & 0.028  &  0.054 & 0.055  &  0.099  & 0.204 \\
\noalign{\smallskip}
&& \multicolumn{6}{c}{B8III} \\
\noalign{\smallskip}
115    & 0.414           &  0.173  & 0.315  &  0.627 & 1.329  &  1.750  & 1.966 \\
182    & 0.261           &  0.062  & 0.116  &  0.240 & 0.274  &  0.499  & 0.946 \\
\noalign{\smallskip}
\hline
\multicolumn{8}{l}{Note: D is the distance and $\theta_o$ is the angular diameter. For a given}\\
\multicolumn{8}{l}{spectral type, the first line is for $V\!=\!4$ mag and the second for $V\!=\!5$}\\ 
\multicolumn{8}{l}{mag.}\\ 
\noalign{\smallskip}
\hline
\end{tabular}
\end{table}
 
 A detailed account of the iteration procedure used to determine $R_{\rm e}/R_{\rm p}$, $\eta_o$, $\alpha$, and $i$ from interferometric data will be given elsewhere for actual stars observed with VEGA/CHARA.\par
\smallskip
 b) \underline{Parameters ($\alpha,i$)}. \citet{celli95a,celli95b} and \citet{petr96} highlighted the advantages of combining Fourier transforms with the differential interferometry. \citet{arm04a} showed that using the wavelength-dependent photometric barycenter, measured in angular units of interferometric intensity maps, it is possible to produce independent observational estimates of the differential rotation parameter $\alpha$ and the inclination angle $i$. The method is fairly independent of the characteristics of the observed spectral line and gives well defined indications of the sign of $\alpha$. It also provides two estimates of $i$. Combining then data from the above class a) with either interferometric or spectroscopic information, it is possible to infer the right value of $i$. The method developed by \citet{arm04a} is used for slow rotators. Its implementation to gravity-darkened and $\alpha$-dependent differential rapidly rotating early-type stars is currently being developed.\par
\smallskip
 c) \underline{Imaging}. Recent imaging of fast rotators using interferometric data has detected deviations from the standard von Zeipel gravity-darkening prescription \citep{monn07,zhao09}. The imaging technique can therefore provide complementary information to solve a possible parameter degeneracy that may include the differential rotation in the stellar surface.\par
  We note that while in the imaging techniques parameters such as $\betaup_{\rm GD}$ and the inclination angle $i$ are inferred mostly from confidence levels in the $\chi^2$ fitting surfaces, in the present attempt we try to infer $\alpha$ and $i$ using direct methods. This does not preclude, however, the use of confidence levels to decide the choice of other parameters, in particular the gravity darkening exponent $\betaup_{\rm GD}$.\par

\subsection{Method to derive the parameters}
\label{methpar}

  We developed GIRFIT, a powerful tool to fit observed spectra of low and high dispersion \citep{fre06} with theoretical fluxes based on model atmospheres calculated with TLUSTY \citep{hubla95} and/or with ATLAS9 \citep{kur93,cagrku97}. Adopting parameterized values of $\betaup_{\rm GD}$, we can thus infer the apparent quantities given in Eq.~(\ref{ltgap}). Corrections for first order rotation effects can then be attempted as in \citet{frem05}, \citet{zor05}, \citet{marta06}, and \citet{marta07} to derive the average parameters in Eq.~(\ref{paramm}) and obtain estimates of $M$ and $t$. Since in most remaining calculations the mass enters as $M^{1/2}$, we can proceed to derive the other quantities with approximate values of $M$.\par
  In principle, from Eq.~(\ref{q2q1l}) and relations of the type shown in Fig.~\ref{w_i}b we can derive $\alpha)$ and $i$. By minimizing the differences between Eq.~(\ref{interfdat}) and interferometric observations, another set of independent estimates of $(\eta_o,\alpha,i)$ can be derived.\par 
  The choice of the iterated parametric set ($M,t,\eta_o,\alpha,i$) as a function of the gravity darkening power $\betaup_{\rm GD}$ can be finally decided by applying a GIRFIT-like residual minimization procedure between the observed and the FASTROT model spectra.\par

\section{The rotational profile in the envelope}
\label{trpe}
 
 From the discussion above, it follows that if $\alpha\neq0$, it will probably be impossible to know anything more about the characteristics of the rotation law in the convective layers beneath the stellar surface. Apart from the rigid or shellular-type rotation profiles being able to be excluded, spectroscopic and interferometric data do not provide any further information. Higher radii ratios $R_{\rm e}/R_{\rm p}$ than allowed by the inferred values of $\alpha$ may be indicative of large values of the energy ratio $\tau=K/|W|$, but no detailed description can be derived about the internal rotation law itself. To establish the properties of this law near the surface, at least whether they are to be likely of the speculated types generated by the condition $S\!=\!S(j^2)$, $S\!=\!S(\Omega^2)$, or $S\!=\!S(\epsilon_{\Omega})$, no information can be obtained otherwise than from the analysis of non-radial pulsation modes excited in these regions, provided they can be unambiguously detected from time series of spectral line variations. \citet{and80} showed that the ``effective depths'' of $p_n$-modes, where $n\!=\!1-5$, for stars with masses from $5M_{\odot}$ to $20M_{\odot}$ range from 0.7 to 0.9$R_{\rm e}$, which means that the detected pulsations, mainly the splitting of modes, may carry some information on the characteristics of the rotation law in the upper envelope layers. In all cases, the interpretation of these modes rely on the knowledge of the stellar fundamental parameters, in particular the inclination angle $i$ of the rotation axis \citep{flo00,flo02,
nei02,leven03,fre06} whose determination can be attempted with the methods outlined in the present paper.\par  

\section{Conclusions}
\label{conclus}

    Recent theoretical works suggest that rapidly rotating early-type stars should have rather deep convective layers in the envelope. This situation may favor some coupling between the convective region and the differential rotation, as happens in the Sun. We have considered the possible properties the iso-rotation curves can have in the envelope immediately under the stellar surface by assuming several conditions for the specific entropy $S$ as a function of the: 1) specific angular momentum $j$, as implied by marginal Solberg-H\o iland stability condition; 2) squared angular velocity $\Omega^2$, which accounts for the Solar differential rotation in the convection zone; 3) specific kinetic energy $\varpi^2\Omega^2$, which reproduces the calculated rotational profiles in radiative envelopes. In all cases, the function that describes the gradient $\partial\Omega/\partial\theta$ in the stellar surface must be specified in advance. We thus assumed a simple Maunder formula with a unique differential rotation-parameter $\alpha$ to represent the imprint on the stellar surface produced by the differential rotation beneath the surface. In spite of the simplicity of this relation, for many theoretical reasons it would be highly interesting to infer from observations reliable orders of magnitude of $\alpha$ and its sign. While a value $\alpha\neq0$ can imply that the rotation beneath the surface is neither rigid nor shellular, neither spectroscopy nor interferometry can say anything about their actual functional nature beneath the surface. Other piece of information is thus needed, perhaps non-radial pulsation data of pulsation modes excited in the intermediate and upper stellar layers.\par
    In early-type fast rotators, we have shown that the surface temperature gradient might induce a positive gradient in the surface angular velocity, i.e., $\partial\Omega/\partial\theta>0$.\par
    In this paper, we have summarized the kind of information that it is possible to extract from the spectroscopic and interferometric data about the differential rotation of rapidly rotating early-type stars. The equation of the surface of a star with a non-conservative rotation law was discussed. It has been shown that a differential rotation in the surface induces measurable stellar deformations. These deformations carry a surface gravitational darkening effect that needs to be studied consistently with the induced geometrical deformations. We have shown that the effects induced by the surface differential rotation can be studied consistently and in a reliable way by a combined analysis of spectroscopic and interferometric data. From spectroscopy, estimates of the differential rotation parameter $\alpha$ with uncertainties $\delta\alpha\lesssim0.15$ and of the inclination angle $i$ with $\delta i\lesssim5^o$ can be obtained. The interferometry can provide information about the true ratio $R_{\rm e}/R_{\rm p}$ of the equatorial to the polar radii to within less than 3\%, provided that a reliable estimate of the inclination angle is given from spectroscopy. The VEGA/CHARA interferometric instrument is capable of providing useful information about rotationally induced stellar deformations for nearly 60 OB rapidly rotating stars ($V\!\sin i>200$ km~s$^{-1}$) whose spectra are not marred by circumstellar emission/absorption. Differential interferometry can help distinguish the $\alpha$ and $i$ parameters and provide reliable independent estimates of each.\par
    By calculating simplified models of stars at different evolutionary stages in the MS with internal rigid and differential conservative rotation laws, we have tested the use of the Roche approximation to represent the gravitational potential of rotating stars.  We have concluded that we can safely use the Roche approximation for most, if not all stellar objects. Only in highly centrifugally deformed objects may this approximation not be applicable, but doubts exist as to whether these cases actually exist in Nature.\par
    Uncertainties and constraints carried by the spectroscopic and interferometric modeling as well as the application of these methods to real stars will be presented in separate papers.\par

\begin{acknowledgements}
We thank the referee whose careful reading, remarks and criticisms allowed to improve the presentation of our results. We are particularly indebted to Dr. Claire Halliday for her
careful language editing of this paper. YF aknowledges the FNRS (Belgium) for travel assistance in the framework of contract 1.5.211.09 ``Cr\'edit aux chercheurs". LC acknowledges financial support from the Agencia de Promoci\'on Cient\'{\i}fica y Tecnol\'ogica (BID 1728 OC/AR PICT 111), from CONICET (PIP 0300) and the Programa de Incentivos G11/089 of the National University of La Plata, Argentina. 
\end{acknowledgements}

\bibliographystyle{aa}
\bibliography{15691}


\begin{appendix}
\section{Approximate analytic representation of $dV^2/\langle V^2\rangle$}
\label{aar}

 We assume that the observed visibility of an oblate source having an equator to polar radii ratio $R_{\rm e}/R_{\rm p}$ to be given for an equivalent uniform disc with angular diameter $\theta_o$ as 

\begin{equation}
\label{aviss0}
V(x) = \left|\frac{2J_1(x)}{x}\right|\ ,
\end{equation}

\noindent where $J_1(x)$ is the Bessel function of the first kind, $x=\pi\theta_o(B/\lambda)$ where $B$ is the baseline of the interferometer projected onto the sky and $\lambda$ is the effective wavelength at which observations are carried out. To achieve an accuracy of better than $0.2\%$, we can use the development

\begin{equation}
\label{aviss1}
V(x) \simeq 1+\sum_{n=1}^3a_nx^n
\end{equation}

\noindent with $a_1=0.0173261$, $a_2=-0.160464$, and $a_3=0.229252$, which is valid for $0.0\leq x \leq3.8327$ [first lobe of $J_1(x)$]. Defining the argument $x$ as a function of the average angular diameter of the source $\theta_o=2\times[(R_{\rm e}+R_{\rm p})/2]/D$ (where $D$ is the distance of the source), the corresponding arguments of the visibility function in the equatorial and polar direction of the same source can be written as

\begin{eqnarray}
\label{aviss2}
x_{\rm p} & = & x\times(1+\delta_{\rm ep}) \nonumber \\
x_{\rm e} & = & x\times(1-\delta_{\rm ep}) \nonumber \\
\delta_{\rm ep} & = & \frac{(R_{\rm e}/R_{\rm p})-1}{(R_{\rm e}/R_{\rm p})+1}.
\end{eqnarray}

   Since $0.0\leq\delta_{\rm ep}\leq1$ we can use a convergent development of Eq.~(\ref{aviss1}) in a series of powers of $\delta_{\rm ep}$. Conserving only the first order terms in $\delta_{\rm ep}$, Eq.~(\ref{aviss1}) becomes

\begin{equation}
\label{aviss3}
V_{\pm\delta}(x) \simeq V(x)\pm \delta_{\rm ep}\Gamma(x),
\end{equation}

\noindent where $\Gamma(x)=\sum_{n=1}^3na_nx^n=dV(x)/d\ln x$. Writing $V_{\rm p}=V[x\times(1+\delta_{\rm ep}]$ and $V_{\rm e}=V[x\times(1-\delta_{\rm ep}]$, we obtain

\begin{equation}
\label{aviss3}
\frac{dV^2}{\langle V^2\rangle} = \frac{V_{\rm p}^2-V_{\rm e}^2}{\frac{1}{2}(V_{\rm p}^2+V_{\rm e}^2)}=
-4\left\{\frac{\delta_{\rm ep}[\Gamma(x)/V(x)]}{1+\delta_{\rm ep}^2[\Gamma(x)/V(x)]^2}\right\}\ ,
\end{equation}

\noindent which is assumed to be used only when both $V_{\rm e}$ and $V_{\rm p}$ are in the respective first lobes of $J_1(x)$, i.e. $x_{\rm p}\!\lesssim3.8327$ and $x_{\rm e}\!\lesssim3.8327$.\par

\end{appendix}

\clearpage
\onecolumn
\scriptsize
\setcounter{table}{4}
\begin{longtable}{ccrrrr|rrrrrrrrrr}
\caption{\label{modkwequi}Zonal harmonic coefficients $\gamma_{2n}$ and residuals $\Delta R(\theta)/R(\theta)$}\\
\hline\hline
\multicolumn{16}{c}{} \\
\multicolumn{7}{r}{M = 5M$_{\odot}$ (ZAMS)} & \multicolumn{9}{c}{$\Delta R(\theta)/R(\theta)=[R(\theta)_{\rm model}-R(\theta)_{\delta_n}]/R(\theta)_{\rm model}\times10^4$} \\
$\beta$ & $\Omega_o/\Omega_{cr}$ & \multicolumn{1}{c}{$\gamma_2$} & \multicolumn{1}{c}{$\gamma_4$} & \multicolumn{1}{c}{$ \gamma_6$} & \multicolumn{1}{c}{$\gamma_8$} & $0^o$ & $5^o$ & $15^o$ & $30^o$ & $40^o$ & $50^o$ & $60^o$ & $70^o$ & $80^o$ & $90^o$ \\
\hline
\endfirsthead
\caption{continued.}\\
\hline\hline
$\beta$& $\Omega_o/\Omega_{cr}$ & \multicolumn{1}{c}{$\gamma_2$} & \multicolumn{1}{c}{$\gamma_4$} & \multicolumn{1}{c}{$\gamma_6$} & \multicolumn{1}{c}{$\gamma_8$} & $0^o$ & $5^o$ & $15^o$ & $30^o$ & $40^o$ & $50^o$ & $60^o$ & $70^o$ & $80^o$ & $90^o$ \\
\hline
\endhead
\hline
\endfoot
\multicolumn{16}{c}{} \\
\multicolumn{16}{c}{} \\
 2  & 1.0  & $-0.119\times10^{-2}$  & $-0.595\times10^{-3}$   & $0.486\times10^{-4}$  & $-0.26\times10^{-3}$  &  9 & 10 & 12 & 20 & 30 & 38 & 43&45&48&49 \\
    & 2.0  & $0.149\times10^{-2}$ & $-0.318\times10^{-3}$ & $0.274\times10^{-3}$ & $-0.104\times10^{-3}$ & -8   & -7 & 3 & 17 & 29 & 35 & 39 & 42 & 46 & 47 \\
    & 3.0  & $0.308\times10^{-2}$ & $0.829\times10^{-4}$ & $0.305\times10^{-4}$ & $-0.729\times10^{-5}$ & -3   & -4 & -5 & 6 & 17 & 25 & 31 & 35 & 37 & 39 \\
    & 3.5  & $0.302\times10^{-2}$ & $-0.986\times10^{-4}$ & $0.103\times10^{-4}$ & $-0.497\times10^{-6}$ &-115&-107& -45 & 12 & 21 & 26 & 38 & 44 & 47 & 48 \\
    & 3.9  & $0.321\times10^{-2}$ & $-0.868\times10^{-4}$ & $0.153\times10^{-5}$ & $-0.112\times10^{-7}$ &-2651&-2416&-907&242& 19& 133& 128& 127 &126& 127 \\
\multicolumn{16}{c}{} \\
 4  & 1.0  & $-0.148\times10^{-2}$ &$-0.757\times10^{-3}$  & $0.660\times10^{-4}$ & $-0.374\times10^{-3}$ & 1 & 1 & 13 & 20 & 31 & 40 & 44 & 45 & 48 & 49 \\
    & 2.0  & $0.761\times10^{-3}$ & $-0.525\times10^{-3}$ & $0.165\times10^{-3}$ & $-0.138\times10^{-3}$ & 2 & 3 &  9 & 20 & 31 & 38 & 42 & 45 & 47 & 49 \\
    & 3.0  & $0.360\times10^{-2}$ & $-0.398\times10^{-3}$ & $0.260\times10^{-3}$ & $-0.525\times10^{-4}$ & 20 & 17 & 1 & 19 & 30 & 36 & 40 & 44 & 46 & 48 \\
    & 4.0  & $0.579\times10^{-2}$ & $-0.277\times10^{-3}$ & $0.189\times10^{-3}$ & $-0.317\times10^{-4}$ &-54 &-49 &-17& 16 & 28 & 33 & 37 & 41 & 43 & 44 \\
    & 5.0  & $0.636\times10^{-2}$ & $-0.771\times10^{-4}$ & $0.101\times10^{-4}$ & $-0.125\times10^{-5}$ &-12 &-13 & -9 & 5 & 18 & 26 & 31 & 35 & 37 & 38 \\
    & 6.2  & $0.615\times10^{-2}$ & $-0.172\times10^{-3}$ & $0.337\times10^{-5}$ & $-0.280\times10^{-7}$ &-3801&-3358 &-1152 &277 & 51 & 63 & 57 &54&53&52 \\
\multicolumn{16}{c}{} \\
 6  & 1.0  & $-0.170\times10^{-2}$ & $-0.682\times10^{-3}$ & $-0.716\times10^{-5}$& $-0.409\times10^{-3}$ & 12 & 12 & 13 & 20 & 31 & 40 & 44 & 46 & 48 & 49 \\
    & 2.0  & $0.146\times10^{-3}$ & $-0.564\times10^{-3}$ & $0.574\times10^{-4}$ & $-0.164\times10^{-3}$ &  7 &  7 & 12 & 21 & 31 & 39 & 43 & 46 & 48 & 49 \\
    & 4.0  & $0.549\times10^{-2}$ & $-0.429\times10^{-3}$ & $0.246\times10^{-3}$ & $-0.309\times10^{-4}$ &-32 &-27 & -1 & 21 & 32 & 37 & 42 & 45 & 48 & 49 \\
    & 5.0  & $0.773\times10^{-2}$ & $-0.403\times10^{-3}$ & $0.230\times10^{-3}$ & $-0.359\times10^{-4}$ &-70 &-63 &-18 & 19 & 31 & 36 & 40 & 44 & 47 & 48 \\
    & 7.0  & $0.905\times10^{-2}$ & $-0.284\times10^{-3}$ & $0.789\times10^{-4}$ & $-0.923\times10^{-5}$ &-82 &-76 &-30 & 14 & 17 & 19 & 37 & 44 & 48 & 49 \\
    & 8.4  &  $0.885\times10^{-2}$ & $-0.233\times10^{-3}$ & $0.404\times10^{-5}$ & $-0.297\times10^{-7}$ &-6993&-6158&-2102&512 & 54 & 23 & 8 & 0 & 5 & 7 \\
\multicolumn{16}{c}{} \\
 8  & 1.0  & $-0.180\times10^{-2}$ & $-0.760\times10^{-3}$ & $-0.140\times10^{-4}$ & $-0.417\times10^{-3}$ & 12 & 12 & 14 & 20 & 31 & 40 & 44 & 46 & 48 &49\\
    & 3.0  & $0.199\times10^{-2}$ & $-0.556\times10^{-3}$ & $0.223\times10^{-3}$ & $-0.138\times10^{-3}$ & -1 &  0 & 10 & 22 & 32 & 39 & 43 & 46 & 49 & 50 \\
    & 5.0  & $0.720\times10^{-2}$ & $-0.453\times10^{-3}$ &  $0.260\times10^{-3}$ & $-0.332\times10^{-4}$ &-45 &-38 & -4 & 22 & 33 & 39 & 43 & 47 & 50 & 50 \\
    & 7.0  & $0.112\times10^{-1}$ & $-0.459\times10^{-3}$ &  $0.139\times10^{-3}$ & $-0.170\times10^{-4}$ &-116&-106&-41 & 18 & 32 & 37 & 42 & 46 & 49 & 50 \\
    & 9.0  & $0.113\times10^{-1}$ & $-0.331\times10^{-3}$ &  $0.101\times10^{-4}$ & $-0.201\times10^{-6}$ &-258&-228&-63 & 17 & 21 & 31 & 36 & 41 & 44 & 45 \\
    &10.0  & $0.106\times10^{-1}$ & $-0.351\times10^{-3}$ &  $0.974\times10^{-5}$ & $-0.126\times10^{-6}$ &-2562&-2316&-712&180& 19 & 23 & 29 & 42 & 47 &49 \\
\multicolumn{16}{c}{} \\
\multicolumn{16}{c}{M = 5 M$_{\odot}$ (TAMS)} \\
\multicolumn{16}{c}{} \\
 2  & 1.0  & $-0.276\times10^{-2}$ & $-0.818\times10^{-3}$ & $0.147\times10^{-3}$ &$-0.413\times10^{-3}$ & 11 & 11 & 14 &  24 & 37 & 47 & 52 & 54 & 58 & 60 \\
    & 2.0  & $-0.231\times10^{-2}$ & $-0.433\times10^{-3}$ & $0.325\times10^{-3}$ & $-0.104\times10^{-3}$ &-13 &-11 & 3 & 21 & 34 & 41 & 45 & 49 & 53 & 54 \\
    & 3.0  & $-0.839\times10^{-3}$ & $-0.114\times10^{-3}$ & $0.603\times10^{-4}$ & $-0.564\times10^{-5}$ &-96 &-88 &-34 & 17 & 28 & 46 & 53 & 58 & 60 & 61 \\
    & 3.5  & $-0.341\times10^{-3}$ & $0.311\times10^{-4}$  &$-0.104\times10^{-5}$ & $0.133\times10^{-7}$ &328 &284 & 80 & -9 & 77 & 96 & 104 & 108 & 110 &111\\
\multicolumn{16}{c}{} \\
 4  & 1.0  & $-0.278\times10^{-2}$ & $-0.928\times10^{-3}$ & $0.920\times10^{-4}$ & $-0.487\times10^{-3}$ & 13 & 13 & 15 & 24 & 38 & 48 & 53 & 55 & 58 & 60 \\
    & 3.0  & $-0.186\times10^{-2}$ & $-0.506\times10^{-3}$ & $0.276\times10^{-3}$ & $-0.359\times10^{-4}$ &-28 &-23 &  1 & 22 & 34 & 40 & 44 & 48 & 51 & 52 \\
    & 4.0  & $-0.709\times10^{-3}$ & $-0.413\times10^{-3}$ & $0.184\times10^{-3}$ & $-0.191\times10^{-4}$ &-131&-113&-27 & 23 & 32 & 44 & 48 & 50 & 52 & 53 \\
    & 5.0  & $-0.589\times10^{-3}$ &  $0.131\times10^{-3}$ &$-0.831\times10^{-5}$ &  $0.193\times10^{-6}$ & 475& 398& 76 & 22 & 58 & 70 & 76 & 81 & 84 & 85 \\
    & 5.5  & $0.141\times10^{-3}$ &  $0.651\times10^{-5}$ &$-0.129\times10^{-6}$ &  $0.726\times10^{-9}$ & 789& 682& 128& 86 & 108& 116& 120& 122& 123& 124 \\
    & 5.7  & $0.417\times10^{-3}$ & $-0.225\times10^{-5}$ & $0.604\times10^{-8}$ & $-0.609\times10{-11}$ &-3913&-3268&-518&80& 92&106 & 105&105 & 104& 104 \\
\multicolumn{16}{c}{} \\
 6  & 1.0  & $-0.280\times10^{-2}$ & $-0.938\times10^{-3}$ & $0.377\times10^{-4}$ & $-0.502\times10^{-3}$ & 14 & 14 & 16 & 25 & 38 & 49 & 53 & 55 & 58 & 60 \\
    & 3.0  & $-0.206\times10^{-2}$ & $-0.674\times10^{-3}$ & $0.211\times10^{-3}$ & $-0.518\times10^{-4}$ & -9 & -6 &  9 & 24 & 35 & 43 & 48 & 52 & 54 & 55 \\
    & 4.0  & $-0.149\times10^{-2}$ & $-0.507\times10^{-3}$ & $0.253\times10^{-3}$ & $-0.166\times10^{-4}$ & -48&-39 & -1 & 23 & 33 & 39 & 43 & 47 & 49 & 50 \\
    & 5.0  & $-0.504\times10^{-3}$ & $-0.482\times10^{-3}$ & $0.207\times10^{-3}$ & $-0.206\times10^{-4}$ &-160&-136& -24& 24 & 32 & 43 & 47 & 49 & 51 & 52 \\
    & 6.0  & $-0.489\times10^{-3}$ &  $0.129\times10^{-3}$ &$-0.609\times10^{-5}$ &  $0.605\times10^{-7}$ & 116& 89 &  9 & 25 & 46 & 55 & 61 & 65 & 68 & 69 \\
    & 7.0  & $0.504\times10^{-3}$ &  $0.222\times10^{-5}$ & $0.156\times10^{-6}$ &  $0.144\times10^{-8}$ & 297&256 & 39 & 44 & 59 & 64 & 66 & 68 & 68 & 69 \\
\multicolumn{16}{c}{} \\
 8  & 1.0  & $-0.282\times10^{-2}$ & $-0.899\times10^{-3}$ &$-0.708\times10^{-4}$ & $-0.480\times10^{-3}$ & 16 & 16 & 17 & 25 & 37 & 49 & 54 & 56 & 59 & 60 \\
    & 3.0  & $-0.219\times10^{-2}$ & $-0.757\times10^{-3}$ & $0.174\times10^{-3}$ & $-0.118\times10^{-3}$ & -0 &  2 & 12 & 25 & 37 & 45 & 50 & 53 & 56 & 57 \\
    & 5.0  & $-0.119\times10^{-2}$ & $-0.505\times10^{-3}$ & $0.259\times10^{-3}$ & $-0.194\times10^{-4}$ &-72 & -57& -3 & 23 & 33 & 39 & 43 & 46 & 48 & 49 \\
    & 6.0  & $-0.306\times10^{-3}$ & $-0.508\times10^{-3}$ & $0.216\times10^{-3}$ & $-0.214\times10^{-4}$ &-199&-164&-23 & 25 & 33 & 43 & 47 & 49 & 50 & 51 \\
    & 8.0  & $0.434\times10^{-4}$ &  $0.886\times10^{-4}$ &$-0.445\times10^{-5}$ &  $0.727\times10^{-7}$ & 932 &755 & 74& 38 & 54 & 61 & 65 & 69 & 71 & 71 \\
    & 9.0  & $0.757\times10^{-3}$ & $-0.292\times10^{-5}$ & $0.526\times10^{-8}$ & $-0.346\times10^{-11}$ &-14389&-11830&-1088&-162&-153&-150&-150&-150&-150&-150\\
\multicolumn{16}{c}{} \\
\multicolumn{16}{c}{M = 15M$_{\odot}$ (ZAMS)} \\
\multicolumn{16}{c}{} \\
 2  & 1.0  & $-0.276\times10^{-3}$ & $-0.598\times10^{-3}$ & $0.139\times10^{-3}$ & $-0.325\times10^{-3}$ & 9 & 9 & 11 & 19 & 30 & 38 & 42 & 44 & 47 & 49 \\
    & 2.0  & $0.438\times10^{-2}$ & $-0.309\times10^{-3}$ & $0.240\times10^{-3}$ & $-0.877\times10^{-4}$ &-8 & -6 & 3 & 17 & 29 & 35 & 39 & 43 & 46 & 47 \\
    & 3.0  & $0.743\times10^{-2}$ & $-0.638\times10^{-4}$ & $0.320\times10^{-4}$ & $-0.766\times10^{-5}$ &-4 & -5 &-4 &  7 & 18 & 26 & 33 & 37 & 40 & 41 \\
    & 3.5  & $0.664\times10^{-2}$ & $-0.201\times10^{-3}$ & $0.147\times10^{-4}$ & $-0.773\times10^{-6}$ &-98&-90 &-35& 12 & 20 & 22 & 32 & 41 & 45 & 46 \\
    & 4.0  & $0.587\times10^{-2}$ & $-0.152\times10^{-3}$ & $0.250\times10^{-5}$ & $-0.170\times10^{-7}$ &-602&-5395&-2079&609&389&342&337&336&336 & 337 \\
\multicolumn{16}{c}{} \\
 4  & 1.0  & $-0.876\times10^{-3}$ & $-0.666\times10^{-3}$ & $0.103\times10^{-3}$ & $-0.408\times10^{-3}$ & 10 & 10 & 12 & 20 & 31 & 39 & 43 & 45 & 47 & 49 \\
    & 2.0  & $0.295\times10^{-2}$ & $-0.478\times10^{-3}$ & $0.218\times10^{-3}$ & $-0.163\times10^{-3}$ &  1 &  2 &  9 & 20 & 31 & 38 & 42 & 45 & 47 & 49 \\
    & 4.0  & $0.118\times10^{-1}$ & $-0.478\times10^{-3}$ & $0.190\times10^{-3}$ & $-0.329\times10^{-4}$ &-50 &-46 &-17 & 16 & 29 & 35 & 39 & 43 & 46 & 47 \\
    & 5.5  & $0.128\times10^{-1}$ & $-0.444\times10^{-3}$ & $0.177\times10^{-4}$ & $-0.489\times10^{-6}$ &-190&-168& -51& 19 & 18 & 25 & 32 & 37 & 40 & 41 \\
    & 6.2  & $0.110\times10^{-}$1 & $-0.416\times10^{-3}$ & $0.137\times10^{-4}$ & $-0.213\times10^{-6}$ &-2217&-210&-688&208 & 89 & 8 & 30 & 41 & 47 & 49 \\
    & 6.4  & $0.107\times10^{-1}$ & $-0.337\times10^{-3}$ & $0.753\times10^{-5}$ & $-0.731\times10^{-7}$ &-5925&-5253&-2038&562&240&-50 &-11&  1 &  8 & 10 \\
\multicolumn{16}{c}{} \\
 6  & 1.0  & $-0.120\times10^{-2}$ & $-0.729\times10^{-3}$ & $0.195\times10^{-4}$ & $0.375\times10^{-3}$  & 11 & 11 & 13 & 20 & 31 & 39 & 43 & 45 & 47 & 48 \\
    & 3.0  & $0.621\times10^{-2}$ & $-0.473\times10^{-3}$ & $0.281\times10^{-3}$ & $-0.943\times10^{-4}$ & -9 & -7 &  6 & 21 & 32 & 39 & 43 & 45 & 48 & 50 \\
    & 5.0  & $0.151\times10^{-1}$ & $-0.666\times10^{-3}$ & $0.232\times10^{-3}$ & $-0.375\times10^{-4}$ &-66 &-60 &-19 & 19 & 33 & 39 & 42 & 47 & 49 & 51 \\
    & 7.0  & $0.187\times10^{-1}$ & $-0.805\times10^{-3}$ & $0.421\times10^{-4}$ & $-0.161\times10^{-5}$ &-268&-233&-67 & 24 & 26 & 33 & 40 & 46 & 49 & 51 \\
    & 8.0  & $0.167\times10^{-1}$ & $-0.722\times10^{-3}$ & $0.296\times10^{-4}$ &  $0.623\times10^{-6}$ &-1477&-1329&-450&130& 54& 23 & 34 & 41 & 45 & 47 \\
    & 8.4  & $0.155\times10^{-1}$ & $-0.620\times10^{-3}$ & $0.205\times10^{-4}$ & $-0.319\times10^{-6}$ &-3411&-3002&-1074&303&137& 7 & 30 & 40 & 45 & 47 \\
\multicolumn{16}{c}{} \\
 8  & 1.0  & $-0.144\times10^{-2}$ & $-0.726\times10^{-3}$ & $0.267\times10^{-4}$ & $-0.402\times10^{-3}$ & 11 & 11 & 13 & 20 & 31 & 39 & 43 & 45 & 47 & 49 \\
    & 2.0  & $0.107\times10^{-2}$ & $-0.598\times10^{-3}$ & $0.167\times10^{-3}$ & $-0.300\times10^{-3}$ &  7 &  8 & 12 & 21 & 32 & 40 & 43 & 45 & 48 & 49 \\
    & 4.0  & $0.922\times10^{-2}$ & $-0.464\times10^{-3}$ & $0.261\times10^{-3}$ & $-0.413\times10^{-4}$ & -20 &-17&  4 & 22 & 33 & 40 & 44 & 47 & 50 & 51 \\
    & 6.0  & $0.177\times10^{-1}$ & $-0.790\times10^{-3}$ & $0.237\times10^{-3}$ & $-0.358\times10^{-4}$ &-81 &-72 &-23 & 21 & 36 & 41 & 46 & 50 & 53 & 54 \\
    & 8.0  & $0.229\times10^{-1}$ & $-0.110\times10^{-2}$ & $0.781\times10^{-4}$ & $-0.450\times10^{-5}$ &-243&-219&-64 & 24 & 33 & 40 & 46 & 53 & 57 & 58 \\
    & 9.6  & $0.219\times10^{-1}$ & $-0.106\times10^{-2}$ & $0.488\times10^{-4}$ & $-0.119\times10^{-5}$ &-1380&-1237&-394&119& 56& 32 & 43 & 51 & 56 & 58 \\
\multicolumn{16}{c}{} \\
\multicolumn{16}{c}{M = 15M$_{\odot}$ (TAMS)} \\
\multicolumn{16}{c}{} \\
 2  & 1.0  & $0.833\times10^{-2}$ & $0.230\times10^{-2}$ & $-0.373\times10^{-3}$ & $0.111\times10^{-2}$ &-33 &-33 &-44 &-73&-113&-144&-158 &-167&-176&-182\\
    & 2.0  & $0.797\times10^{-2}$ & $0.105\times10^{-2}$ & $-0.102\times10^{-2}$ & $0.344\times10^{-3}$ & 39 & 32 &-11 &-71&-116&-1412&-157&-171&-183&-189\\
    & 3.0  & $0.657\times10^{-2}$ & $-0.890\times10^{-3}$ & $0.950\times10^{-5}$ & $0.539\times10^{-5}$ &-176&-141&-12&-11&-82&-120&-141&-156&-166&-169\\
    & 3.5  & $0.260\times10^{-2}$ & $-0.138\times10^{-3}$ & $0.343\times10^{-5}$ &$-0.319\times10^{-7}$ &-3014&-2614&-735&-79&-78&-76&-78&-80&-82&-83\\
\multicolumn{16}{c}{} \\
 4  & 1.0  & $0.836\times10^{-2}$ & $0.251\times10^{-2}$ & $-0.670\times10^{-4}$ & $0.123\times10^{-2}$ &-39 &-40 &-48 &-74 &-112&-144&-159&-166&-175&-180\\
    & 2.0  & $0.820\times10^{-2}$ & $0.205\times10^{-2}$ & $-0.872\times10^{-3}$ & $0.628\times10^{-3}$ & -1 & -6 &-34 &-78 &-117&-144&-157&-168&-178&-183\\
    & 3.0  & $0.774\times10^{-2}$ & $0.125\times10^{-2}$ & $-0.101\times10^{-2}$ & $0.169\times10^{-3}$ & 102& 84 & -1 &-79 &-121&-144&-158&-171&-182&-186\\
    & 4.0  & $0.708\times10^{-2}$ & $0.197\times10^{-3}$ & $-0.545\times10^{-3}$ & $0.780\times10^{-4}$ & 330&304 &105 &-67 &-122&-136&-150&-163&-174&-177\\
    & 5.0  & $0.518\times10^{-2}$ & $-0.515\times10^{-3}$& $0.240\times10^{-4}$ &$-0.414\times10^{-6}$ &-3170&-2691&-481&-53&-90&-103&-115&-126&-132&-135\\
    & 5.5  & $0.173\times10^{-2}$ & $-0.236\times10^{-4}$& $0.145\times10^{-6}$ &$-0.323\times10^{-9}$ &-9898&-8248&-1336&-187&-175&-173&-174&-175&-177&-177\\
\multicolumn{16}{c}{} \\
 6  & 1.0  & $0.836\times10^{-2}$ & $0.270\times10^{-2}$ & $-0.959\times10^{-4}$ & $0.142\times10^{-2}$ &-41&-41&-48&-74&-113&-145&-159&-166&-174&-179\\
    & 3.0  & $0.805\times10^{-2}$ & $0.175\times10^{-2}$ & $-0.837\times10^{-3}$ & $0.176\times10^{-3}$ & 41& 29&-27&-82&-119&-143&-158&-169&-178&-182\\
    & 4.0  & $0.765\times10^{-2}$ & $0.128\times10^{-2}$ & $-0.105\times10^{-2}$ & $0.134\times10^{-3}$ &181&151& 10&-84&-125&-146&-159&-171&-181&-186\\
    & 5.0  & $0.736\times10^{-2}$ & $0.417\times10^{-3}$ & $-0.650\times10^{-3}$ & $0.889\times10^{-4}$ &433&384&97&-79&-126&-138&-152&-165&-175 &-179\\
    & 6.0  & $0.676\times10^{-2}$ & $-0.873\times10^{-3}$ & $0.531\times10^{-4}$ &$-0.115\times10^{-5}$ &-1319&-1044&-112&-63&-97&-115&-132&-147&-157&-160\\
    & 7.0  & $0.258\times10^{-2}$ & $-0.515\times10^{-4}$ & $0.467\times10^{-6}$ &$-0.153\times10^{-8}$ &-10115&-8462&-852&-213&-209&-212&-214&-219&-222&-223\\
\multicolumn{16}{c}{} \\
 8  & 1.0  & $0.840\times10^{-2}$ & $0.276\times10^{-2}$ & $-0.536\times10^{-4}$ & $0.145\times10^{-2}$ &-42&-42&-49&-73&-113&-145&-160&-166&-174&-179\\
    & 3.0  & $0.818\times10^{-2}$ & $0.200\times10^{-2}$ & $-0.749\times10^{-3}$ & $0.346\times10^{-3}$ &14& 5&-37&-82&-118&-142&-157&-168&-177&-181\\
    & 5.0  & $0.768\times10^{-2}$ & $0.120\times10^{-2}$ & $-0.104\times10^{-2}$ & $0.136\times10^{-3}$ &267&218&20&-87&-126&-146&-159&-171&-181&-186\\
    & 6.0  & $0.757\times10^{-2}$ & $0.398\times10^{-3}$ & $-0.615\times10^{-3}$ & $0.797\times10^{-4}$ &499&441&88&-84&-126&-138&-152&-165&-175&-179\\
    & 7.0  & $0.725\times10^{-2}$ & $-0.891\times10^{-3}$ & $0.435\times10^{-4}$ &$-0.301\times10^{-6}$ &-893&-679&-40&-73&-101&-120&-138&-153&-164&-167\\
    & 8.5  & $0.342\times10^{-2}$ & $-0.704\times10^{-4}$ & $0.645\times10^{-6}$ &$-0.211\times10^{-8}$ &-5137&-12194&-990&-271&-273&-278&-284&-290&-294&-295\\
\end{longtable}

\end{document}